\documentclass[12pt,reqno]{amsart}
\usepackage{fullpage}
\usepackage{amsfonts}
\usepackage{amssymb}
\usepackage{times}
\usepackage{graphicx}
\usepackage{subfig}
\usepackage{caption}
\usepackage{xcolor}
\usepackage{indentfirst}
\usepackage{color}
\usepackage{colortbl}
\usepackage{multirow}
\usepackage{booktabs}
\definecolor{tabcolor}{rgb}{0.71,0.49,0.86}
\definecolor{lavender}{rgb}{0.71, 0.49, 0.86}
\definecolor{electricviolet}{rgb}{0.56, 0.0, 1.0}
\graphicspath{{figures/}} 
\newtheorem{theorem}{Theorem}[section]

\newtheorem{lemma}[theorem]{Lemma}
\newtheorem{proposition}[theorem]{Proposition}
{ \theoremstyle{remark} }

\begin{document}

\title[Short version of title]{A robust statistical method for Genome-wide association analysis of human copy number variation}
\author{Han Wang}
\address{(Han Wang) School of Mathematical Sciences and Center for Statistical Science, Peking University}
\email{wanghanmath@pku.edu.cn}
\author{Changhu Wang}
\address{(Changhu Wang) School of Mathematical Sciences, Peking University}
\email{wangch156@pku.edu.cn}
\author{Linjie Wu}
\address{(Linjie Wu) Institute of Hematology and Blood Disease Hospital, Chinese Academy of Medical Sciences and Peking Union Medical College}
\email{wulinjie@ihcams.ac.cn}
\author{Ruibin Xi*}
\address{(Ruibin Xi) School of Mathematical Sciences, Center for Statistical Science and Department of Biostatistics, Peking University}
\email{ruibinxi@math.pku.edu.cn}

\maketitle

\begin{abstract}
Conducting genome-wide association studies (GWAS) in copy number variation (CNV) level is a field where few people involves and little statistical progresses have been achieved, traditional methods suffer from many problems such as batch effects, heterogeneity across genome, leading to low power or high false discovery rate. We develop a new robust method to find disease-risking regions related to CNV's disproportionately distributed between case and control samples, even if there are batch effects between them, our test formula is robust to such effects. We propose a new empirical Bayes rule to deal with overfitting when estimating parameters during testing, this rule can be extended to the field of model selection, it can be more efficient compared with traditional methods when there are too much potential models to be specified. We also give solid theoretical guarantees for our proposed method, and demonstrate the effectiveness by simulation and realdata analysis.
\end{abstract}
	
	\section{Introduction}
	Copy number variation (CNV) detection is of great significance in understanding the mechanism of some severe diseases. Many studies have reported the strong correlation between CNV and diseases ranging from obesity to complex neurological diseases such as Parkinson, Alzheimer, mental retardation and schizophrenia to cancer. Zhang et al. \cite{Zhang F} gave a comprehensive summary about CNV in human health, disease, and evolution. Although studies investigating accurate copy number states have a rapid growth in the past few years \cite{Carter N P, Zhao M}, there is little devotion to developing robust statistical methods for identifying significant associations between CNVs and diseases. Meanwhile, the majority of existing genome-wide association studies (GWAS) paid their attentions to associations between single nucleotide polymorphisms (SNPs) and a certain disease, jumping out of the field of SNP-level seldomly. In this paper, we aim to develop an effective and robust statistical method to conduct GWAS in CNV level, extracting meaningful disease-risking CNV regions.
	
	Statistical methods developed to enhance the power of detecting CNVs emerge prominently in recent years, we can roughly classify them into two types. The former is to perform the detection on one chromosome at a time, for the segments possessing CNVs, the corresponding signals appear to present abrupt rise or drop, thus it can be recognized as a change-point problem. Traditional change-point detection methods like maximum likelihood based, sliding window based, $l_1$-penalization based \cite{Harchaoui Z} or graph based can be applied. Remarkably, one of the most popular maximum likelihood based algorithm is Circular Binary Segmentation (CBS) \cite{Olshen A} , which splice two ends of the chromosome to make it a circle and then test if there exist a CNV segment with a significantly different mean from the remaining part of the circle. The other statistical methods regarding to CNV detection is to pool information of all samples together. By scanning a shared segment simultaneously across multiple sequences, the power can be improved substantially. Zhang et al. \cite{Zhang N R} generalized the CBS statistic into the multiple sample case, the corresponding test statistic is simply a summation of the CBS statistic of a single sample. It's not hard to imagine that despite the enhanced power in finding rare CNV segments, this generalization inherited some shortcomings of the CBS algorithm. Jeng et al. \cite{Jeng X J} modelled each CNV region as a 2-class mixture Gaussian distribution, one class with mean zero represents those samples having normal copy numer 2, the other represents those CNV-carriers in this region with nonzero mean, then they established a test statistic closely relating to standardized uniform empirical process based statistics to pool information. On account of the heterogeneity of CNVs, different samples may possess different kinds of CNVs on the same region, there is credible doubt about the insufficiency of 2 classes in this model. Moreover, these two multiple-sample approaches described above developed their models based on the assumption that the boundaries of a certain CNV region across samples are the same, when handling real data, this assumption is rarely satistified.
	
	Researches focusing on establishing statistical methods to investigate associations between CNVs and diseases are not so active in contrast. One natural way to do GWAS in CNV level is to assign a copy number (CN) state to every case or control sample at a specific region, and then conduct a Fisher's test or chi-square test based on a contingency table. To the best of our knowledge, Barnes’s work \cite{Barnes C} was a representative among the few statistical methods devoted to GWAS in CNV level, their work pointed out shortcomings of doing test based on contingency table described above, and they adopted a likelihood-based approach to test whether there is significant correlation between copy number and case-control status. But it's worth noting that when using generalized linear model (GLM) to describe the connection between signal and copy number, they used the same set of parameters for case and control data, neglecting the possible batch effects, so concise and efficient tools to deal with batch effects are in need under this circumstance. In addition, this method focused on the test of a given CNV region and depended on other methods to pick up suspicious regions.
	
	In this paper, we propose a new testing model to deal with case-control whole genome data, with the aim of finding disease-susceptibility CNV regions and their corresponding genes, handling issues arised from batch effects and heterogeneity across genome simultaneously. When estimating parameters, we develop an empirical Bayes framework to avoid overfitting, this framework can be regarded as a new tool for model seletion, we will show the advantage of our method over Bayesian information criterion (BIC) under some conditions and prove consistency of our resulting estimation. Notably, our model can pool the samples' information together, having the ability to capture short segments without boundaries' accordance assumption. We also show the effectiveness of our method in both simulation and real data analysis.
	
	\section{Methods}
	\subsection{Model formulation}
	
	Suppose we have 2 groups of whole genome data, one for case and the other for control. Both the case and control data can be represented in a matrix form: $\{x_{it}; i=1,\cdots,N_1,t=1,\cdots,T\}$ and  $\{y_{jt}; j=1,\cdots,N_2,t=1,\cdots,T\}$ denote the case and control information separately, where $x_{it}$ and $y_{jt}$ are observations at location $t$ for the $i$th sample of case and $j$th sample of control separately, usually they are in the format of the log 2 transform of the copy number ratio (the ratio of the sample and the reference DNA). Here $N_1$ is the number of case samples and $N_2$ is that of control samples, $T$ is the number of locations in a sequence. Unlike traditional models paying attention to the distribution of a certain row of the data matrix, which is a certain sample sequence, we are interested in the distribution of a certain column of the data matrix, which is a particular location in the sequence for all samples, if at a location the distribution of case is signifiantly different from that of control,  this point may be with suspicion of being disease related. 
	
	As is commonly recognized, we assume that for each sample, taking case sample for an illustration, the observed values $x_i=\{x_{it}, t=1,\cdots,T\}$ are distributed as Gaussian and mutually independent. Furthermore, we assume 5 types of CNVs: 2 copies deletion (CN=0), 1 copy deletion (CN=1), normal (CN=2), 1 copy duplication (CN=3), 2 copies dupilication (CN=4). Since observations of each CNV type accumulate around a particular value, for each location $t$ in the sequence, we obtain a mixture Gaussian distribution: 
	
	\[
X_{it}\sim\displaystyle{\sum_{k=1}^5}\alpha_{tk}^d N(\mu_{tk}^d,(\sigma_{tk}^{d})^2) ,\ i=1,\cdots,N_1\ \mbox{for case samples at location}\ t.
	\]
	\[
	Y_{jt}\sim\displaystyle{\sum_{k=1}^5}\alpha_{tk}^c N(\mu_{tk}^c,(\sigma_{tk}^{c})^2),\ j=1,\cdots,N_2\ \mbox{for control samples at location}\ t.	
	\]

	Where $N(\mu_{tk}^d,(\sigma_{tk}^{d})^2)$ ($k=1,\cdots,5$) correspond to distribution of cluster CN = $1,\cdots,5$ separately for case sample at location $t$, $\alpha_{tk}$ ($k=1,\cdots,5$) are the 5 clusters' proportion at location $t$, we refer to $\alpha_{tk}^d$, $\mu_{tk}^d$ and $(\sigma_{tk}^d)^2$ ($k=1,\cdots,5$) as proportion parameters, location parameters and scale parameters for convenience, illustrations of control model are analogous with that of case.
	
	So far, a natural way of grasping the region we are interested in is to test the equivalence of the two Gaussian mixtures of case and control at a particular point, conventional testing approaches regarding with distribution equivalence can be applied directly. Unfortunately, because of the variety of experimental platforms, unmeasured variables ranging from quality of equipments to technicians and other confounding factors, the Gaussian distributions related to the same cluster for case and control samples may not be the same. Under this circumstance, even for a location without CNV for both case and control samples, the test concerned with distribution equivalence has the potential of claiming a significant outcome. To overcome these effects, we propose a test model as follows:\\
	\begin{equation}
	\begin{split}
	&H_0^t: \alpha_{tk}^d=\alpha_{tk}^c\  \mbox{for all}\ k=1,\cdots,5\ \mbox{at location}\ t\\
	&H_1^t:\alpha_{tk}^d\neq\alpha_{tk}^c\  \mbox{for some}\ k=1,\cdots,5\ \mbox{at location}\ t
	\end{split}
	\end{equation}
	
	Taking notice of the continuity of CNVs,  which means that deletion or duplication occurs on a continuous segment of nucleotide bases but not a single base, we have confidence to believe that the distributions between neighboring locations are the same unless there exists a change point of CN state for some samples in these neighboring locations. So we can increase the testing power by performing test bin by bin rather than location by location, to be specific, we split the whole sequence into bins with the same size, then in each bin of length $p$, there are two multivariate Gaussian mixtures for case and control seperately:
\[
\mathbf{X}_{ib}\sim\displaystyle{\sum_{k=1}^5}\alpha_{bk}^d N(\mu_{bk}^d\mathbf{1}_p,(\sigma_{bk}^{d})^2\mathbf{I}_p),\ i=1,\cdots,N_1\ \mbox{for case samples at bin}\ b.
\]
\[
\mathbf{Y}_{jb}\sim\displaystyle{\sum_{k=1}^5}\alpha_{bk}^c N(\mu_{bk}^c\mathbf{1}_p,(\sigma_{bk}^{c})^2\mathbf{I}_p),\ j=1,\cdots,N_2\ \mbox{for control samples at bin}\ b.
\]
	where $\mathbf{1}_p$ denotes a vector of length $p$ with all elements setting to 1, $\mathbf{I_p}$ is a $p$-dimension identity matrix, $\mathbf{X}_{ib}$ and $\mathbf{Y}_{jb}$ are samples of case and control at bin $b$. The distribution related to each cluster is multivariate Gaussian with mean vector of the same value, mutual independence between locations leads to a diagonal covariance matrix, we further assume equivalence of variance for all univariate Gaussian in this bin regarding to the same CN cluster. The testing rule for each bin is the same as that for each location, which merely tests the equivalence of proportion rather than the whole distribution.
	
	A straightforward way of realizing the proportion test is to perform the likelihood ratio test, calculating the maximum likelihood under $H_0$ and the whole feasible region, getting a likelihood ratio test statistic and claiming a p-value taking chi-square distribution as a reference. 
	\subsection{Setting a prior for CN-specific-mean}
	
	A new obstacle in front is overfitting. For instance, locations at which all samples have normal copy number or only deletions appear, the 5-cluster mixture Gaussian is obviously an overfitting model. To overcome this issue and make our model applicable and robust to all multi-sample CNV situations, we propose an empirical Bayes method to establish a prior for $\mu_{bk}^d$ and $\mu_{bk}^c$ in the mixture Gaussian distributions, pinning the means on their deserved positions in some sense, by which we can also ensure that $\alpha_{bk}^d$ and $\alpha_{bk}^c$ which we used to test equivalence correspond to the same CN state of case and control samples. From here on, unless otherwise specified, we drop the bin index symbol $b$ for simplicity. 
	
	We set $\mu_{k}^c\sim N(\tau_{k}, (\sigma_{\tau k}^{c})^2)$, $\mu_{k}^d\sim N(\tau_{k},(\sigma_{\tau k}^{d})^2)$, $\ k=1,\cdots,5\ $for case and control CN-specific-mean at all bins. Where $\tau_k$ is the prior knowledge of the CN-specific-mean, for example, there are CNV regions declared by some other methods, we can set $\tau_k$ as the mean of all observations in CN cluster $k$ found by a certain method. Here we take $\tau_k=-1.3, -0.5, 0, 0.4, 0.73$ for $k=1,\cdots, 5$ separately according to our experiences. $(\sigma_{\tau k}^{c})^2$, $(\sigma_{\tau k}^{d})^2$ are the prior variance of $\mu_{bk}^c$ and $\mu_{bk}^d$ separately, controlling the ``status'' of prior mean $\tau_k$, when $(\sigma_{\tau k}^{c})^2$, $(\sigma_{\tau k}^{d})^2$ are small, the prior mean displays a dominating status in estimating the CN-specific-mean, whereas weakening the power of sample observations in estimation. When $(\sigma_{\tau k}^{c})^2$, $(\sigma_{\tau k}^{d})^2$ are large, there is little difference between this setting and no prior setting, leading to unavoidable overfitting and estimation bias. So we have to give a moderate prior variance adaptive to various CNV proportion settings, bin length and conceivable batch effects, taking account of sample size simultaneously. It is well-known that as the sample size goes to infinity, any prior distribution of order $O_p(1)$ will be degenerate compared with the distribution explained by data, so the choice of $(\sigma_{\tau k}^{c})^2$ and $(\sigma_{\tau k}^{d})^2$ is indeed of great importance, we will show more in detail in the next section and give some theoretical guarantees on its choice.
	
	Incorporating the empirical Bayes formula, we update the likelihood of case and control observations under $H_1$ separately as: \\
	\begin{equation}
	\begin{split}
	&l(\alpha_{1}^d,\cdots,\alpha_{5}^d,\mu_{1}^d,\cdots,\mu_{5}^d,(\sigma_{1}^{d})^2,\cdots,
	(\sigma_{5}^{d})^2|\mathbf{X}_{1},\cdots,\mathbf{X}_{N_1})\underset{k=1}{\stackrel{5}{\LARGE{\prod}}} g(\mu_{k}^d|\tau_k,(\sigma_{\tau k}^{d})^2)\\
	&=\underset{k=1}{\stackrel{5}{\LARGE{\prod}}} g(\mu_{k}^d|\tau_k,(\sigma_{\tau k}^{d})^2)\underset{i=1}{\stackrel{N_1}{\LARGE{\prod}}}\underset{k=1}{\stackrel{5}{\sum}}\alpha_{k}^d f(\mathbf{X}_{i}|\mu_{k}^d,(\sigma_{k}^{d})^2)\stackrel{\bigtriangleup}{=}\tilde{f}(\mathbf{X}_{1},\cdots,\mathbf{X}_{N_1}|\theta^d)
	\end{split}
	\end{equation}
	\begin{equation}
	\begin{split}
	&l(\alpha_{1}^c,\cdots,\alpha_{5}^c,\mu_{1}^c,\cdots,\mu_{5}^c,(\sigma_{1}^{c})^2,\cdots,
	(\sigma_{5}^{c})^2|\mathbf{Y}_{1},\cdots,\mathbf{Y}_{N_2})\underset{k=1}{\stackrel{5}{\LARGE{\prod}}} g(\mu_{k}^c|\tau_k,(\sigma_{\tau k}^{c})^2)\\
	&=\underset{k=1}{\stackrel{5}{\LARGE{\prod}}} g(\mu_{k}^c|\tau_k,(\sigma_{\tau k}^{c})^2)\underset{j=1}{\stackrel{N_2}{\LARGE{\prod}}}\underset{k=1}{\stackrel{5}{\sum}}\alpha_{k}^c f(\mathbf{Y}_{j}|\mu_{k}^c,(\sigma_{k}^{c})^2) \stackrel{\bigtriangleup}{=}\tilde{f}(\mathbf{Y}_{1},\cdots,\mathbf{Y}_{N_2}|\theta^c)
	\end{split}
	\end{equation}
	Where $g(\mu_{k}^d|\tau_k,\sigma_{\tau k}^2)$ is the density of distribution $N(\tau_k,\sigma_{\tau k}^2)$ valued on $\mu_{k}^d$, $f(\mathbf{X}_{i}|\mu_{k}^d,(\sigma_{k}^{d})^2)$ is the density of distribution $N(\mu_{k}^d\mathbf{1}_p,(\sigma_{k}^{d})^2\mathbf{I}_p)$ valued on $\mathbf{X}_{i}$, symbols under control condition are parallel with that of case samples, $\theta^d$ and $\theta^c$ denotes the overall parameters to be estimated in a bin.
	\subsection{EM updating rule}
	
	Mixture properties of observations in a specific bin make the solution to maximum likelihood estimation(MLE) by first-order derivation intractable. A general approach that can be applied to this kind of dataset is the EM algorithm. Taking case observations under the alternative hypothesis $H_1$ as an instance, there is an unobserved variable $Z_{i}^d$ behind every probe intensity observation $\mathbf{X}_{i}$ for each bin, here $Z_{i}^d$ is the true CN state for sample $i$ in a bin. The E-step of the EM algorithm returns a $Q$-function, which is a conditional expectation of the log-likelihood of the complete data$(\mathbf{X}_{1},\cdots,\mathbf{X}_{N_1},Z_{1}^d,\cdots,Z_{N_1}^d)$ given the observed data $(\mathbf{X}_{1},\cdots,\mathbf{X}_{N_1})$ and a current estimation of parameters $\theta^{(t)}$, which can also be regarded as a parameter-indexed variational lower bound of the likelihood function, we are going to choose a best lower bound, which is indexed by MLE. The M-step returns estimations of parameters which maximize the $Q$-function. Since conclusions are parallel under $H_1$ regarding to case and control, unless otherwise mentioned, we only present those of case sample. Concretely, in our setting, the log-likelihood of the complete data is
	\begin{equation}
	\begin{split}
	&\underset{i=1}{\stackrel{N_1}{\sum}}\{\underset{k=1}{\stackrel{5}{\sum}}[-\frac{p}{2}\ln({\sigma_{k}^{d})^2}-\frac{1}{2(\sigma_{k}^{d})^2}(\mathbf{X}_{i}-\mu_{k}^d\mathbf{1}_p)'(\mathbf{X}_{i}-\mu_{k}^d\mathbf{1}_p)]\mathbf{1}_{\{Z_{i}^d=k\}}\\
	&+\underset{k=1}{\stackrel{5}{\sum}}\mathbf{1}_{\{Z_{i}^d=k\}}\ln\alpha_{k}^d\}+\underset{k=1}{\stackrel{5}{\sum}}[-\frac{p}{2}\ln(\sigma_{\tau k}^d)^2-\frac{1}{2\sigma_{\tau k}^2}(\mu_{k}^d-\tau_k)^2]+C
	\end{split}
	\end{equation}
	Where $C$ is a constant term which is irrelevant to parameter estimation.
	
	The corresponding $Q$-function is simply to replace $\mathbf{1}_{\{Z_{i}^d=k\}}$ with $P(Z_{i}|\mathbf{X}_{i},\theta^{d}(t))$
	\begin{equation}
	\begin{split}
	Q(\theta^d|\theta^{d}(t))=&\underset{i=1}{\stackrel{N_1}{\sum}}\{\underset{k=1}{\stackrel{5}{\sum}}[-\frac{p}{2}\ln{(\sigma_{k}^{d})^2}-\frac{1}{2(\sigma_{k}^{d})^2}(\mathbf{X}_{i}-\mu_{k}^d\mathbf{1}_p)'(\mathbf{X}_{i}-\mu_{k}^d\mathbf{1}_p)]P(Z_{i}|\mathbf{X}_{i},\theta^{d}(t))\\
	&+\underset{k=1}{\stackrel{5}{\sum}}P(Z_{i}|\mathbf{X}_{i},\theta^{d}(t))\ln\alpha_{k}^d\}+\underset{k=1}{\stackrel{5}{\sum}}[-\frac{p}{2}\ln(\sigma_{\tau k}^d)^2-\frac{1}{2(\sigma_{\tau k}^{d})^2}(\mu_{k}^d-\tau_k)^2]
	\end{split}
	\end{equation}
	Hence the EM updating rule is 
	\begin{equation}
	\begin{split}
	&\mu_{k}^{d}(t+1)=\frac{p(\sigma_{\tau k}^d)^2\underset{i=1}{\stackrel{N_1}{\sum}}(\overline{\mathbf{X}}_{i}b_{ik}^{d}(t))+\tau_k (\sigma_{k}^{d}(t))^2}{p(\sigma_{\tau k}^d)^2(\underset{i=1}{\stackrel{N_1}{\sum}}b_{ik}^{d}(t))+(\sigma_{k}^{d}(t))^2}\\
	&(\sigma_{bk}^{d}(t+1))^2=\frac{\underset{i=1}{\stackrel{N_1}{\sum}}(\mathbf{X}_{i}-\mu_{k}^{d}(t+1)\mathbf{1}_p)'(\mathbf{X}_{i}-\mu_{k}^{d}(t+1)\mathbf{1}_p)b_{ik}(t)}{p\underset{i=1}{\stackrel{N_1}{\sum}}b_{ik}^{d}(t)}\\
	&\alpha_{k}^{d}(t+1)=\frac{\underset{i=1}{\stackrel{N_1}{\sum}}b_{ik}^{d}(t)}{N_1}
	\end{split}
	\end{equation}
	Where $p$ is the length of the bin, $\overline{\mathbf{X}}_{i}$ is the mean of the $i$-th case observations in a  bin, parameters updated in the $t$-th step are denoted by $(t)$ in their right side. $b_{ik}^{d}(t)=P(Z_{i}=k|\mathbf{X}_{i},\theta^{d}(t))$ is the conditional probability of the latent variable, i.e. the state of the $i$-th case sample in a bin belongs to CN cluster $k$ given observations under the current parameter estimation. The claim of the control sample is similar to that of case sample.
	
	It is noteworthy that the convergent EM may stop at a local maxima but not a global maxima, there are many literatures discussed about it, for example Jin et.al. \cite{Jin C} focused on theoretical guarantees of uniformly weighted mixtures of M isotropic Gaussians, they pointed out that in order to recover a global maximum with at least constant probability, the EM algorithm
	with random initialization must be executed at least $e^{\Omega(M)}$ times, so efficient initialization methods are strongly needed such as moment-based initialization. In this paper we adopt outcomes from ``\emph{mclust}" \cite{Scrucca L} as the initial value, which is a famous R package for classification. The function ``Mclust" in the package adopted BIC as the model selection criteria, offering us the parameters $(\alpha_k,\mu_k,\sigma_k^2)$ of each corresponding cluster, assigning each sample a cluster label as well. To accommodate our CNV data environment, we initialize our proportion parameter $\alpha_k$ according to the classification output of ``Mclust'', concretely, initialization of $\alpha_k$ is proportion of samples whose ``mclust" cluster's mean parameter is located in a specific interval, i.e. we can initialize a CN state of each sample based on outcomes returned by ``Mclust''. Once initialization of each sample's CN state has been ensured, we can calculate the mean and variance of each initialized cluster as our parameter initialization. In our setting, we initialize a sample to cluster "CN=0" if its cluster's mean parameter returned by ``Mclust" is located in interval $I_1\stackrel{\bigtriangleup}{=}(-3,-0.9]$ analogously, $I_2\stackrel{\bigtriangleup}{=}(-0.9,-0.3]$, $I_3\stackrel{\bigtriangleup}{=}(-0.3,0.25]$, $I_4\stackrel{\bigtriangleup}{=}\in(0.25,0.6]$, $I_5\stackrel{\bigtriangleup}{=}(0.6,3]$.
	
	From the updating rule of $\mu_{k}^d$, we can see that the posterior mean of cluster $k$ is a weighted average of the prior mean and the "data-descriptive" mean of cluster $k$, with weights $(\sigma_{k}^{d})^2/N_1$ and $p(\sigma_{\tau k}^d)^2\hat{\alpha}_{k}^d$ separately, here $\hat{\alpha}_{k}^d=\underset{i=1}{\stackrel{N_1}{\sum}}b_{ik}^{d}/N_1$. So we have confidence to expect that with suitable choice of prior parameters, when the bin size $p\rightarrow\infty$, the estimated parameters have consistent properties, even if some cluster $k$ doesn't exist, we can also expect its corresponding proportion $\alpha_k$ going to 0.
	
	Before showing our theoretical results, we introduce two important lemmas, the first lemma is from Theorem 3.2 proposed by D. Dacuaha-Castelle and E. Gassiat \cite{Dacunha-Castelle D}, due to space limitation, we only list the most important part of this theorem, detailed explanation and proof are available in the original, proof of the second can be found in the appendix section.
	\begin{lemma}
		Suppose $\mathcal{F}=(f_{\xi})_{\xi\in\Gamma}$ is a family of probability densities with respect to $v$, $\Gamma$ is a compact subset of $\mathbf{R}^s$ for some integer $s$, $\mathcal{G}_k$ is the set of all $k$ mixtures of densities of $\mathcal{F}$:
		\[
		\begin{aligned}
		\mathcal{G}_k=\{g_{\alpha,\pi}=&\underset{i=1}{\stackrel{k}{\sum}}\alpha_i\cdot f_{\xi^i}: \alpha=(\alpha_1,\cdots,\alpha_k), \pi=(\xi^1,\cdots,\xi^k),\\
		& \forall i=1,\cdots,k, \xi^i\in\Gamma, 0\leq\alpha_i\leq 1,\underset{i=1}{\stackrel{k}{\sum}}\alpha_i=1
		\}
		\end{aligned}
		\]
		If $X_1,\cdots,X_n$ are $n$ samples from a mixture of $k$ populations; that is $X_1,\cdots,X_n\sim g_0=\underset{l=1}{\stackrel{k}{\sum}}\pi_l^0 f_{\xi^{l,0}}$. Define for any $g$ in $\mathcal{G}_k$, 
		\[
		l_n(g)=\underset{i=1}{\stackrel{n}{\sum}}\log[g(X_i)]
		\]
		and the statistic
		\[
		T_n(k)=\mathop{\sup}\limits_{g\in\mathcal{G}_k}l_n(g)-l_n(g_0)
		\]
		Then under some mild assumptions, $T_n(k)$ converges in distribution to the variable $\frac{1}{2}\mathop{\sup}\limits_{d\in\mathcal{D}}\eta_d^2\mathbf{1}_{\eta_d\geq 0}$, here $\eta_d$ is the Gaussian process indexed by $\mathcal{D}$ with covariance that is the usual hilbertian product in $H$, $\mathcal{D}$ is the subset of the unit sphere of $H$ of functions of a specific form, $H$ is the Hilbert space $L^2(g_0v)$ and does not equal 1 uniformly.
	\end{lemma}
	\begin{lemma}
		We assume $d_k=\tau_{k+1}-\tau_k$ for $k\in\{1,\cdots,4\}$, and suppose there exists an $\zeta>0$ s.t. $|\mu_k^*-\tau_k|<\zeta$ for $k\in\{1,\cdots,5\}$, $\zeta<\mathop{min}\limits_{k\in\{1,\cdots,4\}}d_k/2$, then any nonidentical location permutation rule $\pi$ applied to $\theta^*$ such that $\pi(\pmb{\theta}^*)=\pmb{\theta}^{*\pi}$, we have
		\begin{equation}
		G_n(\pmb{\theta}^*)-G_n(\pmb{\theta}^{*\pi})=O(\frac{1}{\sigma_{\tau k}^2})>0,\ \forall\epsilon_1>0
		\end{equation}
		here we denote $\theta^*$ as the overall underlying true parameters $(\mu_k^*,(\sigma_k^{*})^2,\alpha_k^*)_{k=1}^5$, $G_n(\pmb{\theta})=\\
		G_n(\mu_1,\cdots,\mu_5)=\underset{k=1}{\stackrel{5}{\sum}}\log[g(\mu_k|\tau_k,\sigma_{\tau k}^2)]$ is the logarithm of the prior distribution.
	\end{lemma}
	\begin{theorem}
		Suppose parameters $(\mu_{k}^d,(\sigma_{k}^{d})^2,\alpha_{k}^d)$ for $k=1,\cdots,5$ lie in a closed bounded set $\Theta$. When $p\rightarrow\infty$, sample size $N_1$ is a fixed constant, $(\sigma_{\tau k}^d)^2$ are equivalent for $k=1,\cdots,5$, the order of $(\sigma_{\tau k}^d)^2$ is $O(1/m(p))$, where $m(p)$ is any increasing function of $p$ of order less than $O(p)$;
		\begin{itemize}
			\item[(a)]If the 5 clusters all exist in the sample with size $N_1$, the underlying true location and scale parameters are $(\mu_{k}^{d*},(\sigma_{k}^{d*})^2)_{k=1}^5$, with $\{\mu_{k}^{d*}\}_{k=1}^5$ satisfying $\mu_{1}^{d*}<\mu_{2}^{d*}<\cdots<\mu_{5}^{d*}$, we further assume $d_k=\tau_{k+1}-\tau_k$ for $k\in\{1,\cdots,4\}$, and suppose there exists an $\zeta>0$ s.t. $|\mu_{k}^{d*}-\tau_k|<\zeta$ for $k\in\{1,\cdots,5\}$, and $\zeta<\mathop{min}\limits_{k\in\{1,\cdots,4\}}d_k/2$, then we have estimators $(\hat{\mu}_{k}^d,(\hat{\sigma}_{k}^{d})^2,\hat{\alpha}_{k}^d)$ which maximize $\tilde{f}(\mathbf{X}_{1},\cdots,\mathbf{X}_{N_1}|\pmb{\theta}^d)$ in (2) converge to $(\mu_{k}^{d*},(\sigma_{k}^{d*})^2,\alpha_{k}^{d*})$ in probability for all $k=1, \cdots, 5$, where $\{\alpha_{k}^{d*}\}_{k=1}^5$ are the proportion of 5 clusters in the sample of size $N_1$;
			\item[(b)] If some cluster doesn't exist in the sample with size $N_1$, we can get the corresponding estimator $\hat{\alpha}_{k}^d\xrightarrow{p} 0$, $\hat{\mu}_{k}^{d}\xrightarrow{p}\tau_{k}$. Furthermore, if the existing cluster set is denoted as $K^*$, we have $(\hat{\mu}_{k}^d,(\hat{\sigma}_{k}^{d})^2,\hat{\alpha}_{k}^d)\xrightarrow{p}(\mu_{k}^{d*},(\sigma_{k}^{d*})^2,\alpha_{k}^{d*})$ for $\forall k\in K^*$, where $(\mu_{k}^{d*},(\sigma_{k}^{d*})^2)$ are also the true underlying location and scale parameters, with $\mu_{k}^{d*}$ satisfying smaller $\mu_{k}^{d*}$ corresponding to smaller index $k$, $|\mu_k^{d*}-\tau_k|<\zeta$ as described in (a), $\alpha_{k}^{d*}$ are the proportion of existing cluster $k\in K^*$ in the sample of size $N_1$.
		\end{itemize}
	\end{theorem}
	We also expect to attain some consistency outcomes as sample size tends to infinity under our empirical Bayes framework, thankfully, the consistency of BIC formula under some regularity conditions inspires much regarding to suitable choice of prior variance $(\sigma_{\tau k}^d)^2/(\sigma_{\tau k}^c)^2$. Keribin \cite{Keribin C} discussed the consistent estimation of BIC in selecting the order of mixture models, which implies that the difference between log-likelihood under overfitting setting and log-likelihood under true setting is no more than $O(\log(n))$. Furthermore, from a hypothesis testing perspective, suppose $k_1>k$, the likelihood ratio test (LRT) statistic for testing $H_0$: $k$ mixtures against $H_1$: $k_1$ mixtures can also be expressed as the difference of the 2 log likelihood described above. When $H_0$ holds, this LRT statistic tends to a chi-square distribution as $n\rightarrow\infty$. Intuitively, the order of $1/(\sigma_{\tau k}^d)^2$ or $1/(\sigma_{\tau k}^c)^2$ can be any increasing function of $n$ of order less than $O(n)$, equipping with this order is sufficient to reverse the tendency of overfitting.
	\begin{theorem}
		Suppose parameters $(\mu_{k}^d,(\sigma_{k}^{d})^2,\alpha_{k}^d)$ for $k=1,\cdots,5$ lie in a closed bounded set $\Theta$. When sample size $N_1\rightarrow\infty$, bin size $p$ is a fixed constant, $(\sigma_{\tau k}^d)^2$ are equivalent for $k=1,\cdots,5$, the order of $(\sigma_{\tau k}^d)^2$ is $O(1/m(N_1))$, where $m(N_1)$ is any increasing function of $N_1$ with order less than $O(N_1)$;
		\begin{itemize}
			\item[(a)]If the 5 clusters all exist and the underlying true parameters are $(\mu_{k}^{d*},(\sigma_{k}^{d*})^2,\alpha_{k}^{d*})_{k=1}^5$, with $\{\mu_{k}^{d*}\}_{k=1}^5$ satisfying $\mu_{1}^{d*}<\mu_{2}^{d*}<\cdots<\mu_{5}^{d*}$, we further assume $d_k=\tau_{k+1}-\tau_k$ for $k\in\{1,\cdots,4\}$, and suppose there exists an $\zeta>0$ s.t. $|\mu_k^*-\tau_k|<\zeta$ for $k\in\{1,\cdots,5\}$, $\zeta<\mathop{min}\limits_{k\in\{1,\cdots,4\}}d_k/2$, then estimators $(\hat{\mu}_{k}^d,(\hat{\sigma}_{k}^{d})^2,\hat{\alpha}_{k}^d)_{k=1}^5$ maximize $\tilde{f}(\mathbf{X}_{1},\cdots,\mathbf{X}_{N_1}|\pmb{\theta}^d)$ in (2) converge to true parameters $(\mu_{k}^{d*},(\sigma_{k}^{d*})^2,\alpha_{k}^{d*})$ in probability for all $k=1, \cdots, 5$;
			\item[(b)]If some cluster doesn't exist, the corresponding estimator $\hat{\alpha}_{k}^d\xrightarrow{p} 0$, $\hat{\mu}_{k}^{d}\xrightarrow{p}\tau_{k}$. Furthermore, if the existing cluster set is denoted as $K^*$, we have  $(\hat{\mu}_{k}^d,(\hat{\sigma}_{k}^{d})^2,\hat{\alpha}_{k}^d)\xrightarrow{p}(\mu_{k}^{d*},(\sigma_{k}^{d*})^2,\alpha_{k}^{d*})$ for $\forall k\in K^*$, where $(\mu_{k}^{d*},(\sigma_{k}^{d*})^2)$ are also the true underlying location and scale parameters, with $\mu_{k}^{d*}$ satisfying smaller $\mu_{k}^{d*}$ corresponding to smaller index $k$, $|\mu_k^{d*}-\tau_k|<\zeta$ as described in (a), $\alpha_{k}^{d*}$ are the proportion of existing cluster $k\in K^*$ in the sample.
		\end{itemize}
		
	\end{theorem}
	
	Moreover, when sample size and bin length both tend to infinity, similar consistency results can also be achieved.
	\begin{theorem}
		Under the same conditions as those in Theorem 2, apart from bin size $p\rightarrow\infty$, $N_1/e^p\rightarrow 0$ and the order of $\sigma_{\tau k}^2$ is $O(1/m(N_1p))$, where $m(N_1p)$ is any increasing function of $N_1$ and $p$ with order less than $O(N_1p)$, the same consistency conclusions can also be obtained as Theorem 2.
	\end{theorem}
	From the above 3 theorems we can guarantee consistency properties of our specific empirical Bayes estimators, with the empirical Bayes framework establishing a new prior scheme different from classic prior distributions and robust to overfitting. Furthermore, by arranging the prior mean in increasing order, letting prior variance vary with sample size and bin length, estimators $\{\hat{\mu}_k^d\}_{k=1}^5/\{\hat{\mu}_k^c\}_{k=1}^5$can be arranged increasingly as well, ensuring the proportions of case and control are tested correspondingly. Actually if not using this Bayes formula, BIC can also be applied to achieve satisfying estimators when the sample size is large enough, but when using BIC to implement model selection, the procedure needs to be executed at least $k$ times, where $k$ is the largest possible model size, for example in our model, we need to apply BIC at least 5 times to ensure the most suitable model. We can expect that when there are tens of thousands of models to be specify, as long as we have knowledge about the rough mean, and the means are not too close to each other, doing model selection by our method can be more convenient compared with BIC.
	
	To reduce calculation time, we conduct a conditional test, which means that we suppose the mean and variance of each CN cluster at each bin is known, which have been estimated under $H_1$ by the EM procedure. Consequently, what we only need to estimate under $H_0$ is the proportion parameters $\alpha_{k}^d$ and $\alpha_{k}^c$, which are the maximizers of $\tilde{f}(\mathbf{X}_{i},\cdots,\mathbf{X}_{N_1}|\pmb{\theta}^d)$ and $\tilde{f}(\mathbf{Y}_{i},\cdots,\mathbf{Y}_{N_2}|\pmb{\theta}^c)$ under the constrain that $\alpha_{k}^d=\alpha_{k}^c$ for $k=1,\cdots,5$ in a bin. The corresponding EM updating rule is 
	\begin{equation}
	\alpha_{k}^{d}(t+1)=\alpha_{k}^{c}(t+1)=\frac{\underset{i=1}{\stackrel{N_1}{\sum}}b_{ik}^{d}(t)+\underset{j=1}{\stackrel{N_2}{\sum}}b_{jk}^{c}(t)}{N_1+N_2}
	\end{equation}
	
	In conclusion, we obtain converged parameters $(\hat{\mu}_{k}^d,(\hat{\sigma}_{k}^{d})^2,\hat{\alpha}_{k}^d)_{k=1}^5\stackrel{\bigtriangleup}{=}\hat{\pmb{\theta}}_{H_1}^d$ and $(\hat{\mu}_{k}^c,(\hat{\sigma}_{k}^{c})^2,\hat{\alpha}_{k}^c)_{k=1}^5\\
	\stackrel{\bigtriangleup}{=}\hat{\pmb{\theta}}_{H_1}^c$ following EM updating rule (6) to represent maximum likelihood estimation (MLE) of case and control sample under $H_1$, representation of MLE under $H_0$ is simply substituting $\hat{\alpha}_{k}^d$ in $\hat{\pmb{\theta}}_{H_1}^d$, $\hat{\alpha}_{k}^c$ in $\hat{\pmb{\theta}}_{H_1}^c$ by $\hat{\alpha}_{kH_0}$, which is the converged value following EM updating rule (8), we denote MLE of case and control sample under $H_0$ as $\hat{\pmb{\theta}}_{H_0}^d$ and $\hat{\pmb{\theta}}_{H_0}^c$. For simplicity, we let $h(\mathbf{X}_i|\pmb{\theta})$ denote mixture Gaussian density function $\underset{k=1}{\stackrel{5}{\sum}}\alpha_k f(\mathbf{X}_i|\mu_k\mathbf{1}_p,\sigma_k^2\mathbf{I}_p)$ with $\pmb{\theta}\stackrel{\bigtriangleup}{=}(\mu_k,\sigma_k^2,\alpha_k)_{k=1}^5$, the likelihood ratio statistic which we used to test significant bins now can be formulated as:
	\[
	\begin{split}
	\Lambda&=\underset{i=1}{\stackrel{N_1}{\sum}}\log[h(\mathbf{X}_i|\hat{\pmb{\theta}}_{H_1}^d)]+\underset{j=1}{\stackrel{N_2}{\sum}}\log[h(\mathbf{Y}_j|\hat{\pmb{\theta}}_{H_1}^c)]\\
	&-\underset{i=1}{\stackrel{N_1}{\sum}}\log[h(\mathbf{X}_i|\hat{\pmb{\theta}}_{H_0}^d)]-\underset{j=1}{\stackrel{N_2}{\sum}}\log[h(\mathbf{Y}_j|\hat{\pmb{\theta}}_{H_0}^c)]
	\end{split}
	\]
	
	By the non-negativity of KL divergence, it's not hard to deduce that when sample sizes of case and control both tend to infinity, the power of our likelihood ratio test tends to 1.
	\begin{proposition}
		Suppose the true proportion parameters for case and control are $\alpha_{k}^d$ and $\alpha_{k}^c$, $k=1,\cdots,5$. If there exists a constant $c_b>0$, $\exists k\in\{1,\cdots,5\}$, s.t. $|\alpha_{k}^d-\alpha_{k}^c|>c_b$, when sample size $N_1\rightarrow\infty$, $N_2\rightarrow\infty$, the type II error of likelihood ratio test tends to 0.
	\end{proposition}
	\subsection{Deal with heterogeneity across genome}
	
	Let's revisit our model formulation in section 2.1, where we model 2 isotropous multivariate Gaussian mixtures for case and control samples, which means that for each CN cluster, its mean and variance parameters are the same across locations in the bin. But due to GC-content, sequencing bias or other factors, for each bin of length $p$, it's more reasonable to replace $N(\mu_{k}^d\mathbf{1}_p,(\sigma_{k}^{d})^2\mathbf{I}_p)$ by $N(\pmb{\mu}_{k}^d,Diag((\pmb{\sigma}_{k}^{d})^2))$ to account for heterogeneity across genome, where $\pmb{\mu}_{k}^d$ and $(\pmb{\sigma}_{k}^{d})^2$ are two p-dimensional vectors, thus leading to 2 updated multivariate Gaussian mixtures for case and control sample:
	\[
	\mathbf{X}_{i}\sim\displaystyle{\sum_{k=1}^5}\alpha_{k}^d N(\pmb{\mu}_{k}^d,Diag((\pmb{\sigma}_{k}^{d})^2)),\ i=1,\cdots,N_1\ for\ case\ samples\ in\ a\ bin
    \]
    \[
	\mathbf{Y}_{j}\sim\displaystyle{\sum_{k=1}^5}\alpha_{k}^c N(\pmb{\mu}_{k}^c,Diag((\pmb{\sigma}_{k}^{c})^2)),\ j=1,\cdots,N_2\ for\ control\ samples\ in\ a\ bin  
    \]
    	
	Accordingly, we set a Gaussian prior for every element in the mean parameter $\pmb{\mu}_{k}^d$, $\pmb{\mu}_{k}^c$ for $k=1,\cdots,5$, take case samples for an example, the prior density of $\pmb{\mu}_{k}^d$ is 
	\[
	g(\pmb{\mu}_{k}^d)=\underset{j=1}{\stackrel{p}{\LARGE{\prod}}} g(\mu_{kj}^d|\tau_k,(\sigma_{\tau k}^{d})^2)
	\]
	
	Under this updated empirical Bayes framework, take case samples as an instance, the likelihood used to conduct parameter estimation in a bin of length $p$ can be reformulated as:
	\begin{equation}
	\begin{split}
	&\tilde{f}(\mathbf{X}_{1},\cdots,\mathbf{X}_{N_1}|\pmb{\theta}^d)\stackrel{\bigtriangleup}{=}l(\theta^d|\mathbf{X}_{1},\cdots,\mathbf{X}_{N_1})\underset{k=1}{\stackrel{5}{\LARGE{\prod}}}g(\pmb{\mu}_{k}^d)\\
	=&\underset{j=1}{\stackrel{p}{\LARGE{\prod}}}\underset{k=1}{\stackrel{5}{\LARGE{\prod}}}g(\mu_{kj}^d|\tau_k,(\sigma_{\tau k}^{d})^2)\underset{i=1}{\stackrel{N_1}{\LARGE{\prod}}}\underset{k=1}{\stackrel{5}{\sum}}\alpha_{k}^d f(\mathbf{X}_{i}|\pmb{\mu}_{k}^d,(\pmb{\sigma}_{k}^{d})^2)\\
	\end{split}
	\end{equation}
	Analogous to what we have reached in section 2.2 and 2.3, the EM updating rule is
	\begin{equation}
	\begin{split}
	&\mu_{kj}^{d}(t+1)=\frac{(\sigma_{\tau k}^{d})^2\underset{i=1}{\stackrel{N_1}{\sum}}(X_{ij}b_{ik}^{d}(t))+\tau_k(\sigma_{kj}^{d}(t))^2}{(\sigma_{\tau k}^{d})^2\underset{i=1}{\stackrel{N_1}{\sum}}b_{ik}^{d}(t)+(\sigma_{kj}^{d}(t))^2}\\
	&(\sigma_{kj}^{d}(t+1))^2=\frac{\underset{i=1}{\stackrel{N_1}{\sum}}(X_{ij}-\mu_{kj}^{d}(t+1))^2b_{ik}^{d}(t)}{\underset{i=1}{\stackrel{N_1}{\sum}}b_{ik}^{d}(t)}\\
	&\alpha_{k}^{d}(t+1)=\frac{\underset{i=1}{\stackrel{N_1}{\sum}}b_{ik}^{d}(t)}{N_1}\\
	\end{split}
	\end{equation}
	for $k=1,\cdots,5$, $j=1,\cdots,p$, where $X_{ij}$ is the $j$-th observation of case sample $i$ in a bin of length $p$, $b_{ik}^{d}(t)=P(Z_{i}=k|\mathbf{X}_{i},\pmb{\theta}^{d}(t))$ is denoted as the posterior probability of sample $i$ belongs to CN cluster $k$ under the current parameter estimation.
	
	Furthermore, under this anisotropic multivariate Gaussian mixture setting, with suitable choice of prior distribution, we can also obtain consistency conclusions as described in section 2.3.
	
	\begin{theorem}
		Suppose parameters $(\pmb{\mu}_{k}^d,(\pmb{\sigma}_{k}^{d})^2,\alpha_{k}^d)$ for $k=1,\cdots,5$ lie in a closed bounded set $\Theta$, $(\sigma_{\tau k}^d)^2$ are equivalent for $k=1,\cdots,5$, the order of $(\sigma_{\tau k}^d)^2$ is $O(1/m(N_1))$, where $m(N_1)$ is any increasing function of $N_1$ with order less than $O(N_1)$, when sample size $N_1\rightarrow\infty$, for $\forall p=1,2,3,\cdots$
		\begin{itemize}
			\item[(a)]if the 5 clusters all exist in the sample, the underlying true location and scale parameters are $(\pmb{\mu}_{k}^{d*},(\pmb{\sigma}_{k}^{d*})^2)_{k=1}^5$ with $\{\pmb{\mu}_{k}^{d*}\}_{k=1}^5$ satisfy $\mu_{1j}^{d*}<\mu_{2j}^{d*}<\cdots<\mu_{5j}^{d*}$ for $\forall j=1,\cdots,p$, we further assume $d_k=\tau_{k+1}-\tau_k$ for $k\in\{1,\cdots,4\}$, and suppose there exists an $\zeta>0$ s.t. $|\mu_{kj}^{d*}-\tau_k|<\zeta$ for $k\in\{1,\cdots,5\}$, $\forall j=1,\cdots,p$ and $\zeta<\mathop{min}\limits_{k\in\{1,\cdots,4\}}d_k/2$, then we have estimators $(\hat{\pmb{\mu}}_{k}^d,(\hat{\pmb{\sigma}}_{k}^d)^{2},\hat{\alpha}_{k}^d)$ maximize $\tilde{f}(\mathbf{X}_{1},\cdots,\mathbf{X}_{N_1}|\pmb{\theta}^d)$ in (9) converge to $(\pmb{\mu}_{k}^{d*},(\pmb{\sigma}_{k}^{d*})^2,\alpha_{k}^{d*})$ in probability for all $k=1, \cdots, 5$, where $\{\alpha_{k}^{d*}\}_{k=1}^5$ are the true proportions of 5 clusters in the sample; 
			
			\item[(b)]if some cluster doesn't exist in the sample, which forms a set $K^{*c}$, we can get the corresponding $\hat{\alpha}_{k}^d\xrightarrow{p} 0$, $\hat{\mu}_{kj}^{d}\xrightarrow{p}\tau_{k}$ for $\forall j=1,\cdots,p$ and $\forall k\in K^{*c}$. Furthermore, if the existing cluster set is denoted as $K^*$, we have  $(\hat{\mu}_{kj}^d,(\hat{\sigma}_{kj}^{d})^2,\hat{\alpha}_{k}^d)\xrightarrow{p}(\mu_{kj}^{d*},(\sigma_{kj}^{d*})^2,\alpha_{k}^{d*})$ for $\forall k\in K^*$, $j=1,\cdots,p$, where $(\pmb{\mu}_{k}^{d*},(\pmb{\sigma}_{k}^{d*})^2)$, $k\in K^*$ are also the true underlying location and scale parameters with $\mu_{kj}^{d*}$ satisfy smaller $\mu_{kj}^{d*}$ corresponding to smaller index $k$, $|\mu_{kj}^{d*}-\tau_k|<\zeta$ for $k\in K^*$, $\forall j=1,\cdots,p$ as described in (a), $\alpha_{k}^{d*}$ are the proportion of existing cluster $k\in K^*$ in the sample.
		\end{itemize}
	\end{theorem}
	\subsection{GWAS accounting for deletion or duplication only}
	
	In practice, when encountering significant regions after conducting GWAS in CNV level, we often want to know which type of variation accounts for the significance most. For example, in a certain significant region, if duplication is predominante compared with deletion and duplication carriers in cancer samples are more than those in normal samples, we suspect this region has potential characteristics of proto-oncogene. In this section we propose a new test formula which accounts for differences between case and control samples induced by deletion or duplication only.
	
	If we consider influence induced only by deletion, we can perform a likelihood ratio test of the following:
	\begin{equation}
	\begin{split}
	&H_0: \alpha_{1}^d=\alpha_{1}^c, \alpha_{2}^d=\alpha_{2}^c \\
	&H_1:\alpha_{1}^d\neq\alpha_{1}^c\  \mbox{or}\ \alpha_{2}^d\neq\alpha_{2}^c
	\end{split}
	\end{equation}
	where $H_1$ is the same as before, whereas $H_0$ only restricts equivalence of proportion on clusters corresponding to deletion, proportion parameters on other 3 clusters are unconstrained, so the likelihood ratio value stands for extent of deletion's influence. Duplication can be processed similar as deletion:
	\begin{equation}
	\begin{split}
	&H_0: \alpha_{4}^d=\alpha_{4}^c, \alpha_{5}^d=\alpha_{5}^c \\
	&H_1:\alpha_{4}^d\neq\alpha_{4}^c\  \mbox{or}\ \alpha_{5}^d\neq\alpha_{5}^c
	\end{split}
	\end{equation}
	where $\alpha_4^d$/$\alpha_4^c$, $\alpha_5^d$/$\alpha_5^c$ are proportion parameters corresponding to 2  duplication clusters. 
	
	Parameter estimation under $H_1$ can be obtained by iterating according to (10), for ease of calculation, we still conduct a conditional likelihood ratio test, assuming location parameters $\pmb{\mu}_{k}^d$, $\pmb{\mu}_{k}^c$ and scale parameters $(\pmb{\sigma}_{k}^{d})^2$, $(\pmb{\sigma}_{k}^{c})^2$ for $k=1,\cdots,5$ are the same under $H_0$ and $H_1$. The EM updating rule of proportion parameters under $H_0$ of (11) is:
	\begin{equation}
	\begin{split}
	&\alpha_{k}^{d}(t+1)=\alpha_{k}^{c}(t+1)=\frac{\underset{i=1}{\stackrel{N_1}{\sum}}b_{ik}^{d}(t)+\underset{j=1}{\stackrel{N_2}{\sum}}b_{jk}^{c}(t)}{N_1+N_2}; k=1,2\\
	&\alpha_{k}^{d}(t+1)=\frac{\underset{i=1}{\stackrel{N_1}{\sum}}b_{ik}^{d}(t)}{\lambda_1}, \lambda_1=(N_1+N_2)\frac{\underset{i=1}{\stackrel{N_1}{\sum}}\underset{k=3}{\stackrel{5}{\sum}}b_{ik}^{d}(t)}{\underset{i=1}{\stackrel{N_1}{\sum}}\underset{k=3}{\stackrel{5}{\sum}}b_{ik}^{d}(t)+\underset{j=1}{\stackrel{N_2}{\sum}}\underset{k=3}{\stackrel{5}{\sum}}b_{jk}^{c}(t)}; k=3,4,5\\
	&\alpha_{k}^{c}(t+1)=\frac{\underset{j=1}{\stackrel{N_2}{\sum}}b_{jk}^{c}(t)}{\lambda_2}, \lambda_2=(N_1+N_2)\frac{\underset{j=1}{\stackrel{N_2}{\sum}}\underset{k=3}{\stackrel{5}{\sum}}b_{jk}^{c}(t)}{\underset{i=1}{\stackrel{N_1}{\sum}}\underset{k=3}{\stackrel{5}{\sum}}b_{ik}^{d}(t)+\underset{j=1}{\stackrel{N_2}{\sum}}\underset{k=3}{\stackrel{5}{\sum}}b_{jk}^{c}(t)}; k=3,4,5
	\end{split}
	\end{equation}
	As for duplication, the updating rule is analogous to deletion:
	\begin{equation}
	\begin{split}
	&\alpha_{k}^{d}(t+1)=\alpha_{k}^{c}(t+1)=\frac{\underset{i=1}{\stackrel{N_1}{\sum}}b_{ik}^{d}(t)+\underset{j=1}{\stackrel{N_2}{\sum}}b_{jk}^{c}(t)}{N_1+N_2}; k=4,5\\
	&\alpha_{k}^{d}(t+1)=\frac{\underset{i=1}{\stackrel{N_1}{\sum}}b_{ik}^{d}(t)}{\lambda'_1}, \lambda'_1=(N_1+N_2)\frac{\underset{i=1}{\stackrel{N_1}{\sum}}\underset{k=1}{\stackrel{3}{\sum}}b_{ik}^{d}(t)}{\underset{i=1}{\stackrel{N_1}{\sum}}\underset{k=1}{\stackrel{3}{\sum}}b_{ik}^{d}(t)+\underset{j=1}{\stackrel{N_2}{\sum}}\underset{k=3}{\stackrel{5}{\sum}}b_{jk}^{c}(t)}; k=1,2,3\\
	&\alpha_{k}^{c}(t+1)=\frac{\underset{j=1}{\stackrel{N_2}{\sum}}b_{jk}^{c}(t)}{\lambda'_2}, \lambda'_2=(N_1+N_2)\frac{\underset{j=1}{\stackrel{N_2}{\sum}}\underset{k=1}{\stackrel{3}{\sum}}b_{jk}^{c}(t)}{\underset{i=1}{\stackrel{N_1}{\sum}}\underset{k=1}{\stackrel{3}{\sum}}b_{ik}^{d}(t)+\underset{j=1}{\stackrel{N_2}{\sum}}\underset{k=3}{\stackrel{5}{\sum}}b_{jk}^{c}(t)}; k=1,2,3
	\end{split}
	\end{equation}
	here $b_{ik}^{d}(t)=P(Z_i=k|\mathbf{X}_i,\theta^{d}(t))$ and $b_{jk}^{c}(t)=P(Z'_j=k|\mathbf{Y}_j,\theta^{c}(t))$ are posterior probability of sample $i$ in case and sample $j$ in control belong to CN cluster $k$ under the current parameter estimation correspondingly.
	\subsection{Merge}
	
	So far we can implement the "equivalence of proportion between case and control" test bin by bin along the whole chromosome, claiming the significant bins we are interested in. From theorem 1 we can see that in the isotropous multivariate Gaussian mixture case, the longer the bin is, the more accurate parameter estimation we can get, under the premise that each sample don't have CN state change point at each bin. Specifically, when it comes to the sparse signal detection problem, we can prove the following theorem.
	\begin{theorem}
		When the bin size $p$ and sample size $n$ tend to $\infty$, and assume $\frac{\log(n)}{p}\rightarrow 0$, $\epsilon=n^{-\beta}$ for some $\beta\in (0,1)$, $A=\sqrt{2r\log(n)}$ for $r\in (0,1)$, regarding to the signal detection test 
		\begin{equation}
		\begin{split}
		&H_0: \mathbf{X}_i\sim N(0\mathbf{1}_p,\sigma_1^2\mathbf{I}_p),\quad 1\leq i\leq n\\
		&H_1:\mathbf{X}_i\sim (1-\epsilon)N(0\mathbf{1}_p,\sigma_1^2\mathbf{I}_p)+\epsilon N(A\mathbf{1}_p,\sigma_2^2\mathbf{I}_p),\quad 1\leq i\leq n
		\end{split}
		\end{equation}
		the sum of type I and type II error of the likelihood ratio test tends to 0. 
	\end{theorem}
	It concludes that even if the signal is very sparse, when the bin size tends to $\infty$, the likelihood under $H_0$ and $H_1$ can be separated completely. The detection boundary of this kind of test has been analysed by Tony et al. \cite{Tony 1} , whereas they focused on univariate conditions, and the variance under $H_0$ by their assumption is simply set to 1.
	
	Owing to the increased accuracy brought by the longer bin, we want to further enhance our testing performance by merging. The adjacent bins possessing similar distributions can be merged together forming a larger bin, we expect test carried out in the larger bin will be more powerful. The intuition of evaluating whether two adjacent bins can be merged is summarized as following: we consider case samples first, suppose the $b$-th bin of length $p_1$ and the $(b+1)$-th bin of length $p_2$ can be merged together, we can estimate parameters in the newly formed bin $b-(b+1)$ by maximizing $\tilde{f}(\mathbf{X}_{1,b-(b+1)},\cdots,\mathbf{X}_{N_1,b-(b+1)}|\pmb{\theta}_{b-(b+1)}^d)$ in (9), here $\mathbf{X}_{1,b-(b+1)},\cdots,\mathbf{X}_{N_1,b-(b+1)}$ are $(p_1+p_2)$-dimensional sample, representing observations in the $b$-th and ($b+1$)-th bin. We further suppose the estimators we obtained in the merged bin is $\hat{\pmb{\theta}}_{b-(b+1)}^d=(\hat{\alpha}_{b-(b+1),1}^d,\cdots,\hat{\alpha}_{b-(b+1),5}^d,\hat{\pmb{\mu}}_{b-(b+1),1}^{d},\cdots,\\
	\hat{\pmb{\mu}}_{b-(b+1),5}^{d},(\hat{\pmb{\sigma}}_{b-(b+1),1}^{d})^2,\cdots,(\hat{\pmb{\sigma}}_{b-(b+1),5}^{d})^2)$, and pick up parameters corresponding to the $b$-th bin and $(b+1)$-th bin separately, suppose they are denoted as $\hat{\pmb{\theta}}_{b-(b+1)}^{d(1)}$ and $\hat{\pmb{\theta}}_{b-(b+1)}^{d(2)}$, with $\hat{\pmb{\theta}}_{b-(b+1)}^{d(i)}=(\hat{\alpha}_{b-(b+1),1}^d,\cdots,\hat{\alpha}_{b-(b+1),5}^d,\hat{\pmb{\mu}}_{b-(b+1),1}^{d(i)},\cdots,\hat{\pmb{\mu}}_{b-(b+1),5}^{d(i)},(\hat{\pmb{\sigma}}_{b-(b+1),1}^{d(i)})^2,\cdots,(\hat{\pmb{\sigma}}_{b-(b+1),5}^{d(i)})^2)$, $i=1,2$, where\\ $\hat{\pmb{\mu}}_{b-(b+1),1}^{d(1)},\cdots,\hat{\pmb{\mu}}_{b-(b+1),5}^{d(1)},(\hat{\pmb{\sigma}}_{b-(b+1),1}^{d(1)})^2,\cdots,(\hat{\pmb{\sigma}}_{b-(b+1),5}^{d(1)})^2$ are the first $p_1$-dimension of $\hat{\pmb{\mu}}_{b-(b+1),1}^{d},\cdots,\\
	\hat{\pmb{\mu}}_{b-(b+1),5}^d,(\hat{\pmb{\sigma}}_{b-(b+1),1}^{d})^2,\cdots,(\hat{\pmb{\sigma}}_{b-(b+1),5}^{d})^2$ and $\hat{\pmb{\mu}}_{b-(b+1),1}^{d(2)},\cdots,\hat{\pmb{\mu}}_{b-(b+1),5}^{d(2)},(\hat{\pmb{\sigma}}_{b-(b+1),1}^{d(2)})^2,\cdots, (\hat{\pmb{\sigma}}_{b-(b+1),5}^{d(2)})^2$ are the last $p_2$-dimension part. If not merge, we can also get estimators $\hat{\pmb{\theta}}_b^d$ and $\hat{\pmb{\theta}}_{b+1}^d$ by maximizing $\tilde{f}(\mathbf{X}_{1b},\cdots,\mathbf{X}_{N_1b}|\pmb{\theta}_{b}^d)$ and $\tilde{f}(\mathbf{X}_{1,b+1},\cdots,\mathbf{X}_{N_1,b+1}|\pmb{\theta}_{b+1}^d)$ separately. If for every case sample, their background CN states are the same in these 2 adjacent bins, which is ideal for merging, we can expect the following value can be small:
	
	\[
	\begin{split}
	M_{b,b+1}^d&=\underset{i=1}{\stackrel{N_1}{\sum}}\log[h(\mathbf{X}_{i,b}|\hat{\pmb{\theta}}_{b}^d)]-\underset{i=1}{\stackrel{N_1}{\sum}}\log[h(\mathbf{X}_{i,b}|\hat{\pmb{\theta}}_{b-(b+1)}^{d(1)})]\\
	&+\underset{i=1}{\stackrel{N_1}{\sum}}\log[h(\mathbf{X}_{i,b+1}|\hat{\pmb{\theta}}_{b+1}^d)]-\underset{i=1}{\stackrel{N_1}{\sum}}\log[h(\mathbf{X}_{i,b+1}|\hat{\pmb{\theta}}_{b-(b+1)}^{d(2)})]
	\end{split}
	\]
	as mentioned above, $h(\mathbf{X}|\pmb{\theta})$ denotes mixture Gaussian distribution with parameter $\pmb{\theta}$.
	Situations in control samples can be analogous, we can obtain a similar $M_{b,b+1}^c$ determining whether it's suitable to merge or not in control samples. 
	
	If the background CN states are different in the 2 adjacent bins for some sample, the $(p_1+p_2)$-dimensional mixture Gaussian is the misspecified model, resulting in $\hat{\theta}_{bm}^{d(1)}$ and $\hat{\theta}_{bm}^{d(2)}$ are not fine estimators, the value $M_{b,b+1}^d$ or $M_{b,b+1}^c$ described above can be large. Theoretically, take case samples as an instance, we can give the order of $M_{b,b+1}^d$ when the $b$-th and $b+1$-th bins can or can't be merged together.
	
	\begin{theorem}
		Suppose proportion parameters $(\alpha_{b1}^d,\cdots,\alpha_{b5}^d)$ and $(\alpha_{(b+1),1}^d,\cdots,\alpha_{(b+1),5}^d)$ are the same in the $b$-th and $(b+1)$-th bin if and only if the background CN states are the same in these 2 bins for every sample. 
		\begin{itemize}
			\item[(a)]When the proportion parameters are the same in the $b$-th and $(b+1)$-th bin, the order of $M_{b,b+1}^d$ is $O_p(m(N_1p))$, where $p=p_1+p_2$ is the length of the merged bin $b-(b+1)$, $m(N_1p)$ is any increasing function of $N_1,p$ with order less than $O(N_1p)$.
			\item[(b)]If there exists a constant $c_b>0$, $\exists k\in\{1,\cdots,5\}$, s.t. $|\alpha_{bk}^d-\alpha_{(b+1),k}^d|>c_b$, the order of $M_{b,b+1}^d$ is $O_p(N_1)$.
		\end{itemize}
	\end{theorem}
	Form the above order estimation, we set the criterion of whether the 2 adjacent bins can or can't be merged together as: $M_{b,b+1}^d<\lambda_d \log(N_1p)$ and $M_{b,b+1}^c<\lambda_c \log(N_2p)$, then we will merge, else we will not merge. In practice, $\lambda_c$ and $\lambda_d$ can be chosen proportional to sample size, for example, if $N_2=2N_1$, we can set $\lambda_c=2\lambda_d$.
	
	In general, we carry out the merge process by the following iterating steps:
	\begin{itemize}
		\item[(1)]Calculate $M_{b,b+1}^d$ and $M_{b,b+1}^c$ for every adjacent bins after the bin-by-bin test.
		\item[(2)]Select adjacent bins with the smallest $M_{b,b+1}^d+M_{b,b+1}^c$ value to merge.
		\item[(3)]Update parameters in the newly formed bin by rule (10). 
		\item[(4)]Update $M_{b,b+1}^d$ and $M_{b,b+1}^c$ with respect to the newly formed bin and its adjacent bins.
		\item[(5)]Repeat steps 2-4 until there are no adjacent bins satisfying $M_{b,b+1}<\lambda_d \log(N_1p)$ and $M_{b,b+1}^c<\lambda_c \log(N_2p)$.
	\end{itemize}
	
	Notably, when there exists samples whose underlying CN states are not the same in the 2 adjacent bins, mixture Gaussian is no longer suitable, under the premise that each Gaussian corresponds with only one CN state. For instance, if in the $b$-th bin, $\mathbf{X}_{ib}\sim N(\mathbf{0}_{p_1},\mathbf{I}_{p_1})$, while in the ($b+1$)-th bin, $\mathbf{X}_{i,b+1}\sim 0.8N(\mathbf{0}_{p_2},\mathbf{I}_{p_2})+0.2N(0.4\mathbf{1}_{p_2},\mathbf{I}_{p_2})$, then $0.8N(\mathbf{0}_{p_1+p_2},\mathbf{I}_{p_1+p_2})+0.2N((\mathbf{0}_{p_1},0.4\mathbf{1}_{p_2}),\mathbf{I}_{p_1+p_2})$ is the true distribution for $\mathbf{X}_{i,b-(b+1)}$, but we regard this form as meaningless. When dealing with parameter estimation in merging, we can obtain a meaningful mixture Gaussian by enhancing the power of prior, until location parameter in each CN cluster is located in a reasonable region. According to our experience, location parameters of CN=0 and 1 are restricted to be smaller than -0.9 and -0.4; location parameters of CN=3 and 4 are supposed to be larger than 0.35 and 0.65.
	
	After merging, what we have in hand are longer bins with their corresponding parameters estimated simultaneously in the merging process, which means that parameters under $H_1$ have been well prepared. We only need to estimate proportion parameters under the constraint of $H_0$, that is we do the conditional test again as mentioned in section 2.3——conduct the "equivalence of proportion between case and control" test at each newly formed bin conditional on the known mean and variance parameters of each CN cluster. Estimation of proportion parameters under $H_0$ is analogous with the EM updating rule mentioned in (8). If we focus on single factor---deletion or duplication, rule (13) or (14) can be executed until convergence.
	\section{Simulation}
	We undertake simulation on 1000 case samples and 2000 control samples on $10^4$ base pairs (bp) sites, after generating 2 matrices of size 1000*$10^4$ and 2000*$10^4$, with each element sampled from $N(0,0.3^2)$, we pick up 30 segments to add signals as our interested CNV-GWAS regions. These 30 segments are equally separated into 5 different lengths---10 bp, 30bp, 50bp, 100 bp, 500 bp and 2 types of CNV---deletion and duplication.
	
	We designed 3 different case-control CNV carriers' proportion contrast settings for both deletion and duplication (table 1-6). Scenario 1 of deletion or duplication stands for a rare CNV setting, with low frequency of observing a CNV carrier, scenario 2 belongs to a middle frequency setting, while scenario 3 stands for a common CNV setting, with high frequency of observing a CNV carrier. We select carriers with a certain CN state at random and satisfy the proportion settings on table 1-6.
	
	To explore performance affected by batch effects between case and control, we generate some case and control sample with batch effects among them. Specifically, in a CNV-GWAS region, for a certain CN state, case and control are sampled from different Gaussian distribution. Parameters of the Gaussian distribution under different CN state are shown in table 7. If we add no batch effects, parameters under different CN state of case and control are sampled both according to the second row in table 7.
	
	We also want to kown the robustness of our method under different variance settings, for example, if there's no CNV carrier in a region, but the variance of this region is very large, there may be a tendency of increased false discovery rate. In the whole chromosome, besides the 30 CNV-GWAS regions, what's left are 31 normal regions, we  select 2 regions of length 20 bp and 200 bp on the 1st and 2nd normal regions separately,  they have no CNV carrier both in case and control samples, but we simulate case sample from a Gaussian distribution with larger variance. Concretely, in these 2 regions, case sample are from $N(0, 0.9^2)$, while control sample are from $N(0, 0.6^2)$, we call these 2 regions vary-var1 regions. For the remaining 29 normal regions, we pick up 14 regions to let variance vary not only between case and control, but also among different locations of this region, we call these 14 regions vary-var2 regions. For each vary-var2 region, we split it into equal 10 bp-length bins with index $1,\cdots,B$ if it has 10B bp, bins are drawn from the same distribution if they have the same remainder when their indexes divided by 3. Table 8 shows detailed sampling rules on 14 vary-var2 regions. If we add no different variance rule, elements from both vary-var1 and vary-var2 regions are sampled from $N(0,0.3^2)$.
	
	In total, we can do simulation under 12 different scenarios (CNV carrier's frequency\\ low/middle/high$\times$have/not have batch effects$\times$have/not have different variance settings). For each scenario, we randomly simulate 50 case and control data sets, they have the same 30 CNV-GWAS regions.
	
	We compared our method with several CNV calling methods---CBS \cite{Olshen A} , Median \cite{Tony 2} , multiCBS \cite{Zhang N R} . Circular binary segmentation (CBS) \cite{Olshen A} is a very popular and classic method in detecting CNVs on a single chromosome, Cai et al. (2012) \cite{Tony 2} performed a median transformation and then use the transformed data to establish a test statistic to call CNV. Zhang et al. \cite{Zhang N R} generalized CBS into multiple sample setting, aiming at detecting CNV regions shared between multiple samples. We applied these methods on our simulated case and control data sets. 
	
	Figure 1 shows the performance in terms of sensitivity and FDR of our method and other 3 methods mentioned above. As is well-known that sensitivity is the proportion of true positives among all real positive incidents, and FDR is the proportion of false positives among all detected incidents. We use the overlap of detected CNVs and designed CNVs to infer true and false discovered CNVs for each sample. For a putative CNV detected by a given method, if the overlapping region is more than half of a designed CNV and more than half of this putative CNV, this designed CNV is claimed to be detected in the current sample and we define it to be a true positive. If the overlapping region with any of the designed CNVs is less than half of the discovered CNV, we define it to be a false positive. 
	
	From figure 1 we can see that our method outperforms CBS and Median in all simulation scenarios, it performs better than multiCBS or comparable with multiCBS in terms of sensitivity. Our method is robust to batch effects and variance influences. In terms of sensitivity, multiCBS is comparable with our method in some scenarios, but it can be severely impacted by varying variance, when the variance of some regions are very large, it misaligns normal samples to other abnormal CN states and results in large FDR value. The power of CBS and Median is lower than multiCBS and our method in every scenario, but these 2 methods seems to be more robust to varying variance than multiCBS as shown in the FDR plot.
	
	As for our aim of finding significant regions in CNV-GWAS, we pick up regions with P value less than 0.05 after merging and regard these as CNV-GWAS significant regions. We compared these discovered regions with 30 designed CNV-GWAS regions, and evaluate our method's performance in terms of sensitivity and FDR as defined above, figure 2 shows sensitivity of our method with different lengths, batches and variance settings. Figure 3 shows performance in terms of FDR under different settings.
	
	As can be seen from figure 2, apart from some short CNV-GWAS regions in low frequency setting, our method can detect almost all significant regions, when there is no batch effect between case and control sample, no region has varying variance, short regions can be detected with larger power. Figure 3 shows that FDR under all circumstances approaches 0, varying variance seems to have a larger impact on the performance than batch effects. This result verifies effectiveness of our method again.

\begin{figure}
	\centering         
	\includegraphics[width=1.0\textwidth]{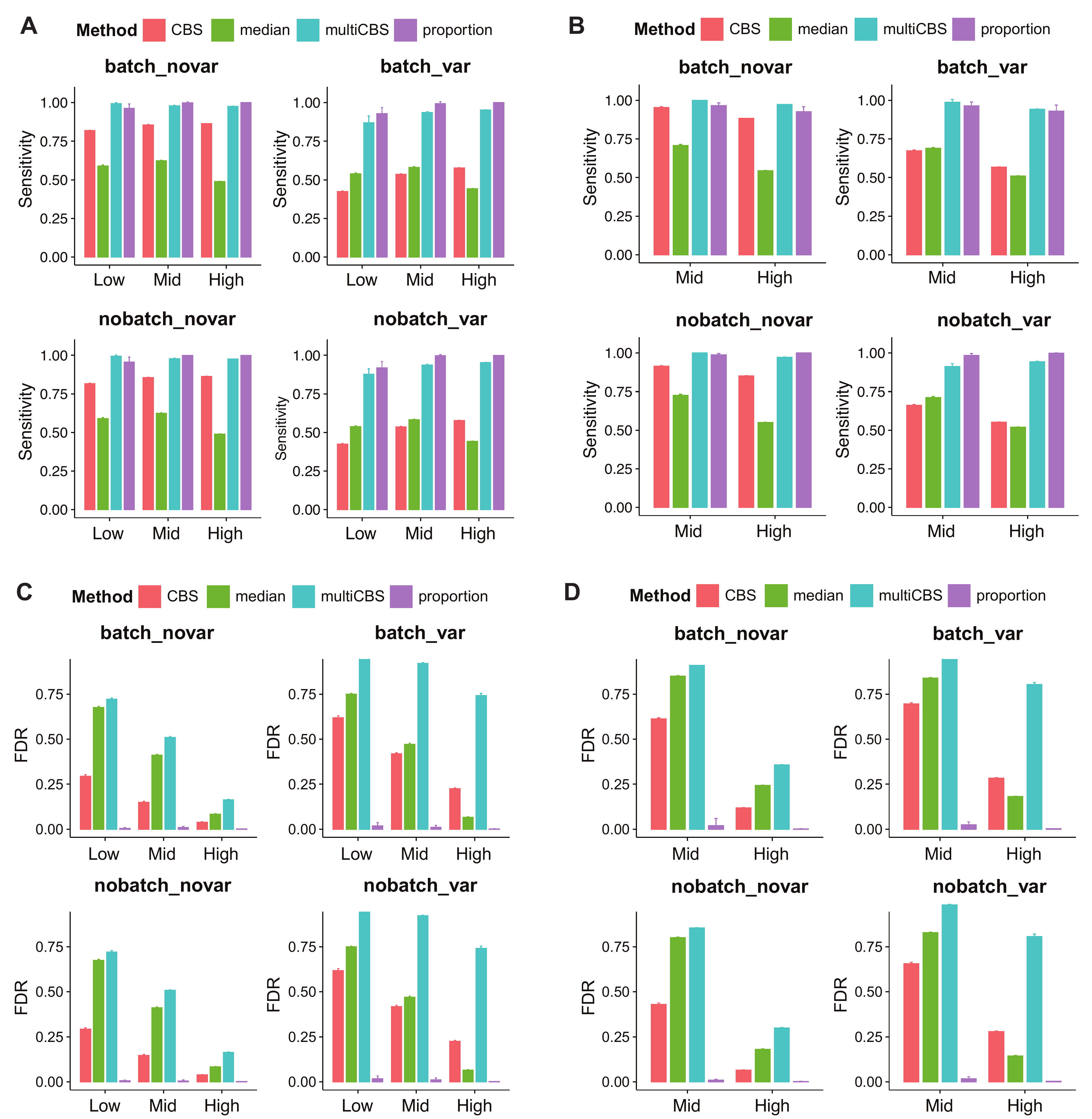}   
	\caption{Average sensitivity and FDR with error bar of different method under different settings, height of column denotes mean of sensitivity/FDR, height of error bar denotes mean+standard deviation of sensitivity/FDR. ``low", ``mid" and ``high" refer to low frequency, middle frequency and high frequency. (A) Sensitivity of different methods under different batches or variance settings of case samples. (B) Sensitivity of different methods under different batch or variance settings of control samples. (C) FDR of different methods under different batch or variance settings of case samples. (D) FDR of different methods under different batch or variance settings of control samples. }	
\end{figure}

\begin{figure}
	\centering         
	\includegraphics[width=1.0\textwidth]{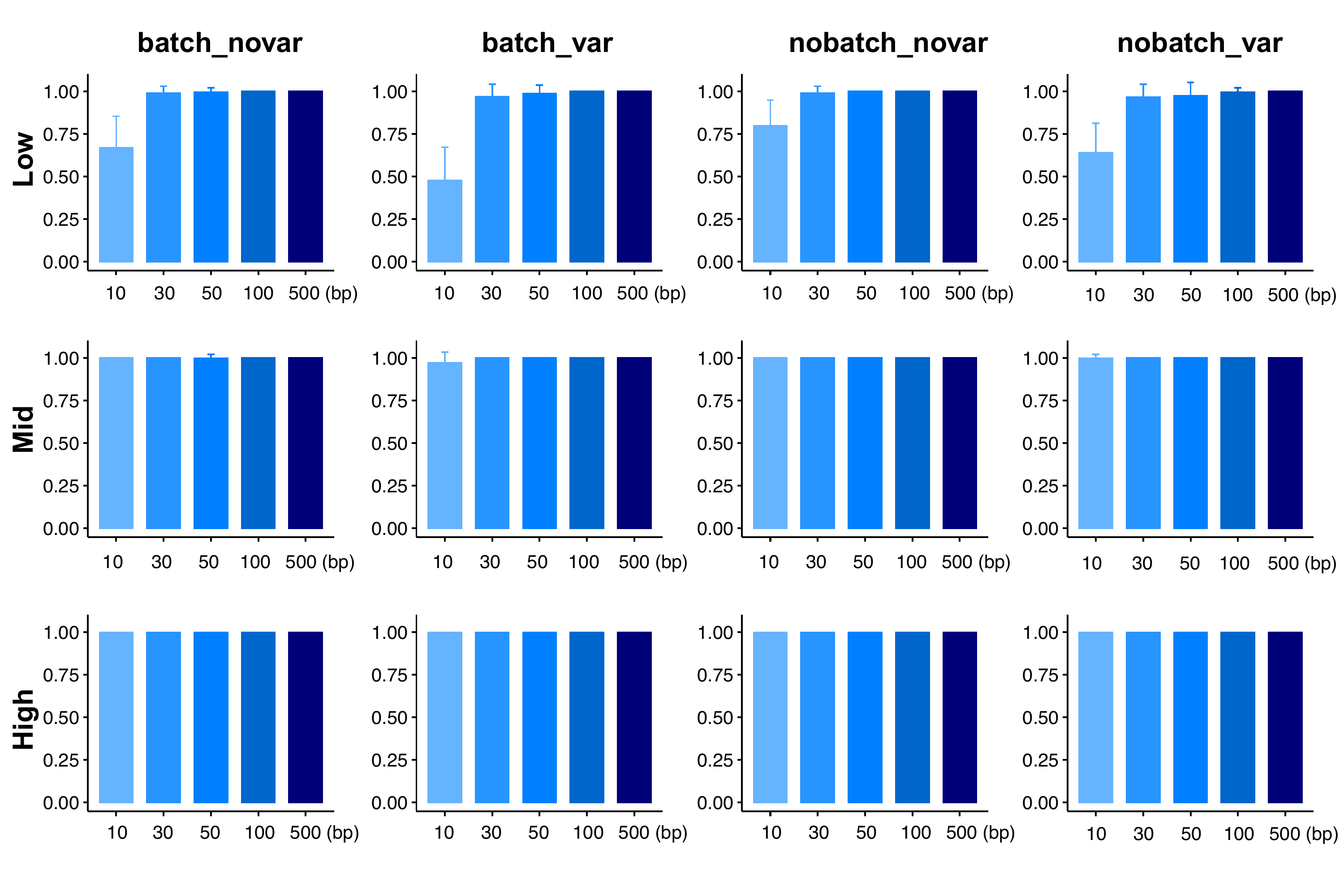}   
	\caption{Average sensitivity with error bar in detecting significant regions in CNV-GWAS under different frequencies, batches and variance settings of our method. Height of column denotes mean of sensitivity, height of error bar denotes mean+standard deviation of sensitivity. x-axis denotes different lengths of CNV, y-axis denotes sensitivity. ``low", ``mid'', and ``high'' refer to low frequency, middle frequency and high frequency.}	
\end{figure}

\begin{figure}
	\centering         
	\includegraphics[width=1.0\textwidth]{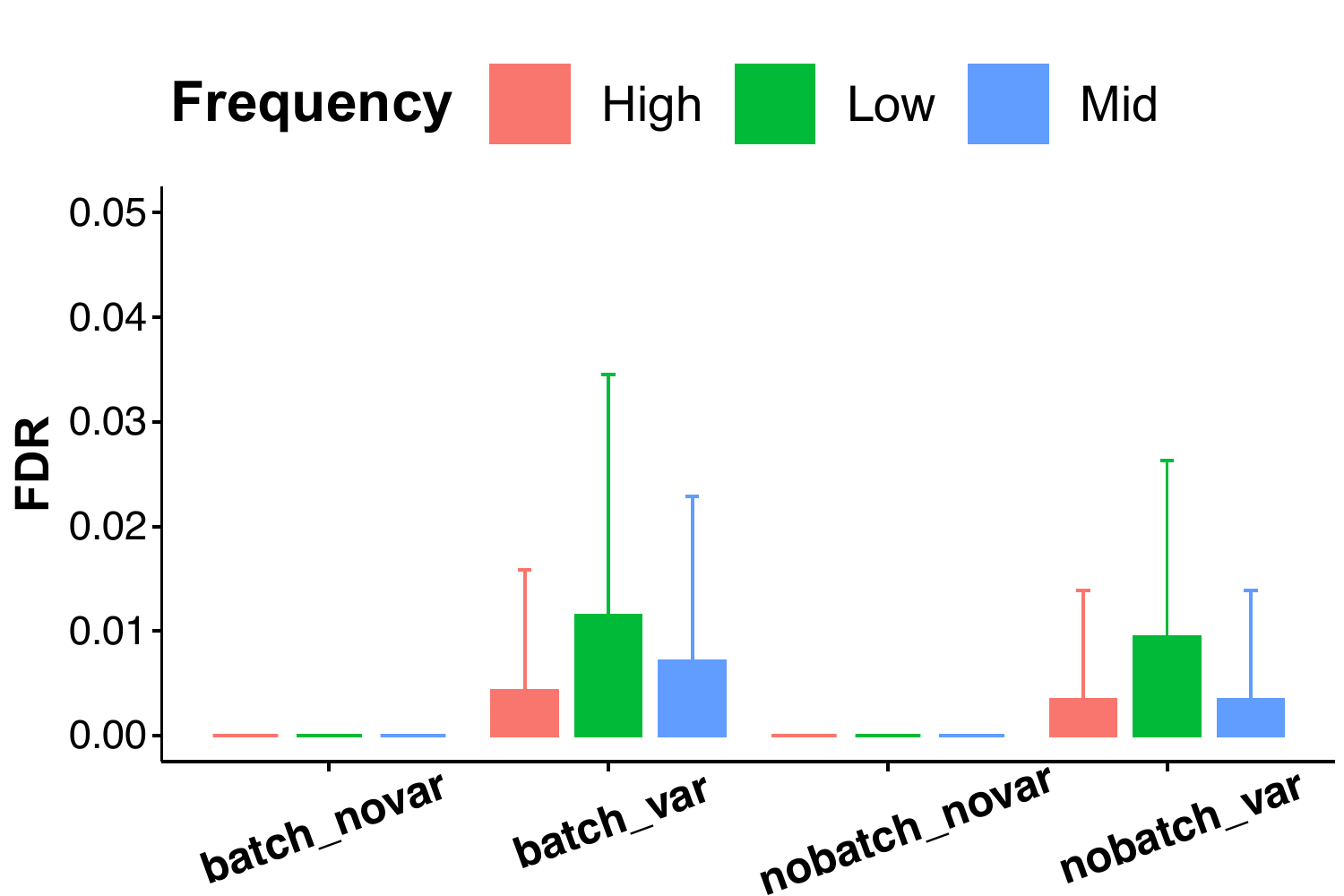}   
	\caption{Average FDR with error bar in detecting significant regions in CNV-GWAS under different frequencies, batches and variance settings of our method. Height of column denotes mean of FDR, height of error bar denotes mean+standard deviation of FDR. ``low", ``mid'', and ``high'' refer to low frequency, middle frequency and high frequency.}	
\end{figure}

\begin{figure}
	\centering         
	\includegraphics[width=1.0\textwidth]{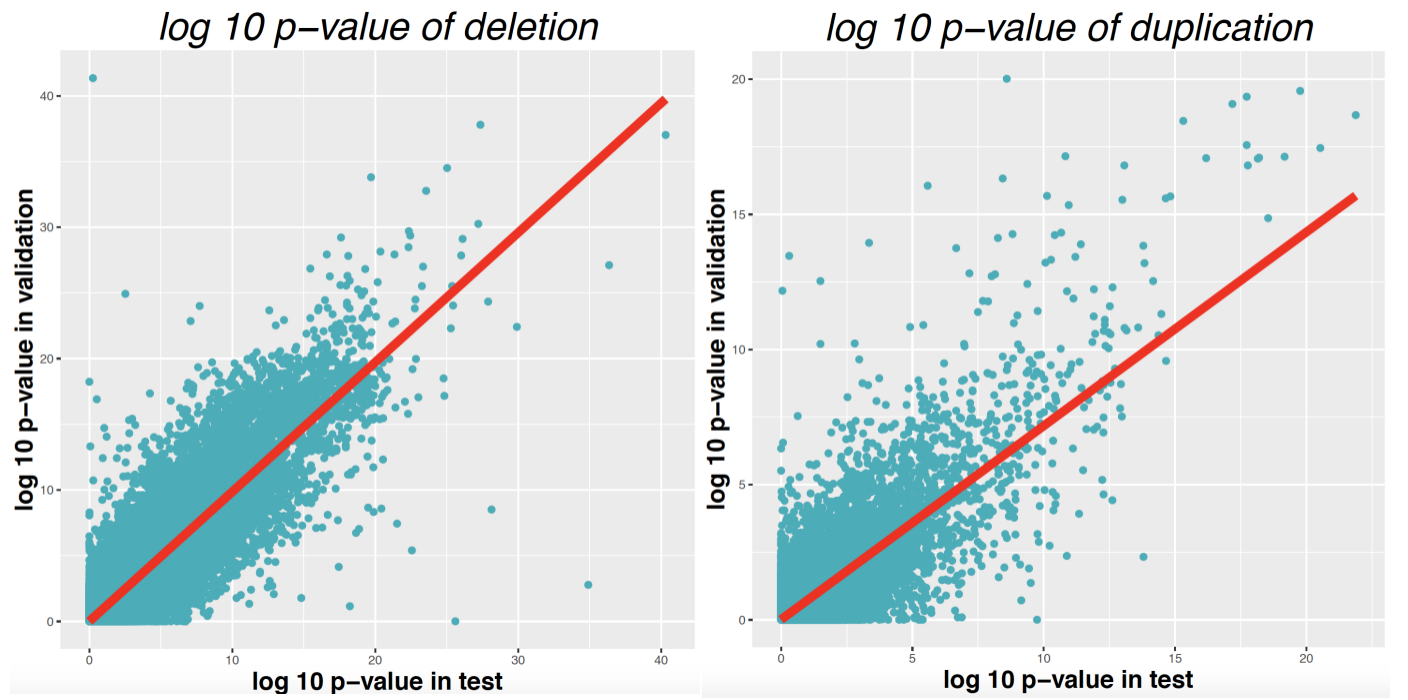}   
	\caption{Scatter plot of log 10 P-value of deletion or duplication on each segment after merging, left plot is deletion, right plot is duplication. Each point refers to a segment with abscissa and ordinate corresponding to log 10 P-value in test and validation set separately.}	
\end{figure}

\begin{table}[!htb]
	\begin{minipage}{.5\textwidth}
		\centering
		\label{tab:first}
		\begin{tabular}{ l l l l l l }	
			\hline
			{} & CN:0 & CN:1 & CN:2 & CN:3 & CN:4\\
			\hline
			case & 0 & 0.05 & 0.95 & 0 &0 \\
			\hline
			control& 0 & 0 & 1 & 0 &0 \\
			\hline
		\end{tabular}
		\caption{case-control CNV carrier's proportion contrast, deletion's scenario 1}
	\end{minipage}%
	\begin{minipage}{.5\textwidth}
		\centering
		\label{tab:second}
		\begin{tabular}{ l l l l l l }	
			\hline
			{} & CN:0 & CN:1 & CN:2 & CN:3 & CN:4\\
			\hline
			case & 0 & 0 & 0.95 & 0.05 &0 \\
			\hline
			control& 0 & 0 & 1 & 0 &0 \\
			\hline
		\end{tabular}
		\caption{case-control CNV carrier's proportion contrast, duplication's scenario 1}
	\end{minipage}
\end{table}

\begin{table}[!htb]
	\begin{minipage}{.5\textwidth}
		\centering
		\label{tab:first}
		\begin{tabular}{ l l l l l l }	
			\hline
			{} & CN:0 & CN:1 & CN:2 & CN:3 & CN:4\\
			\hline
			case & 0.03 & 0.1 & 0.87 & 0 &0 \\
			\hline
			control& 0.01 & 0.01 & 0.98 & 0 &0 \\
			\hline
		\end{tabular}
		\caption{case-control CNV carrier's proportion contrast, deletion's scenario 2}
	\end{minipage}%
	\begin{minipage}{.5\textwidth}
		\centering
		\label{tab:second}
		\begin{tabular}{ l l l l l l }	
			\hline
			{} & CN:0 & CN:1 & CN:2 & CN:3 & CN:4\\
			\hline
			case & 0 & 0 & 0.95 & 0.1 &0.03 \\
			\hline
			control& 0 & 0 & 0.98 & 0.01 &0.01 \\
			\hline
		\end{tabular}
		\caption{case-control CNV carrier's proportion contrast, duplication's scenario 2}
	\end{minipage}
\end{table}

\begin{table}[!htb]
	\begin{minipage}{.5\textwidth}
		\centering
		\label{tab:first}
		\begin{tabular}{ l l l l l l }	
			\hline
			{} & CN:0 & CN:1 & CN:2 & CN:3 & CN:4\\
			\hline
			case & 0.17 & 0.46 & 0.37 & 0 &0 \\
			\hline
			control & 0.08 & 0.28 & 0.64 & 0 &0 \\
			\hline
		\end{tabular}		
		\caption{case-control CNV carrier's proportion contrast, deletion's scenario 3}
	\end{minipage}%
	\begin{minipage}{.5\textwidth}
		\centering
		\label{tab:second}
		\begin{tabular}{ l l l l l l }	
			\hline
			{} & CN:0 & CN:1 & CN:2 & CN:3 & CN:4\\			
			\hline
			case & 0 & 0 & 0.37 & 0.46 &0.17 \\
			\hline
			control & 0 & 0 & 0.64 & 0.28 &0.08 \\
			\hline
		\end{tabular}
		\caption{case-control CNV carrier's proportion contrast, duplication's scenario 3}
	\end{minipage}
\end{table}

\begin{table}[h]
	\begin{tabular}{ l l l l l l }	
		\hline
		{} & CN=0 & CN=1 & CN=2 & CN=3 & CN=4\\		
		\hline
		$\mu$ of case &-1.3 & -0.4 & -0.13 & 0.3 & 0.63 \\
		\hline
		$\mu$ of control & -1.2 & -0.3 & 0.15 & 0.44 & 0.73 \\
		\hline
		$\sigma$ of case &0.5 & 0.13 & 0.3 & 0.18 & 0.4\\
		\hline
		$\sigma$ of control &0.52 & 0.14 & 0.31 & 0.2 & 0.42\\
		\hline
	\end{tabular}
	\caption{Gussian parameters on CNV-GWAS regions when there's batch effect}
\end{table}

\begin{table}[h]
	\begin{tabular}{ l l l l }	
		\hline
		{} & remainder=0 & remainder=1 & remainder=2 \\		
		\hline
		case & $N(0,0.3^2)$  & $N(0,0.6^2)$ & $N(0,0.9^2)$ \\
		\hline
		control & $N(0,0.6^2)$ & $N(0,0.9^2)$ & $N(0,0.3^2)$  \\
		\hline
	\end{tabular}
	\caption{Different Gaussian distributions from which locations in a bin are sampled, choice of distributions depend on the remainder of bin index divided by 3, this rule adapts to vary-var2 regions only.}
\end{table}
	\section{Real Data}
We applied our method to real data with 2042 esophagus cancer (ESCC) samples and 2060 normal (control) samples, these data are from array omparative genomic hybridization (aCGH) and their original form are CEL files. We pool case and control samples together and run the PennCNV-Affy procedure \cite{Wang K} to process raw CEL files, PennCNV-Affy can generate canonical genotype clusters and then convert signal intensity for each sample to log2 ratio values, we also implement a wave adjustment procedure \cite{Diskin} to filter biases caused by GC content. After processed by PennCNV-Affy, we have case and control matrices with each element a wave-adjusted log2 ratio value, which can be used to do downstream CNV-GWAS analysis.
	
	To increase the reliability of results, we split both case and control samples into 2 sets with equal size---test set and validation set. On the test set we implement the whole procedure of our method and find out 
	suspicious disease-risking regions, for the validation set, we only conduct testing procedure on each segment returned by the merge outcome on test set, if the test also returns a significant p-value, we claim this region a disease-risking region. We perform test returning impact value explained by deletion or duplication only, i.e. we implement test (11) and (12) for every segment in test and validation set. 
	
	Figure 4 depicts the log 10 p-value for every segment after merging, the Pearson's correlation of log 10 p-value returned by test set and validation set are 0.91 and 0.82 for deletion and duplication, indicating the robustness of our method. Figure 5 depicts the signals corresponding to the most significant point in the deletion's log 10 P-value plot, with (a) and (b) referring to signals of case and control separately. For the overall 2060 control samples, we find 875 copy number deletion carriers, figuer 4(b) shows CNV signals of all these 875 samples, the ``mid" part is the segment of interest with largest p-value in terms of deletion, we also plot the adjacent ``left" and ``right" segments to demonstrate reliability of our merge method. For ease of comparison, in figure 5(A) we also plot 875 case sample's CNV signals, which correspond to the smallest 875 signal intensities in this segment, here we define mean of log 2 ratio value as the signal intensity in a segment. This significant segment is not  contained in the well-known RefSeqGene set, and this deletion might play a protective role.
	
	In the realdata analysis, we come across many situations that some CNV regions are very long, but due to small proportion of carriers, it's very hard to detect them, these long regions may have important biological implications. So besides carrying out the likelihood ratio test as mentioned in our method, we propose a new test statistic to deal with long segments, aiming of detecting long CNV regions even if carriers are rare.
	
	For a certain segment, suppose we have estimated parameters for both case and control samples under $H_0$ and $H_1$, all case and control parameters under $H_0$ are denoted as $\hat{\theta}_{H_0}^d$ and $\hat{\theta}_{H_0}^c$, $\hat{\theta}_{H_1}^d$ and $\hat{\theta}_{H_1}^c$ are parameters under $H_1$. Suppose there are $B$ bins in this segment, we can assign each bin $b$ a group of parameters $\hat{\theta}_{bH_0}^d$,  $\hat{\theta}_{bH_0}^c$,  $\hat{\theta}_{bH_1}^d$ and  $\hat{\theta}_{bH_1}^c$ by splitting $\hat{\theta}_{H_0}^d$, $\hat{\theta}_{H_0}^c$, $\hat{\theta}_{H_1}^d$ and $\hat{\theta}_{H_1}^c$. For example, if $\hat{\theta}_{H_0}^d=(\hat{\alpha}_{H_0,1}^d,\cdots,\hat{\alpha}_{H_0,1}^d,\hat{\pmb{\mu}}_{H_0,1}^{d},\cdots,\hat{\pmb{\mu}}_{H_0,5}^{d},(\hat{\pmb{\sigma}}_{H_0,1}^{d})^2,\cdots,(\hat{\pmb{\sigma}}_{H_0,5}^{d})^2)$, we can denote $\hat{\theta}_{bH_0}^d$ as $(\hat{\alpha}_{H_0,1}^d,\cdots,\hat{\alpha}_{H_0,1}^d,\hat{\pmb{\mu}}_{bH_0,1}^{d},\cdots,\hat{\pmb{\mu}}_{bH_0,5}^{d},(\hat{\pmb{\sigma}}_{bH_0,1}^{d})^2,\cdots,(\hat{\pmb{\sigma}}_{bH_0,5}^{d})^2)$, where $\hat{\pmb{\mu}}_{bH_0,1}^{d}$, $\cdots,\hat{\pmb{\mu}}_{bH_0,5}^{d},(\hat{\pmb{\sigma}}_{bH_0,1}^{d})^2,\cdots,(\hat{\pmb{\sigma}}_{bH_0,5}^{d})^2$ are the ($5b-4$)-th to $5b$-th elements of $\hat{\pmb{\mu}}_{H_0,1}^{d},\cdots,\hat{\pmb{\mu}}_{H_0,5}^{d}$,\\ $(\hat{\pmb{\sigma}}_{H_0,1}^{d})^2,\cdots,(\hat{\pmb{\sigma}}_{H_0,5}^{d})^2$ correspondingly, under this premise that the whole chromosome is splitted into bins with equal length 5. If case and control samples in this segment are $\mathbf{X}_1,\cdots,\mathbf{X}_{N_1}$; $\mathbf{Y}_1,\cdots,\mathbf{Y}_{N_2}$ with $\mathbf{X}_i=(\mathbf{X}_{i1},\cdots,\mathbf{X}_{iB})$, $\mathbf{Y}_j=(\mathbf{Y}_{j1},\cdots,\mathbf{Y}_{jB})$ for case sample $i=1,\cdots,N_1$, control sample $j=1,\cdots,N_2$, we define the new test statistic as:
	\[
	\Lambda_{sum}=\underset{i=1}{\stackrel{N_1}{\sum}}\underset{b=1}{\stackrel{B}{\sum}}\large{[}h(\mathbf{X}_{ib}|\hat{\pmb{\theta}}_{bH_1}^d)-h(\mathbf{X}_{ib}|\hat{\pmb{\theta}}_{bH_0}^d)\large{]}+\underset{j=1}{\stackrel{N_2}{\sum}}\underset{b=1}{\stackrel{B}{\sum}}\large{[}h(\mathbf{Y}_{jb}|\hat{\pmb{\theta}}_{bH_1}^c)-h(\mathbf{Y}_{jb}|\hat{\pmb{\theta}}_{bH_0}^c)\large{]}
	\]
	here $h(\mathbf{X}_{ib}|\hat{\theta}_{bH_0})$ is the mixture Gaussian distribution with parameters $\hat{\theta}_{bH_0}^d$, other functions $h$ are analogous. It's obvious that this test statistic accumulates case-control difference bin by bin, increasing the power of detecting long CNVs with rare carriers, we call this statistic as sumed likelihood ratio statistic. Here we can also replace proportion parameters under $H_0$ by parameters derived from (13) or (14) to explicate the influence of deletion or duplication only.
	
	Figure 6 shows an example of a CNV segment that can't be detected by the likelihood ratio test but returns a significant value using the sumed likelihood ratio statistic. This region contains 105 probes and overlaps with 116kb length genome region, we detect 43 and 10 carriers among case and control samples separately, after intersecting with RefSeqGene set, it overlaps with 3 genes---PKP1, TNNT2, LAD1. High abundance of LAD1 have been reported to be associated with breast cancer \cite{Roth} , PKP1 is also regarded as a biomarker in several cancers \cite{Haase, Kaz} . Both these 2 genes are highly expressed in esophagus, in accordance with cancer type of case samples.
	
   To give an overall picture of the resulting significant regions, we find out genes corresponding to each significant region by intersecting with RefSeqGene set, and reserve those genes whose proportion parameter of deletion or duplication in case is larger than that in control samples. We calculate p-value using both likelihood ratio statistic and sumed likelihood ratio statistic, focusing on influence derived by deletion or duplication only. For each CNV type, 2 significant gene sets $G_1$ and $G_2$ can be obtained with p-value returned by 2 statistics, we then perform enrichment analysis by computing overlaps of these 2 gene sets with known pathway genes separately, giving us the knowledge that on which pathways gene set are enriched. We select the top 20 pathways with largest gene set enrichment for $G_1$ and $G_2$, then we pick up pathways emerged in both top 20 pathways of $G_1$ and $G_2$, table 9, 10 presents the overlap of enriched pathways for $G_1$ and $G_2$ in each CNV type. 

   From table 9, we find that genes which have more deletions in cancer samples are enriched in pathways such as cell cycle and immune system, indicating destruction of cell cycle and immune system promotes tumor progression. As for table 10, pathways indicating activities during cancer process have been found, such as cell motility, cell projection origanization and cell part morphogenesis, in accordance with the duplication CNV type.
\begin{figure}[htbp] 
	\centering  	
	\subfloat[] 
	{
		\begin{minipage}[t]{0.5\textwidth}
			\centering         
			\includegraphics[width=1.0\textwidth]{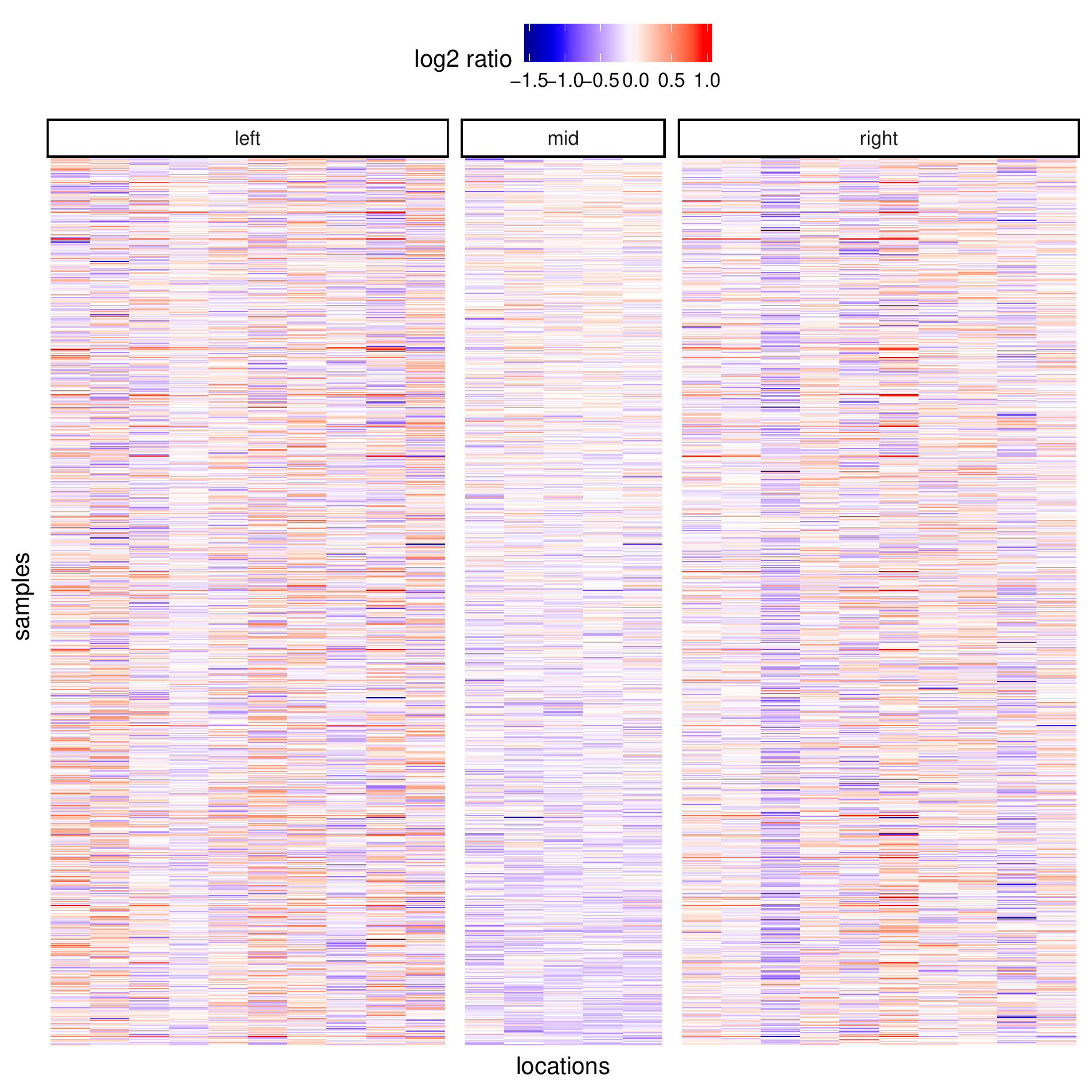}   
		\end{minipage}%
	}	
	\subfloat[] 
	{
		\begin{minipage}[t]{0.5\textwidth}
			\centering     
			\includegraphics[width=1.0\textwidth]{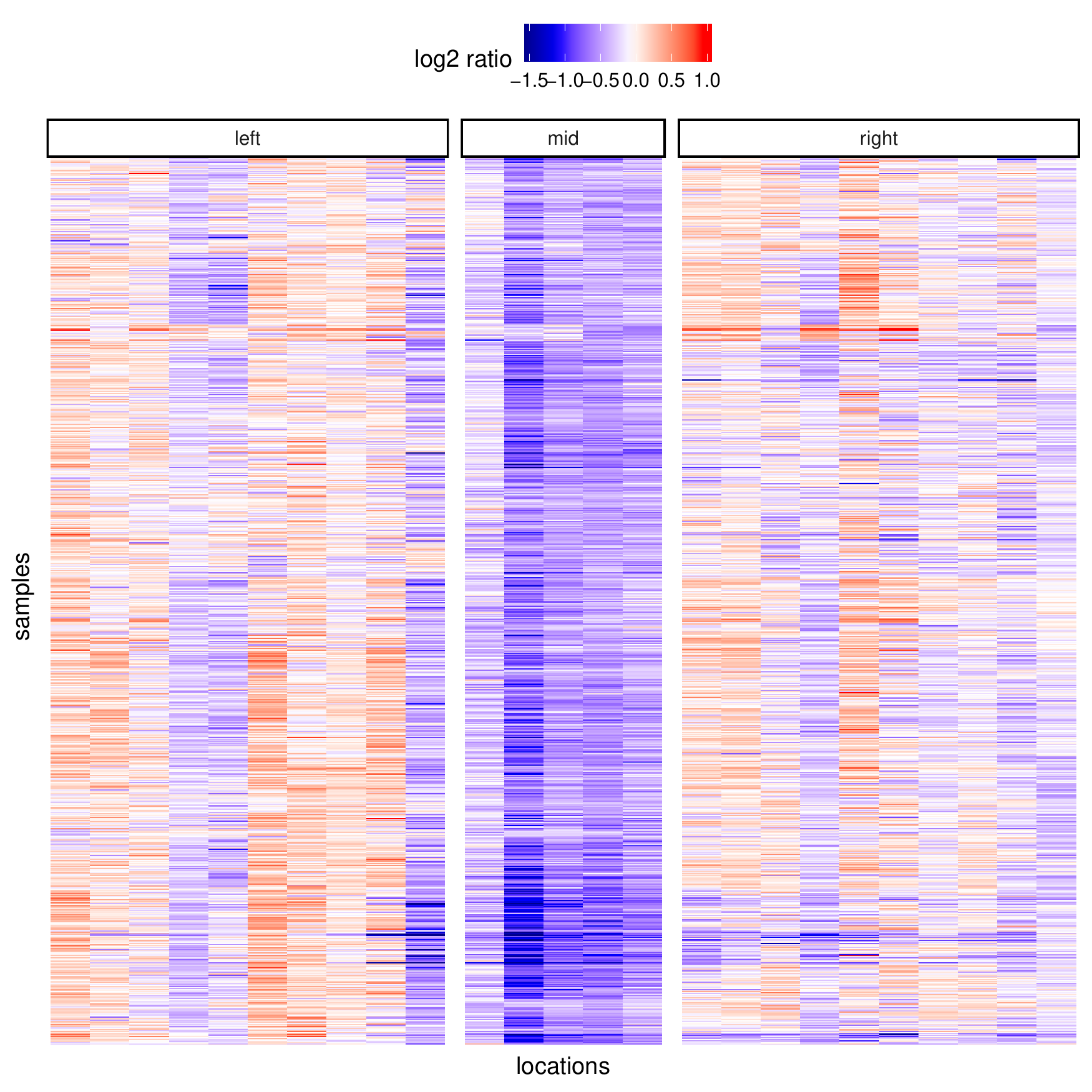}   
		\end{minipage}
	}%
	
	\caption{Signals corresponding to the most significant point in the deletion's log 10 P-value plot. (A) and (B) refer to signals of case and control, "mid" part depicts signals of the most significant segment, we also plot its left and right segments for comparison.} 
	\label{fig1}  
\end{figure}

\begin{figure}[htbp] 
	\centering  	
	\subfloat[] 
	{
		\begin{minipage}[t]{0.5\textwidth}
			\centering         
			\includegraphics[width=1.0\textwidth]{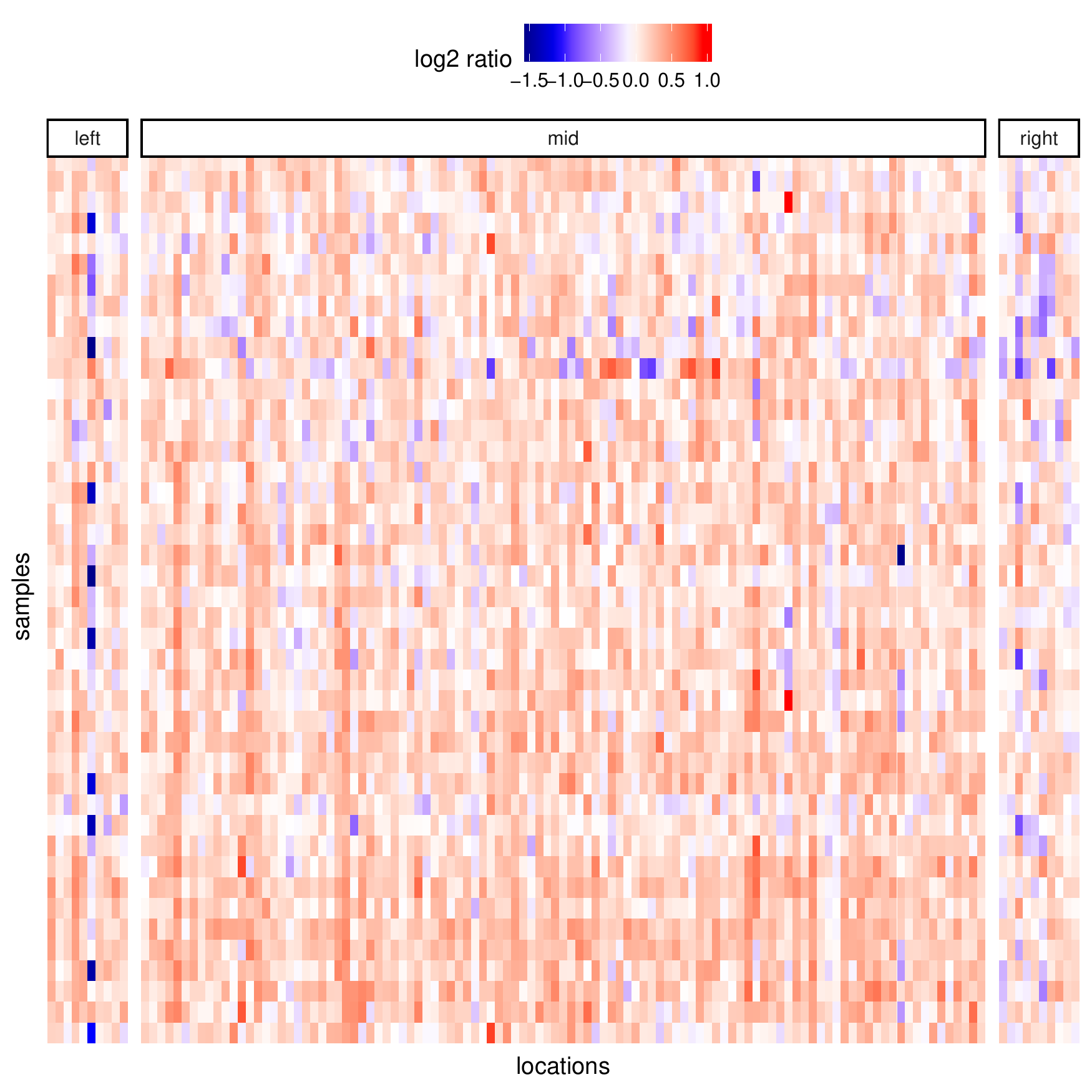}   
		\end{minipage}%
	}	
	\subfloat[] 
	{
		\begin{minipage}[t]{0.5\textwidth}
			\centering     
			\includegraphics[width=1.0\textwidth]{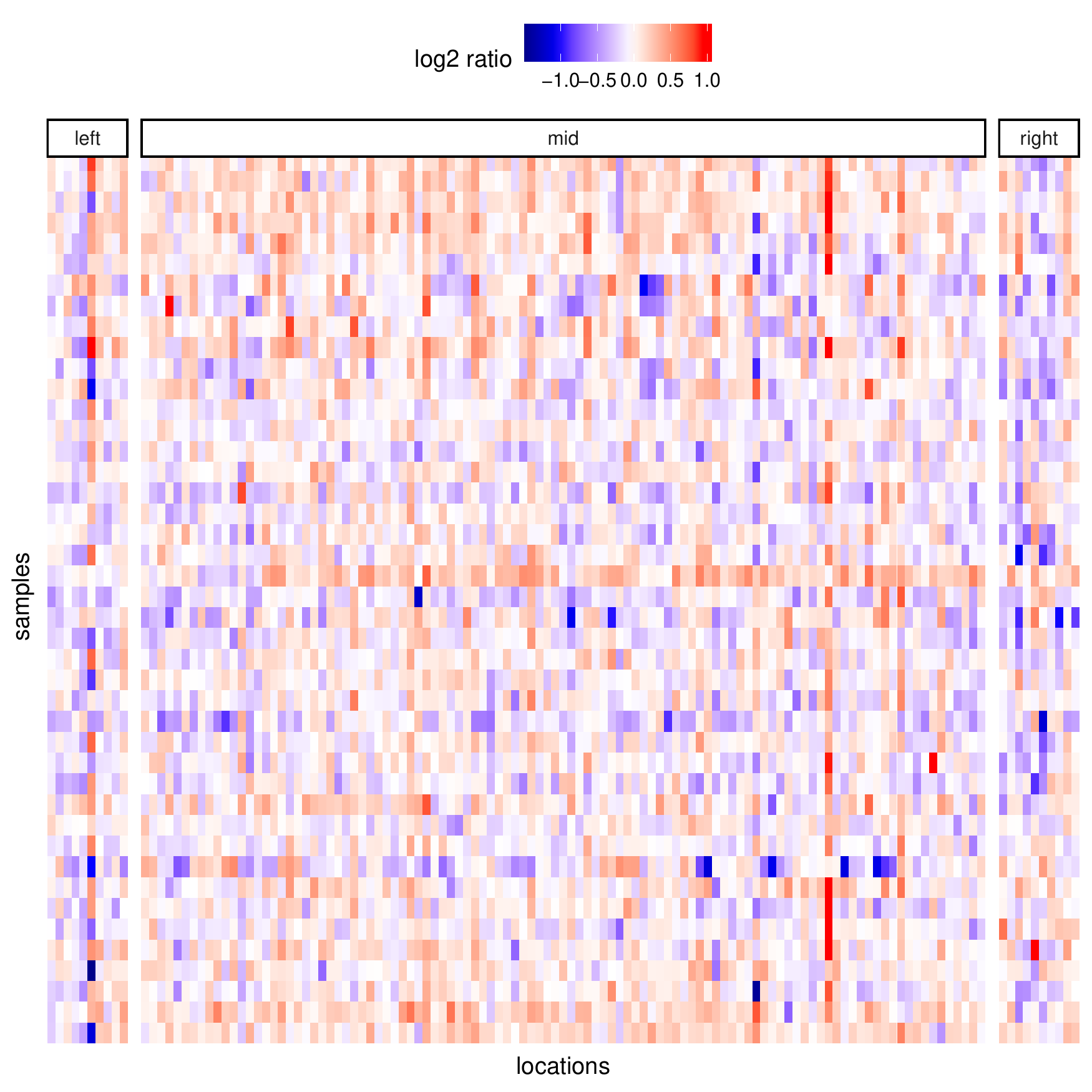}   
		\end{minipage}
	}%
	
	\caption{A segment contains genes PKP1, TNNT2, LAD1 which can be detected by sumed likelihood ratio statistics but can't be detected by likelihood ratio test.} 
	\label{fig1}  
\end{figure}

\begin{table}[h] 
	\begin{tabular}{p{7.0cm}|p{5.6cm}} 
		\hline
		\hline
		Gene Set Name & Representative genes \\ 
		\hline 
		GO catalytic complex & ERC1, DNAH12, PPP2CB\\
		\hline
		GO microtubule cytoskeleton & PRKAR2A, CSNK1D, SPAST\\
		\hline
		GO positive regulation of cellular biosynthetic process &TET1, KLF5, CDH13  \\
		\hline		
		GO ribonucleotide binding & ATAD5, LATS2, APC\\
		\hline
		GO adenyl nucleotide binding &RAD51, ACTR2, HSPD1 \\
       \hline		
		GO cell cycle process & BRCA1, KAT2B, DNA2\\
		\hline
		GO RNA binding & CDC40, XPO1, SMG6\\
		\hline
		GO chromosome &PRIM1, KAT8, RIF1 \\
		\hline
		GO chromosome organization & KMT2C, TAF9, PCGF6 \\
		\hline
		GO microtubule organizing center &MAPRE1, PPP4R2, CTNNBL1 \\
		\hline
		GO peptidyl amino acid modification &ERBB4, ARID4B, TGFB1 \\
		\hline
		GO protein containing complex assembly & TAF12, AXIN1, JAK2 \\
		\hline
		GO immune system development &FOXO3, RHOA, PPP2R3C  \\
		\hline
		\hline
	\end{tabular}
	\caption{Top pathways in deletion's significant gene set.} 
\end{table}

\begin{table}[h] 
	\begin{tabular}{p{7.0cm}|p{5.6cm}} 
		\hline
		\hline
		Gene Set Name & Representative genes \\ 
		\hline 
		GO intrinsic component of plasma membrane & EPHB1, ASIC2, PTPRN2\\
		\hline
		GO synapse& CHRNA4, MAP1B, GABRB3, \\
		\hline
		GO cell motility & ERBB2, PALLD, BCL2\\
		\hline
		GO locomotion & PLXNA4, MET, SDC4\\
		\hline		
		GO neuron projection & ANKS1B, STON2, TENM4 \\
		\hline	
		GO neurogenesis & CDH4, CAMK1D, SH3GL3\\
		\hline
		GO cellular component morphogenesis & PALLD, KIT, GRB7\\
		\hline			
		GO neuron differentiation & EYA1, ERCC6, KIRREL3\\
		\hline
		GO neuron development & NTM, ERCC6, CAMK1D \\
		\hline
		GO cell projection organization & ADGRB1, TMEFF2, ABL2 \\
		\hline		
		GO synaptic signaling & PRKCE, ADRA1A, KCNQ2 \\
		\hline
		GO cell part morphogenesis & EIF2AK4, MAP1B, DAB1 \\
		\hline
		\hline
	\end{tabular}
	\caption{Top pathways in duplication's significant gene set.} 
\end{table}
	
\section{Discussion}
	In this paper, we developed a method aiming at finding disease-risking regions related to CNV's disproportionately distributed between case and control samples. To overcome the batch effect and heterogeneity, we proposed a test formula only focusing on testing the equivalence of proportion of each CN state between case and control. We didn't make an assumption of accordance of CNV boundaries among all samples, cutting chromosome into bins and then merging bins to enhance power instead.
	
	Furthermore, we raised a new empirical Bayes formula to overcome overfitting, making sure the proportions of CN states between case and control are tested correspondingly. By means of the prior variance varying with sample size and bin length, the prior distribution can still play an important role when the sample size or bin size going to infinity, making it a suitable choice of model selection especially for those situations where there are many models to be specified. We have also proposed the theoretical guarantees of our empirical Bayes estimators under the circumstance of large sample size or bin size and the prior didn't require the CN specific prior mean locates quite near by the sample mean. 
	
	We demonstrate the effectiveness of our method by simulation and realdata analysis, our method behaves well in term of sensitivity and FDR performance in simulation. In realdata, besides finding out many famous cancer genes, we pick up some genes that haven't reported before, which show significant difference between case and control samples. Outcomes in the pathway analysis further validate the meaningfulness of our method.
	
	We have to admit that when sample size is very large, due to heterogeneity across samples, there's large probability that samples' CN state are not all the same in the adjacent bins, thereby we can't merge too much, leading to over-segmentation and low power in detecting large CNVs, under this circumstance, our method needs to accompany with other CNV dedection methods  which behave well in large CNV detection.
	
	\section{Appendix}
	\subsection{Proof of theorem 2.3}
	\begin{proof}
		We separate our proof into 2 main parts: existence of all 5 clusters and nonexistence of some clusters, and we consider the first circumstance first. For convinence we abbreviate "with probability tending to 1" as "wpt 1".
		
		For notation convenience, we drop the symbol of bin location $b$, case status symbol $d$ and let $\theta$ denote the overall parameters $(\mu_k, \sigma_k^2, \alpha_k)_{k=1}^5$to be estimated, let $\pmb{\xi}$ denote location and scale parameters $(\mu_k, \sigma_k^2)_{k=1}^5$, let $\pmb{\alpha}$ denote proportion parameters $(\alpha_1,\cdots,\alpha_5)$, simplify $N_1$ to $n$, let $h(\mathbf{X}_i|\pmb{\theta})$ denote density function $\underset{k=1}{\stackrel{5}{\sum}}\alpha_{k} f(\mathbf{X_{i}}|\mu_{k},\sigma_{k}^{2})$, $F_{n,p}(\mathbf{X}_1,\cdots,\mathbf{X}_{n}|\pmb{\theta})\stackrel{\bigtriangleup}{=}\underset{i=1}{\stackrel{n}{\sum}}\log[h(\mathbf{X}_i|\pmb{\theta})]$, $G_n(\mu_1,\cdots,\mu_5)\stackrel{\bigtriangleup}{=}
		\underset{k=1}{\stackrel{5}{\sum}}\log[g(\mu_k|\tau_k,\sigma_{\tau k}^2)]$, $L_{n,p}(\mathbf{X}_1,\cdots,\mathbf{X}_{n}|\pmb{\theta})\stackrel{\bigtriangleup}{=}
		F_{n,p}(\mathbf{X}_1,\cdots,\mathbf{X}_{n}|\pmb{\theta})+\\
		G_n(\mu_1,\cdots,\mu_5)\stackrel{\bigtriangleup}{=}F_{n,p}(\pmb{\theta})+G_n(\pmb{\theta})$.
		
		We define a set $B$ such that $\forall\pmb{\theta}\in B$, $h(\mathbf{X}|\pmb{\theta})$ is the "true" underlying distribution, which is equivalent to $h(\mathbf{X}|\{\mu_{k}^{*},(\sigma_{k}^{*})^2,\alpha_{k}^{*}\}_{k=1}^5)$ . If for each element $\pmb{\theta}$ in $B$, we drop the proportion part $\{\alpha_k\}_{k=1}^5$ and denote the left part $(\mu_k,\sigma_k^2)_{k=1}^5$ as $\pmb{\xi}$, all such $\pmb{\xi}$ formed a new set $B_{\pmb{\xi}}$. In the case of existence of all 5 clusters, $B$ consists of $5!$ elements obviously, any element in $B$ is a location permutation of true parameters $({\mu}_k^*,({\sigma}_{k}^{*})^2,{\alpha}_{k}^*)_{k=1}^5\stackrel{\bigtriangleup}{=}\pmb{\theta}^*\stackrel{\bigtriangleup}{=}(\pmb{\xi}^*,\pmb{\alpha}^*)$ (for example, $0.2N(-1,1)+0.8N(1,1)$ and $0.8N(1,1)+0.2N(-1,1)$ denote the same mixture Gaussian distribution). If we define estimators $(\hat{\mu}_{k},\hat{\sigma}_{k}^{2},\hat{\alpha}_{k})_{k=1}^5$ maximize $\tilde{f}(\mathbf{X_{1}},\cdots,\mathbf{X_{n}}|\pmb{\theta})$ in (2) (i.e. maximize $L_{n,p}(\mathbf{X}_1,\cdots,\mathbf{X}_{n}|\pmb{\theta})$) as $\hat{\pmb{\theta}}_{p}\stackrel{\bigtriangleup}{=}(\hat{\pmb{\xi}}_p,\hat{\pmb{\alpha}}_p)$, and denote a set $\{\pmb{\theta}=(\pmb{\xi},\pmb{\alpha}):|\pmb{\xi}-\pmb{\xi}^*|>\epsilon\ or\ |\pmb{\alpha}-\pmb{\alpha}^*|>\epsilon\}$ for some $\epsilon>0$ as $H$, we want to first claim that 
		\begin{equation}
		\forall\epsilon>0,\exists P>0,\ when\ p>P,\ \mathop{\sup}\limits_{\pmb{\theta}\in H}L_{n,p}(\pmb{\xi},\pmb{\alpha})<L_{n,p}(\hat{\pmb{\xi}}_p,\hat{\pmb{\alpha}}_p)\ wpt\ 1
		\end{equation}
		To complete (16), we separate $H$ into 3 sets: $H=H_1\cup H_2\cup H_3$ with 
		$H_1=\{\pmb{\theta}:|\pmb{\xi}-\pmb{\xi}^*|>\epsilon\ and\ \forall\alpha_k>0, k=1,\cdots,5\}$, $H_2=\{|\pmb{\xi}-\pmb{\xi}^*|>\epsilon\ and\ \exists\alpha_k=0, k=1,\cdots,5\}$, $H_3=\{|\pmb{\alpha}-\pmb{\alpha}^*|>\epsilon\}$, in the following we will discuss (16) under these 3 circumstances. 
		
		With regard to $H_1$, we want to show that for all fixed $\pmb{\alpha}$ with all its $\alpha_k>0, k=1,\cdots,5$, we can attain
		\begin{equation}
		\forall\epsilon>0,\ \exists P>0,\ when\ p>P, \mathop{\sup}\limits_{\{\pmb{\xi}:|\pmb{\xi}-\pmb{\xi}^*|>\epsilon\}}L_{n,p}(\pmb{\xi},\pmb{\alpha})<L_{n,p}(\hat{\pmb{\xi}}_p,\pmb{\alpha})\ wpt\ 1
		\end{equation}
		Now we divide $\{\pmb{\xi}:|\pmb{\xi}-\pmb{\xi}^*|>\epsilon\}$ into 2 parts:\\
		$(\bigcup\limits_{\check{\pmb{\xi}}\in B_{\pmb{\xi}}{\backslash}\pmb{\xi}^*}\{\pmb{\xi}:|\pmb{\xi}-\check{\pmb{\xi}}|<\epsilon\})\bigcup(\{\pmb{\xi}:|\pmb{\xi}-\pmb{\xi}^*|>\epsilon\}\backslash\bigcup\limits_{\check{\pmb{\xi}}\in B_{\pmb{\xi}}{\backslash}\pmb{\xi}^*}\{\pmb{\xi}:|\pmb{\xi}-\check{\pmb{\xi}}|<\epsilon\})$.\\
		For convenience $\bigcup\limits_{\check{\pmb{\xi}}\in B_{\pmb{\xi}}{\backslash}\pmb{\xi}^*}\{\pmb{\xi}:|\pmb{\xi}-\check{\pmb{\xi}}|<\epsilon\}$ is denoted as $Q$, $Q^c$ is denoted as its complement in set $\{\pmb{\xi}:|\pmb{\xi}-\pmb{\xi}^*|>\epsilon\}$. In order to reach the conclusion described in (17), firstly we show that 
		\begin{equation}
		\mathop{\sup}\limits_{\pmb{\xi}\in Q}L_{n,p}(\pmb{\xi},\pmb{\alpha})>\mathop{\sup}\limits_{\pmb{\xi}\in Q^c}L_{n.p}(\pmb{\xi},\pmb{\alpha})\ \ wpt\ 1
		\end{equation}
		
		Suppose $\tilde{\pmb{\xi}}_p=\mathop{argmax}\limits_{\pmb{\xi}\in Q}F_{n,p}(\pmb{\xi},\pmb{\alpha})$, $\tilde{\tilde{\pmb{\xi}}}_p=\mathop{argmax}\limits_{\pmb{\xi}\in Q^c}L_{n,p}(\pmb{\xi},\pmb{\alpha})$, according to  the defination of $Q$ and randomicity of $\epsilon$ we can get that there is a location permutation rule $\pi$ such that $\pi(\tilde{\pmb{\xi}}_p)=\tilde{\pmb{\xi}}_p^{\pi}=(\tilde{\mu}_{pk}^{\pi},(\tilde{\sigma}_{pk}^{\pi})^2)_{k=1}^5$ and $\tilde{\pmb{\xi}}_p^{\pi}\xrightarrow{}\pmb{\xi}^*$ a.s., we further assume the number of samples belonging to cluster $k$ is $n_k$ (here we suppose clusters of smaller index $k$ are equipped with smaller location parameter $\mu_k$ ), and for simplicity, samples $\mathbf{X}_1,\cdots,\mathbf{X}_{n_1}$ belong to cluster 1, $\cdots$, samples $\mathbf{X}_{n_1+\cdots+n_4+1},\cdots,\mathbf{X}_n$ belong to cluster 5. Take samples $\mathbf{X}_1,\cdots,\mathbf{X}_{n_1}$ for an example, it is obvious that $X_{it}\sim N(\mu_1^*,(\sigma_1^{*})^2), \forall i\in1,\cdots,n_1; \forall t\in 1,\cdots,p$, by law of large numbers and $\tilde{\pmb{\xi}}_p^{\pi}\xrightarrow{}\pmb{\xi}^*$ a.s., we have for $\forall i\in 1,\cdots,n_1$ and $\forall k\neq 1$,
		\begin{equation}
		\begin{split}
		&\ \ \ \frac{1}{p}(\log[f(\mathbf{X}_i|\tilde{\mu}_{p1}^{\pi},(\tilde{\sigma}_{p1}^{\pi})^2)]-\log[f(\mathbf{X}_i|\tilde{\mu}_{pk}^{\pi},(\tilde{\sigma}_{pk}^{\pi})^2)])\\
		&=\frac{1}{p}(\underset{t=1}{\stackrel{p}{\sum}}\log[\phi(X_{it}|\tilde{\mu}_{p1}^{\pi},(\tilde{\sigma}_{p1}^{\pi})^2)]-\underset{t=1}{\stackrel{p}{\sum}}\log[\phi(X_{it}|\tilde{\mu}_{pk}^{\pi},(\tilde{\sigma}_{pk}^{\pi})^2)])\\
		&\xrightarrow{p}\int \log[\frac{\phi(x|\mu_1^*,\sigma_1^{2*})}{\phi(x|\mu_k^*,(\sigma_k^{*})^2)}]\phi(x|\mu_1^*,(\sigma_1^{*})^2)dx=O(1)>0
		\end{split}
		\end{equation}
		thereby leading to an order estimation 
		\begin{equation}
		\log[f(\mathbf{X}_i|\tilde{\mu}_{p1}^{\pi},(\tilde{\sigma}_{p1}^{\pi})^2)]-\log[f(\mathbf{X}_i|\tilde{\mu}_{pk}^{\pi},(\tilde{\sigma}_{pk}^{\pi})^2)]=O_p(p), \ \forall k\neq 1
		\end{equation} 
		hence $\log[h(\mathbf{X}_i|\tilde{\pmb{\xi}}_p,\pmb{\alpha})]-\log(\alpha_1^{\hat{\pi}})-\underset{t=1}{\stackrel{p}{\sum}}\log[\phi(X_{it}|\tilde{\mu}_{p1}^{\pi},(\tilde{\sigma}_{p1}^{\pi})^2)]=O_p(e^{-p})$ for $\forall i \in 1,\cdots,n_1$, here $\alpha_1^{\hat{\pi}}$ represents the location of $\alpha_1^{\hat{\pi}}$ is in accordance with the location of $\{\tilde{\mu}_{p1}^{\pi},(\tilde{\sigma}_{p1}^{\pi})^2\}$ in $\tilde{\pmb{\xi}}_p$. Samples belonging to clusters $2,\cdots,5$ follow similar rules as described above, combining these conclusions we can obtain
		\begin{equation}
		\begin{split}
		&\ \ \ F_{n,p}(\mathbf{X}_1,\cdots,\mathbf{X}_n|\tilde{\pmb{\xi}}_p,\pmb{\alpha})=F_{n,p}(\mathbf{X}_1,\cdots,\mathbf{X}_n|\tilde{\pmb{\xi}}_p^{\pi},\pmb{\alpha}^{\hat{\pi}})\\
		&=\underset{k=1}{\stackrel{5}{\sum}}n_k \log(\alpha_k)+\underset{k=1}{\stackrel{5}{\sum}}\underset{i\in C_k}{\stackrel{}{\sum}}\underset{t=1}{\stackrel{p}{\sum}}\log[\phi(X_{it}|\tilde{\mu}_{pk}^{\pi},(\tilde{\sigma}_{pk}^{\pi})^2)]+O_p(\frac{n}{e^p})
		\end{split}
		\end{equation}
		where $C_k$ is a set containing index of samples belonging to cluster $k$.
		
		Next we shift to $\tilde{\tilde{\pmb{\xi}}}_p$, for any  realization $\mathbf{x}_1,\cdots,\mathbf{x}_n$ of sample $\mathbf{X}_1,\cdots,\mathbf{X}_n$, there is a realization $\tilde{\tilde{\pmb{\xi}}}_p$, by a permutation rule $\pi_1$, we can transform $\tilde{\tilde{\pmb{\xi}}}_p$ into $\tilde{\tilde{\pmb{\xi}}}_p^{\pi_1}$ with location parameters in increasing order, i.e. $\pi_1(\tilde{\tilde{\pmb{\xi}}}_p)=\tilde{\tilde{\pmb{\xi}}}_p^{\pi_1}=(\tilde{\tilde{\mu}}_{pk}^{\pi_1},(\tilde{\tilde{\sigma}}_{pk}^{\pi_1})^2)_{k=1}^5$ with $\tilde{\tilde{\mu}}_{p1}^{\pi_1}\leq\cdots\leq\tilde{\tilde{\mu}}_{p5}^{\pi_1}$. By the defination of $Q^c$, it is easy to see that there exists as least one $(\mu_{\tilde{k}}^*,(\sigma_{\tilde{k}}^{*})^2)$, $\tilde{k}\in \{1,\cdots,5\}$ such that events $|\tilde{\tilde{\mu}}_{pk}^{\pi_1}-\mu_{\tilde{k}}^*|>\epsilon$ and $|(\tilde{\tilde{\sigma}}_{pk}^{\pi_1})^2-(\sigma_{\tilde{k}}^{*})^2|>\epsilon$ for $\forall k\in \{1,\cdots,5\}$ happen at least once. So by a similar argument as (19), samples fall into cluster $\tilde{k}$ must satisfy
		\begin{equation}
		\log[f(\mathbf{x}_i|\tilde{\mu}_{p\tilde{k}}^{\pi},(\tilde{\sigma}_{p\tilde{k}}^{\pi})^2)]-\log[f(\mathbf{x}_i|\tilde{\tilde{\mu}}_{pk}^{\pi_1},(\tilde{\tilde{\sigma}}_{pk}^{\pi_1})^2)]=O(p), \  k= 1,\cdots,5, \  i\in C_{\tilde{k}}
		\end{equation}
		hence we have
		\begin{equation} \log[h(\mathbf{x}_i|\tilde{\tilde{\pmb{\xi}}}_p,\alpha)]=\underset{t=1}{\stackrel{p}{\sum}}\log[\phi(x_{it}|\tilde{\mu}_{p\tilde{k}}^{\pi},(\tilde{\sigma}_{p\tilde{k}}^{\pi})^2)]-O(p)\ for\ \forall i\in C_{\tilde{k}}
		\end{equation}
		Regarding to those samples not belonging to cluster $\tilde{k}$, if $\pmb{\xi}$ doesn't have any constraints, $h(\mathbf{x}_i|\pmb{\xi},\pmb{\alpha})$ with fixed $\pmb{\alpha}$ must attains its maximum at $\pmb{\xi}_M^i$ with $\mu_1=\cdots=\mu_5=\mu_{Mk^*}^i$, $\sigma_1^2=\cdots=\sigma_5^2=(\sigma_{Mk^*}^{i})^2$, here $(\mu_{Mk^*}^i,(\sigma_{Mk^*}^{i})^2)$ is the realization of MLE $(\hat{\mu}_{Mk^*}^i,(\hat{\sigma}_{Mk^*}^{i})^2)$ of sample $X_{i1},\cdots,X_{ip}$ following certain normal distribution with $i\in C_{k^*}\neq C_{\tilde{k}}$. By the properties of MLE, we know that $\hat{\mu}_{Mk^*}^i\xrightarrow{p}\mu_{k^*}^*$, $(\hat{\sigma}_{Mk^*}^{i})^2\xrightarrow{p}(\sigma_{k^*}^{*})^2$, combining with $\tilde{\pmb{\xi}}_p^{\pi}\xrightarrow{p}\pmb{\xi}^*$ and the continuity of Gaussian density function we can get for $i\in C_{k^*}$
		\begin{equation}
		\big|\underset{t=1}{\stackrel{p}{\sum}}\log[\phi(x_{it}|\tilde{\mu}_{pk^*}^{\pi},(\tilde{\sigma}_{pk^*}^{\pi})^2)]-\underset{t=1}{\stackrel{p}{\sum}}\log[\phi(x_{it}|\mu_{Mk^*}^i,(\sigma_{Mk^*}^{i})^2)]\big|=o(p)
		\end{equation}
		\begin{equation}
		\log[h(\mathbf{x}_i|\tilde{\tilde{\pmb{\xi}}}_p,\pmb{\alpha})]\leq \log[h(\mathbf{x}_i|\pmb{\xi}_M^i,\pmb{\alpha})]=\underset{t=1}{\stackrel{p}{\sum}}\log[\phi(x_{it}|\mu_{Mk^*}^i,(\sigma_{Mk^*}^{i})^2)]
		\end{equation}
		Combining (23), (24) and (25), a result in analogy with (21) can be achieved
		\begin{equation}
		\begin{split}
		&F_{n,p}(\mathbf{X}_1,\cdots,\mathbf{X}_n|\tilde{\tilde{\pmb{\xi}}}_p,\pmb{\alpha})\leq\\
		&\underset{i\in C_{\tilde{k}}}{\stackrel{}{\sum}}\underset{t=1}{\stackrel{p}{\sum}}\log[\phi(X_{it}|\tilde{\mu}_{p\tilde{k}}^{\pi},(\tilde{\sigma}_{p\tilde{k}}^{\pi})^2)]+\underset{k^*\neq\tilde{k}}{\stackrel{}{\sum}}\underset{i\in C_{k^*}}{\stackrel{}{\sum}}\underset{t=1}{\stackrel{p}{\sum}}\log[\phi(X_{it}|\hat{\mu}_{Mk^*}^i,(\hat{\sigma}_{Mk^*}^{i})^2)]-O_p(p)=\\
		&\underset{i\in C_{\tilde{k}}}{\stackrel{}{\sum}}\underset{t=1}{\stackrel{p}{\sum}}\log[\phi(X_{it}|\tilde{\mu}_{p\tilde{k}}^{\pi},(\tilde{\sigma}_{p\tilde{k}}^{\pi})^2)]+\underset{k^*\neq\tilde{k}}{\stackrel{}{\sum}}\underset{i\in C_{k^*}}{\stackrel{}{\sum}}\underset{t=1}{\stackrel{p}{\sum}}\log[\phi(X_{it}|\tilde{\mu}_{pk^*}^{\pi},(\tilde{\sigma}_{pk^*}^{\pi})^2)]-O_p(p)+o_p(p)
		\end{split}
		\end{equation}
		where the sign of the last term $o_p(p)$ is uncertain, but no matter what the sign is, it is clear that $o_p(p)-O_p(p)=-O_p(p)\ll0$. Now compare (26) with (21) leading to $F_{n,p}(\mathbf{X}_1,\cdots,\mathbf{X}_n|\tilde{\tilde{\pmb{\xi}}}_p,\pmb{\alpha})= F_{n,p}(\mathbf{X}_1,\cdots,\mathbf{X}_n|\tilde{\pmb{\xi}}_p,\pmb{\alpha})-O_p(p)$, adding the fact that $\forall
		\pmb{\theta}_1$, $\pmb{\theta}_2$, $|G_n(\pmb{\theta}_1)-G_n(\pmb{\theta}_2)|<O_p(p)$, we can reach that
		\begin{equation}
		\begin{split}
		&\ \ \ \mathop{\sup}\limits_{\pmb{\xi}\in Q}L_{n,p}(\pmb{\xi},\pmb{\alpha})-\mathop{\sup}\limits_{\pmb{\xi}\in Q^c}L_{n,p}(\pmb{\xi},\pmb{\alpha})\\
		&\geq(F_{n,p}(\tilde{\pmb{\xi}}_p,\pmb{\alpha})-F_{n,p}(\tilde{\tilde{\pmb{\xi}}}_p,\pmb{\alpha}))+(G_n(\tilde{\pmb{\xi}}_p)-G_n(\tilde{\tilde{\pmb{\xi}}}_p))=O_p(p)>0
		\end{split}
		\end{equation}
		which means that (18) is achieved, with this in hand, what's left is to show that for all  fixed $\pmb{\alpha}$ with $\alpha_k>0, k=1,\cdots,5$, $\mathop{\sup}\limits_{\pmb{\xi}\in Q}L_{n,p}(\pmb{\xi},\pmb{\alpha})<L_{n,p}(\hat{\pmb{\xi}}_p,\pmb{\alpha})$ wpt 1 when $p$ is large enough.
		
		We assume $\tilde{\pmb{\xi}}_p'=\mathop{argmax}\limits_{\pmb{\xi}\in Q}L_{n,p}(\pmb{\xi},\pmb{\alpha})$, similar as described before there exists a location permutation rule $\pi_2$ such that $\pi_2(\tilde{\pmb{\xi}}_p')=\tilde{\pmb{\xi}}_p^{'\pi_2}\xrightarrow{p}\xi^*$, we further assume $\pi_2(\pmb{\alpha})=\pmb{\alpha}^{\pi_2}$, noticing that permutating the location of the 5 groups of parameters $(\mu_k,\sigma_k^2,\alpha_k)_{k=1}^5$ doesn't change values of $F_{n,p}(\pmb{\xi},\pmb{\alpha})$, which means that $F_{n,p}(\tilde{\pmb{\xi}}_p',\pmb{\alpha})=F_{n,p}(\tilde{\pmb{\xi}}_p^{'\pi_2},\pmb{\alpha}^{\pi_2})$. Recall the result presented in (21), we have
		\begin{align*} F_{n,p}(\tilde{\pmb{\xi}}_p',\pmb{\alpha})-F_{n,p}(\tilde{\pmb{\xi}}_p^{'\pi_2},\pmb{\alpha})\leq\underset{k=1}{\stackrel{5}{\sum}}n_k |\log(\alpha_k)-\log(\alpha_k^{\pi_2})|+O_p(\frac{n}{e^p})
		\end{align*}
		In addition, by the defination of $Q$, we can draw that location parameters $\{\tilde{\mu}_{kp}'\}_{k=1}^5$ aren't arranged in increasing order as $\{\tilde{\mu}_{kp}^{'\pi_2}\}_{k=1}^5$, by lemma 2,
		\begin{align*} G_n(\tilde{\pmb{\xi}}_p^{'\pi_2})-G_n(\tilde{\pmb{\xi}}_p')\geq O_p(m(p))>0
		\end{align*}
		where $m(p)$ is any increasing function of order less than $O(p)$, tending to infinity as long as $p\rightarrow\infty$. Comparing this order estimation with its above inequality, the conclusion $\mathop{\sup}\limits_{\pmb{\xi}\in Q}L_{n,p}(\pmb{\xi},\pmb{\alpha})<L_{n,p}(\hat{\pmb{\xi}}_p,\pmb{\alpha})$ wpt 1 is available, meaning that we have completed the proof of (17).
		
		Now let's turn to $H_2$, there exists at least one $\alpha_k=0$ for $k=1,\cdots,5$, so no matter what values the corresponding $(\mu_k,\sigma_k^2)$ take, the part $\alpha_k f(\mathbf{x}|\mu_k,\sigma_k^2)$ doesn't account for $F_{n,p}(\pmb{\theta})$ any more, i.e. if there is a $\alpha_k=0$, we only have 4 groups of parameters $(\alpha_k,\mu_k,\sigma_k^2)$ to fit a 5-cluster $p$-dimensional mixture Gaussian model. For example if we assume $(\tilde{\pmb{\xi}}_p'',\tilde{\pmb{\alpha}}_p'')=\mathop{argmax}\limits_{\alpha_1=0}L_{n,p}(\pmb{\xi},\pmb{\alpha})$, there must exist at least one cluster $k_1$ such that 
		\begin{equation}
		\log[h(\mathbf{X}_i|\tilde{\pmb{\xi}}_p'',\tilde{\pmb{\alpha}}_p'')]=\underset{t=1}{\stackrel{p}{\sum}}\log[\phi(X_{it}|\mu_{k_1}^*,(\sigma_{k_1}^{*})^2)]-O_p(p)\ for\ \forall i\in C_{k_1}
		\end{equation}
		For sample $i\in C_{k_2}\neq C_{k_1}$, we bound $\log[h(\mathbf{x}|\tilde{\pmb{\xi}}_p'',\tilde{\pmb{\alpha}}_p'')]$ with $\tilde{\pmb{\xi}}_p''$ replaced by MLE $\pmb{\xi}_M^i$ of sample $X_{i1},\cdots,X_{ip}$ as introduced above
		\begin{equation}
		\begin{split}
		\log[h(\mathbf{X}_i|\tilde{\pmb{\xi}}_p'',\tilde{\pmb{\alpha}}_p'')]\leq \log[f(\mathbf{X}_i|\pmb{\xi}_M^i,\tilde{\pmb{\alpha}}_p'')]=\underset{t=1}{\stackrel{p}{\sum}}\log[\phi(X_{it}|\mu_{Mk_2}^i,(\sigma_{Mk_2}^{i})^2)],\\
		\underset{t=1}{\stackrel{p}{\sum}}\log[\phi(X_{it}|\mu_{Mk_2}^i,(\sigma_{Mk_2}^{i})^2)]=\underset{t=1}{\stackrel{p}{\sum}}\log[\phi(X_{it}|\mu_{k_2}^*,(\sigma_{k_2}^{*})^2)]+o_p(p)
		\end{split}
		\end{equation}
		Combining (28) and (29) we can give an upper bound of $F_{n,p}(\tilde{\pmb{\xi}}_p'',\tilde{\pmb{\alpha}}_p'')$
		\begin{equation}
		\begin{split}
		F_{n,p}(\tilde{\pmb{\xi}}_p'',\tilde{\pmb{\alpha}}_p'')&\leq\underset{k=1}{\stackrel{5}{\sum}}\underset{i\in C_{k}}{\stackrel{}{\sum}}\underset{t=1}{\stackrel{p}{\sum}}\log[\phi(x_{it}|\mu_k^*,(\sigma_k^{*})^2)]-O_p(p)+o_p(p)\\
		&=F_{n,p}(\pmb{\xi}^*,\pmb{\alpha}^*)-\underset{k=1}{\stackrel{5}{\sum}}n_k\log(\alpha_k^*)-O_p(p)+o_p(p)-o_p(n)
		\end{split}
		\end{equation}
		In addition to the fact that $G_n(\tilde{\pmb{\xi}}_p'')-G_n(\pmb{\xi}^*)<O_p(p)$, we can obtain
		\begin{equation}
		\mathop{\sup}\limits_{\alpha_1=0}L_{n,p}(\pmb{\xi},\pmb{\alpha})=L_{n,p}(\tilde{\pmb{\xi}}_p'',\tilde{\pmb{\alpha}}_p'')<L_{n,p}(\pmb{\xi}^*,\alpha^*)\leq L_{n,p}(\hat{\pmb{\xi}}_p,\hat{\pmb{\alpha}}_p)
		\end{equation}
		Scenarios belonging to $H_2$ apart from only $\alpha_1=0$ can be discussed in analogy with the case of $\alpha_1=0$. Furthermore, in the discussion of $H_2$, we didn't make use of the condition $|\pmb{\xi}-\pmb{\xi}^*|>\epsilon$, so actually we have proved (16) under the constrain of $\pmb{\theta}\in\{\pmb{\theta}:\exists\alpha_k=0,k=1,\cdots,5\}\stackrel{\bigtriangleup}{=}H_2'$. 
		
		In the following we move to $H_3$. Observing that $H_3$ has intersection with $H_1$ and $H_2'$, what's left is to prove (16) with the constrain $\pmb{\theta}\in\{\pmb{\theta}:|\pmb{\alpha}-\pmb{\alpha}^*|>\epsilon,\forall\alpha_k>0,k=1,\cdots,5\ and\ |\pmb{\xi}-\pmb{\xi}^*|<\epsilon\}\stackrel{\bigtriangleup}{=}H_3'$.
		
		Suppose $\tilde{\pmb{\theta}}_p'''=(\tilde{\pmb{\xi}}_p''',\tilde{\pmb{\alpha}}_p''')=\mathop{argmax}\limits_{\pmb{\theta}\in H_3'}L_{n,p}(\pmb{\xi},\pmb{\alpha})$, by the randomicity of $\epsilon$, $\tilde{\pmb{\xi}}_p'''\xrightarrow{p}\pmb{\xi}^*$, resulting in similar conclusions as (19), (20) and (21), i.e. we can get
		\begin{equation}
		F_{n,p}(\mathbf{X}_1,\cdots,\mathbf{X}_n|\tilde{\pmb{\theta}}_p''')=\underset{k=1}{\stackrel{5}{\sum}}n_k \log(\tilde{\alpha}_{pk}''')+O_p(\frac{n}{e^p})+\underset{k=1}{\stackrel{5}{\sum}}\underset{i\in C_k}{\stackrel{}{\sum}}\underset{t=1}{\stackrel{p}{\sum}}\log[\phi(X_{it}|\tilde{\mu}_{pk}''',(\tilde{\sigma}_{pk}^{'''})^2)]
		\end{equation}
		If we replace $\tilde{\pmb{\alpha}}_p'''$ in $\tilde{\pmb{\theta}}_p'''$ with $\pmb{\alpha}^*$, analogous to (32), estimation of $F_{n,p}(\tilde{\pmb{\xi}}_p''',\pmb{\alpha}^*)$ can be obtained. With the addition of $\pmb{\alpha}^*=\mathop{argmax}\limits_{\pmb{\alpha}}\underset{k=1}{\stackrel{5}{\sum}}n_k \log(\alpha_k)$ and $G_n(\tilde{\pmb{\xi}}_p''',\tilde{\pmb{\alpha}}_p''')=G_n(\tilde{\pmb{\xi}}_p''',\pmb{\alpha}^*)$ we have 
		\begin{equation}
		\begin{split}
		L_{n,p}(\tilde{\pmb{\theta}}_p''')-L_{n,p}(\tilde{\pmb{\xi}}_p''',\pmb{\alpha}^*)&=\underset{k=1}{\stackrel{5}{\sum}}n_k [\log(\tilde{\alpha}_{pk}''')-\log(\alpha_k^*)]+O_p(\frac{n}{e^p})\\
		&=-O_p(n)+o_p(n)<0\\
		\end{split}
		\end{equation}
		thus we have the final conclusion under $H_3'$: $\mathop{\sup}\limits_{\pmb{\theta}\in H_3'}L_{n,p}(\pmb{\xi},\pmb{\alpha})<L_{n,p}(\tilde{\pmb{\xi}}_p''',\pmb{\alpha}^*)\leq L_{n,p}(\hat{\pmb{\xi}}_p,\hat{\pmb{\alpha}}_p)$. By now we have completed the proof of consistency under the condition of existence of all 5 clusters, we will go on to the second main part.
		
		We continue to use the symbol $B$ to denote a set containing all parameters $\pmb{\theta}$ such that $h(\mathbf{x}|\pmb{\theta})$ is the true underlying distribution, in this circumstance, there exist innumerable possible $\pmb{\theta}$ in $B$. For example, if the true underlying distribution is $N(0,1)$, then all distributions $\underset{k=1}{\stackrel{5}{\sum}}\alpha_{k} N(0,1)$ satisfying $\underset{k=1}{\stackrel{5}{\sum}}\alpha_{k}=1$ is the same as $N(0,1)$. For notation simplification, we denote parameters only corresponding to set $K\subseteq\{1,\cdots,5\}$ in $\pmb{\theta}$ as $\pmb{\theta}_K=(\pmb{\xi}_K,\pmb{\alpha}_K)$, for example, paramters only corresponding to clusters 1 and 2 ($K=\{1,2\}$) $(\mu_k,\sigma_k^2,\alpha_k)_{k=1}^2$ can be denoted as $\pmb{\theta}_K$. Suppose the existing clusters form a set $K^*$, maximizer of $L_{n,p}(\mathbf{X}_1,\cdots,\mathbf{X}_n|\pmb{\theta})$ is also denoted as $\hat{\pmb{\theta}}_p\stackrel{\bigtriangleup}{=}(\hat{\pmb{\xi}}_p,\hat{\pmb{\alpha}}_p)$, and we define a set $S$ as $\{\pmb{\theta}:\ |\pmb{\theta}_{K^*}-\pmb{\theta}_{K^*}^*|>\epsilon\ or \ |\mu_k-\tau_k|>\epsilon\ for \ \exists k\in K^{*c}\ or \ |\alpha_k|>\epsilon\ for\ \exists k\in K^{*c}\}$, under this circumstance we want to first show that
		\begin{equation}
		\forall\epsilon>0,\exists P>0,\ when\ p>P,\ \mathop{\sup}\limits_{\pmb{\theta}\in S}L_{n,p}(\pmb{\xi},\pmb{\alpha})<L_{n,p}(\hat{\pmb{\xi}}_p,\hat{\pmb{\alpha}}_p)\ wpt\ 1
		\end{equation}
		
		As in the first main circumstance, we divide $S$ into 2 parts: $\bigcup\limits_{\check{\pmb{\theta}}\in B{\backslash}\pmb{\theta}^*}\{\pmb{\theta}:|\pmb{\theta}-\check{\pmb{\theta}}|<\epsilon\}\bigcap S\stackrel{\bigtriangleup}{=}V$ and its complement in $S$ denoted as $V^c$. Furthermore, $V^c=V_1^c\bigcup V_2^c$ with $V_1^c=\{\pmb{\theta}\in V^c:\exists\check{\pmb{\theta}}\in B,s.t.\ |\pmb{\xi}-\check{\pmb{\xi}}|<\epsilon\ for\ \forall\epsilon>0,\ and\ \exists\epsilon>0,s.t.\ |\pmb{\alpha}-\check{\pmb{\alpha}}|>\epsilon\}$, $V_2^c=\{\pmb{\theta}\in V^c:\forall\check{\pmb{\theta}}\in B,\exists\epsilon>0,s.t.|\pmb{\xi}-\check{\pmb{\xi}}|>\epsilon\}$, in the following we will show the correctness of (34) with set $S$ replaced by $V_2^c$, $V$, $V_1^c$ orderly.
		
		Firstly, we asume $\ddot{\tilde{\pmb{\theta}}}_p=\mathop{argmax}\limits_{\pmb{\theta}\in V_2^c}L_{n,p}(\pmb{\xi},\pmb{\alpha})$, by the defination of $V_2^c$ it is not hard to deduce that either of the following 2 circumstances holds:
		\begin{itemize}
			\item[(a)]$\exists\hat{k}\in K^*$, $\epsilon>0$, s.t. $\forall k\in\{1,\cdots,5\}$, $|\ddot{\tilde{\pmb{\xi}}}_{pk}-\pmb{\xi}_{\hat{k}}^*|>\epsilon$
			\item[(b)]$\forall\hat{k}\in K^*$, $\forall\epsilon>0$, $\exists k\in\{1,\cdots,5\}$, s.t. $|\ddot{\tilde{\pmb{\xi}}}_{pk}-\pmb{\xi}_{\hat{k}}^*|<\epsilon$ and $\exists k\in\{1,\cdots,5\}$, $\exists\epsilon>0$, s.t. $\forall\hat{k}\in K^*$, $|\ddot{\tilde{\pmb{\xi}}}_{pk}-\pmb{\xi}_{\hat{k}}^*|>\epsilon$, $\ddot{\tilde{\alpha}}_{pk}>\epsilon$
		\end{itemize}
		Circumstance (a) implies any sample $i$ belonging to cluster $\hat{k}$ has the property similar with (22) and (23)
		\begin{equation}
		\begin{split}
		\log[f(\mathbf{X}_i|&\mu_{\hat{k}}^*,(\sigma_{\hat{k}}^{*})^2)]-\log[f(\mathbf{X}_i|\ddot{\tilde{\mu}}_{pk},\ddot{\tilde{\sigma}}_{pk}^2)]=O_p(p),\ \forall k\in\{1,\cdots,5\},\\
		&\log[h(\mathbf{X}_i|\ddot{\tilde{\pmb{\xi}}}_p,\ddot{\tilde{\pmb{\alpha}}}_p)]=\underset{t=1}{\stackrel{p}{\sum}}\log[\phi(X_{it}|\mu_{\hat{k}}^*,(\sigma_{\hat{k}}^{*})^2)]-O_p(p)
		\end{split}
		\end{equation}
		Therefore we can get 
		\begin{equation}
		\begin{split}
		&F_{n,p}(\mathbf{X}_1,\cdots,\mathbf{X}_n|\ddot{\tilde{\pmb{\xi}}}_p,\ddot{\tilde{\pmb{\alpha}}}_p)\leq\\
		&\underset{i\in C_{\hat{k}}}{\stackrel{}{\sum}}\underset{t=1}{\stackrel{p}{\sum}}\log[\phi(X_{it}|\mu_{\hat{k}}^*,(\sigma_{\hat{k}}^{*})^2)]+\underset{k^*\neq\hat{k}}{\stackrel{}{\sum}}\underset{i\in C_{k^*}}{\stackrel{}{\sum}}\underset{t=1}{\stackrel{p}{\sum}}\log[\phi(X_{it}|\mu_{Mk^*}^i,(\sigma_{Mk^*}^{i})^2)]-O_p(p)=\\
		&\underset{i\in C_{\hat{k}}}{\stackrel{}{\sum}}\underset{t=1}{\stackrel{p}{\sum}}\log[\phi(X_{it}|\mu_{\hat{k}}^*,(\sigma_{\hat{k}}^{*})^2)]+\underset{k^*\neq\hat{k}}{\stackrel{}{\sum}}\underset{i\in C_{k^*}}{\stackrel{}{\sum}}\underset{t=1}{\stackrel{p}{\sum}}\log[\phi(X_{it}|\mu_{k^*}^*,(\sigma_{k^*}^{*})^2)]-O_p(p)+o_p(p)
		\end{split}
		\end{equation}
		i.e. $F_{n,p}(\ddot{\tilde{\pmb{\xi}}}_p,\ddot{\tilde{\pmb{\alpha}}}_p)\leq F_{n,p}(\pmb{\theta}^*)-O_p(p)$, adding the fact $G_n(\ddot{\tilde{\pmb{\xi}}}_p)-G_n(\pmb{\xi}^*)<O_p(p)$ we can draw the conclusion
		\begin{equation}
		\mathop{\sup}\limits_{\pmb{\theta}\in V_2^c}L_{n,p}(\pmb{\xi},\pmb{\alpha})=L_{n,p}(\ddot{\tilde{\pmb{\xi}}}_p,\ddot{\tilde{\pmb{\alpha}}}_p)<L_{n,p}(\pmb{\xi}^*,\pmb{\alpha}^*)\leq L_{n,p}(\hat{\pmb{\xi}}_p,\hat{\pmb{\alpha}}_p)
		\end{equation}
		In regard to circumstance (b), we can draw a similar conclusion as (32):
		\[
		F_{n,p}(\mathbf{X}_1,\cdots,\mathbf{X}_n|\ddot{\tilde{\pmb{\xi}}}_{pk})=\underset{k\in K^*}{\stackrel{}{\sum}}n_{k}\log(\ddot{\tilde{\alpha}}_{pk})+\underset{k\in K^*}{\stackrel{}{\sum}}\underset{i\in C_k}{\stackrel{}{\sum}}\underset{t=1}{\stackrel{p}{\sum}}\log[\phi(X_{it}|\ddot{\tilde{\mu}}_{pk},\ddot{\tilde{\sigma}}_{pk}^2)]+O_p(\frac{n}{e^p})
		\]
		Combining with the fact that $\pmb{\alpha}^*=\mathop{argmax}\limits_{\pmb{\alpha}}\underset{k\in K^*}{\stackrel{}{\sum}}n_{k}\log(\alpha_k)$ and $\exists k\in K^{*c}$, $\epsilon>0$, s.t. $\ddot{\tilde{\alpha}}_{pk}>\epsilon$, an upper bound under circumstance (b) can be obtained:
		\begin{equation}
		L_{n,p}(\ddot{\tilde{\pmb{\theta}}}_p)<L_{n,p}(\ddot{\tilde{\pmb{\xi}}}_p,\pmb{\alpha}^*)\leq L_{n,p}(\hat{\pmb{\xi}}_p,\hat{\pmb{\alpha}}_p)\ \ wpt 1
		\end{equation}
		
		Secondly, we assume $\ddot{\tilde{\pmb{\theta}}}_p'=\mathop{argmax}\limits_{\pmb{\theta}\in V}L_{n,p}(\pmb{\xi},\pmb{\alpha})$, noticing that $\ddot{\tilde{\pmb{\theta}}}_p'$ lies in $V$, a set satisfying $\forall\pmb{\theta}\in V$, $h(\mathbf{x}|\pmb{\theta})$ approximates $h(\mathbf{x}|\pmb{\theta}^*)$ in arbitrary close distance, then $\forall k\in K^*$, there must exist at least one $k'\in\{1,\cdots,5\}$ such that $\ddot{\tilde{\mu}}_{pk'}'\xrightarrow{p}\mu_k^*$, based on this fact, we next discuss (34) on set $V$ under 2 conditions:
		\begin{itemize}
			\item[(c)]$\exists\epsilon>0,\ s.t.\ |\ddot{\tilde{\alpha}}_{pk}'|>\epsilon$ for $k\in K_1$, where $|K_1|>|K^*|$, here we use $"|\cdot|"$ to denote cardinality of a set.
			\item[(d)]$\forall\epsilon>0$, $\exists P>0$, when $p>P, |\ddot{\tilde{\alpha}}_{kp}'|<\epsilon$ for $k\in K_1^c$, where $|K_1|=|K^*|.$
		\end{itemize}
	
		With regard to condition (c), $\forall k\in K_1$, there exists a $k^*\in K^*$, s.t. $\ddot{\tilde{\pmb{\xi}}}_{pk}'\xrightarrow{p}\pmb{\xi}_{k^*}^*$, and we introduce a set $D(k^*)$ to denote those $k\in K_1$ with the property $\ddot{\tilde{\pmb{\xi}}}_{pk}'\xrightarrow{p}\pmb{\xi}_{k^*}^*$ holds. Furthermore, if $c(i)$ represents the cluster to which sample $i$ belongs to, we have the following approximation of $F_{n,p}(\mathbf{X}_1,\cdots,\mathbf{X}_n|\ddot{\tilde{\pmb{\theta}}}_p')$:
		\begin{equation}
		F_{n,p}(\mathbf{X}_1,\cdot,\mathbf{X}_n|\ddot{\tilde{\pmb{\theta}}}_p')=\underset{k^*\in K^*}{\stackrel{}{\sum}}\underset{c(i)=k^*}{\stackrel{}{\sum}}\log\Big{[}\underset{k\in D(k^*)}{\stackrel{}{\sum}}\alpha_k f(\mathbf{X}_i|\ddot{\tilde{\pmb{\xi}}}_{pk}')+O_p(\frac{1}{e^p})\Big{]}
		\end{equation}
		If we pick up a specific term $\underset{c(i)=k^*}{\stackrel{}{\sum}}\log\Big{[}\underset{k\in D(k^*)}{\stackrel{}{\sum}}\alpha_k f(\mathbf{X}_i|\ddot{\tilde{\pmb{\xi}}}_{pk}')+O_p(\frac{1}{e^p})\Big{]}\stackrel{\bigtriangleup}{=}F_{n,p}^{k^*}(\ddot{\tilde{\pmb{\theta}}}_p')$, by direct calculation it's not hard to deduce that neglecting terms of smaller order, $exp\big{[} F_{n,p}^{k^*}(\ddot{\tilde{\pmb{\theta}}}_p')\big{]}$ is a summation of terms like $\underset{c(i)=k^*}{\stackrel{}{\LARGE{\prod}}}\alpha_{k_i} f(\mathbf{X}_i|\ddot{\tilde{\pmb{\xi}}}_{pk_i}')$, where $k_i$ traverses any value in $D(k^*)$ resulting in overall $|D(k^*)|^{n_{k^*}}$ terms in the summation, here $n_{k^*}$ is the number of samples belonging to cluster $k^*$. In the following we want to bound $F_{n,p}^{k^*}(\ddot{\tilde{\pmb{\theta}}}_p')$ from a hypothesis testing viewpoint. Noticing that 
		\[
		\underset{c(i)=k^*}{\stackrel{}{\LARGE{\prod}}}\alpha_{k_i} f(\mathbf{X}_i|\ddot{\tilde{\pmb{\xi}}}_{pk_i}')=\underset{k_d\in D(k^*)}{\stackrel{}{\LARGE{\prod}}}\underset{\substack{c(i)=k^*\\k_i=k_d}}{\stackrel{}{\LARGE{\prod}}}\alpha_{k_i} f(\mathbf{X}_i|\ddot{\tilde{\pmb{\xi}}}_{pk_i}'),
		\]
		and we denote the set $\{i|k_i=k_d\}$ as $D_1,\cdots,D_{D(k^*)}$ for $k_d\in D(k^*)$. Suppose samples from the same set $D_{k_d}$, $k_d\in D(k^*)$ follow the same normal distribution with parameters indexed by $\pmb{\xi}_{k_d}$, consider the following hypothesis testing:
		\begin{equation}
		\begin{split}
		& H_0: \pmb{\xi}_{k_d}\  are\ equivalent\  for\ all \ k_d\in D(k^*)\\
		& H_1: \pmb{\xi}_{k_d}\ are\ not\ all\ equivalent\ for\ k_d\in D(k^*)
		\end{split}
		\end{equation}
		The likelihood under $H_1$ is $\underset{k_d\in D(k^*)}{\stackrel{}{\LARGE{\prod}}}\underset{i\in D_{k_d}}{\stackrel{}{\LARGE{\prod}}}f(\mathbf{X}_i|\hat{\pmb{\xi}}_{Lk_d})\stackrel{\bigtriangleup}{=}L_{H_1}$, where $\hat{\pmb{\xi}}_{Lk_d}$ is the likelihood of sample $\{X_{i1},\cdots,X_{ip}\}_{i\in D_{k_d}}$. The likelihood under $H_0$ is $\underset{k_d\in D(k^*)}{\stackrel{}{\LARGE{\prod}}}\underset{i\in D_{k_d}}{\stackrel{}{\LARGE{\prod}}}f(\mathbf{X}_i|\hat{\pmb{\xi}}_{Lc(i)})\stackrel{\bigtriangleup}{=}L_{H_0}$, whereas $\hat{\pmb{\xi}}_{Lc(i)}$ is the likelihood of all sample $\{X_{i1},\cdots,X_{ip}\}$ with $i$ satisfying $c(i)=k^*$. From theory of hypothesis testing we can draw that 
		\[
		\log(L_{H_1})-\log(L_{H_0})=O_p(1)
		\]
		For every $\underset{k_d\in D(k^*)}{\stackrel{}{\LARGE{\prod}}}\underset{\substack{c(i)=k^*\\k_i=k_d}}{\stackrel{}{\LARGE{\prod}}}\alpha_{k_i} f(\mathbf{X}_i|\ddot{\tilde{\pmb{\xi}}}_{pk_i}')$, there is a particular testing formula like (40) corresponding to it, equipping with the property that its value is not larger than the corresponding $\log(L_{H_1})$, i.e. $\log(L_{H_0})+O_p(1)$. So by now we have attain the result that every term in $exp[F_{n,p}^{k^*}(\ddot{\tilde{\pmb{\theta}}}_p')]$ is bounded by $\underset{c(i)=k^*}{\stackrel{}{\LARGE{\prod}}}f(\mathbf{X}_i|\hat{\pmb{\xi}}_{Lc(i)})\cdot O_p(1)$, therefore we can update the approximation of $F_{n,p}(\mathbf{X}_1,\cdots,\mathbf{X}_n|\ddot{\tilde{\pmb{\theta}}}_p')$ in (39):
		\begin{equation}
		\begin{split}
		F_{n,p}(\mathbf{X}_1,\cdots,\mathbf{X}_n|\ddot{\tilde{\pmb{\theta}}}_p')=&\underset{k^*\in K^*}{\stackrel{}{\sum}}F_{n,p}^{k^*}(\ddot{\tilde{\pmb{\xi}}}_p')\\
		&\leq\underset{k^*\in K^*}{\stackrel{}{\sum}}\underset{c(i)=k^*}{\stackrel{}{\sum}}\underset{t=1}{\stackrel{p}{\sum}}\log[\phi(X_{it}|\hat{\pmb{\xi}}_{Lc(i)})]+O_p(1)
		\end{split}
		\end{equation}
		If we denote the set $\{\hat{\pmb{\xi}}_{Lc(i)}|c(i)=k^*,\forall k^*\in K^*\}\bigcup\{\pmb{\xi}_k^*,\forall k\in K^{*c}\}$ as $\hat{\pmb{\xi}}_{L_c}$ and replace $\pmb{\xi}_{K^*}^*$ in $\pmb{\theta}^*$ with it, we call the new formed $\pmb{\theta}^*$ as $\hat{\pmb{\theta}}_{L_c}$, thus
		\begin{equation}
		\begin{split}
		F_{n,p}(\mathbf{X}_1,\cdots,\mathbf{X}_n|\hat{\pmb{\theta}}_{L_c})&=\underset{k^*\in K^*}{\stackrel{}{\sum}}n_{k^*} \log(\alpha_{k^*}^*)+O_p(\frac{n}{e^p})\\
		&+\underset{k^*\in K^*}{\stackrel{}{\sum}}\underset{i\in C_{k^*}}{\stackrel{}{\sum}}\underset{t=1}{\stackrel{p}{\sum}}\log[\phi(X_{it}|\hat{\pmb{\xi}}_{Lc(i)})]
		\end{split}
		\end{equation}
		So we can reach that $F_{n,p}(\ddot{\tilde{\pmb{\theta}}}_p')\leq F_{n,p}(\hat{\pmb{\theta}}_{L_c})+O_p(1)$.
		
		Taking the between-cluster distance into account, $|\ddot{\tilde{\mu}}_{pk}'-\tau_k|=O_p(1)$ for $\forall k\in K_1\backslash K^*$, adding with lemma 2 can lead to 
		\begin{equation}
		G_n(\ddot{\tilde{\pmb{\xi}}}_p')-G_n(\hat{\pmb{\xi}}_{L_c})=-O_p(m(p))<0
		\end{equation}
		By now combining the facts we have reached in (41), (42) and (43) leads to $L_{n,p}(\ddot{\tilde{\pmb{\theta}}}_p')<L_{n,p}(\hat{\pmb{\xi}}_{L_c},\pmb{\alpha}^*)\\
		\leq L_{n,p}(\hat{\pmb{\xi}}_p,\hat{\pmb{\alpha}}_p)\ wpt\ 1$.
		
		With regard to condition (d), we separate it into 2 smaller cases: (d1) $K_1=K^*$ and (d2) $K_1\neq K^*$.
		
		Considering (d1) first, if $\ddot{\tilde{\pmb{\theta}}}_{pK_1}'\xrightarrow{p}\pmb{\theta}_{K^*}^*$, since $\ddot{\tilde{\pmb{\theta}}}_p'$ lies in $S$, there exists at least one $k\in K^{*c}$ such that $|\ddot{\tilde{\mu}}_{pk}'-\tau_k|>\epsilon$ for some $\epsilon>0$, resulting in $G_n(\ddot{\tilde{\pmb{\xi}}}_p')-G_n(\pmb{\xi}^*)=-O_p(m(p))<0$. On the other hand, if we replace $\ddot{\tilde{\pmb{\xi}}}_{pK_1^c}'$ in $\ddot{\tilde{\pmb{\theta}}}_p'$ by $\pmb{\xi}_{K_1^c}^*$ and denote the new formed $\ddot{\tilde{\pmb{\theta}}}_p'$ as $\ddot{\tilde{\pmb{\theta}}}_{pc^*}'$, it is obvious that $|G_n(\pmb{\theta}^*)-G_n(\ddot{\tilde{\pmb{\theta}}}_{pc^*}')|\ll O_p(m(p))$, what's more, by similar arguments as (42) we can show the outcome $F_{n,p}(\ddot{\tilde{\pmb{\theta}}}_p')-F_{n,p}(\ddot{\tilde{\pmb{\theta}}}_{pc^*}')= O_p(\frac{n}{e^p})$, thereby we have $L_{n,p}(\ddot{\tilde{\pmb{\xi}}}_p')<L_{n,p}(\ddot{\tilde{\pmb{\theta}}}_{pc^*}')\leq L_{n,p}(\hat{\pmb{\xi}}_p,\hat{\pmb{\alpha}}_p)\ wpt\ 1$; if $\ddot{\tilde{\pmb{\theta}}}_{pK_1}'$ doesn't converge in probability to $\pmb{\theta}_{K^*}^*$, since $\ddot{\tilde{\pmb{\theta}}}_p'$ lies in $V$, by some location permutation $\ddot{\tilde{\pmb{\theta}}}_{pK_1}'$ must converge in probability to $\pmb{\theta}_{K^*}^*$, we call $\ddot{\tilde{\pmb{\theta}}}_p'$ after such permutation as $\ddot{\tilde{\pmb{\theta}}}_{p\pi}'$. So by lemma 2, $G_n(\ddot{\tilde{\pmb{\theta}}}_p')-G_n(\ddot{\tilde{\pmb{\theta}}}_{p\pi}')=-O_p(m(p))$ can be obtained, again by similar arguments as (42), we can also reach $L_{n,p}(\ddot{\tilde{\pmb{\theta}}}_p')<L_{n,p}(\ddot{\tilde{\pmb{\theta}}}_{p\pi}')\leq L_{n,p}(\hat{\pmb{\xi}}_p,\hat{\pmb{\alpha}}_p)\ wpt\ 1$.
		
		Next we consider (d2), only by location permutation can $\ddot{\tilde{\pmb{\theta}}}_{pK_1}'$ converge in probability to $\pmb{\theta}_{K^*}^*$, we call the permutated $\ddot{\tilde{\pmb{\theta}}}_{p}'$ as $\ddot{\tilde{\pmb{\theta}}}_{p\pi}'$ too, thereby following lemma 2, we can attain $G_n(\ddot{\tilde{\pmb{\theta}}}_p')-G_n(\ddot{\tilde{\pmb{\theta}}}_{p\pi}')=-O_p(m(p))$, the claim of $F_{n,p}(\ddot{\tilde{\pmb{\theta}}}_p')-F_{n,p}(\ddot{\tilde{\pmb{\theta}}}_{p\pi}')=O_p(\frac{n}{e^p})$ can be achieved similar with (42) as well, i.e. under case (b2), $L_{n,p}(\ddot{\tilde{\pmb{\xi}}}_p')<L_{n,p}(\ddot{\tilde{\pmb{\theta}}}_{p\pi}')\leq L_{n,p}(\hat{\pmb{\xi}}_p,\hat{\pmb{\alpha}}_p)\ wpt\ 1$. By now we have completed the proof of (34) with $S$ replaced by $V_2^c$. 
		
		Last but not least, we consider the set $V_1^c$ and assume $\ddot{\tilde{\pmb{\theta}}}_p''=\mathop{argmax}\limits_{\pmb{\theta}\in V_1^c}L_{n,p}(\pmb{\xi},\pmb{\alpha})$, we will show $L_{n,p}(\ddot{\tilde{\pmb{\theta}}}_p'')<L_{n,p}(\hat{\pmb{\theta}}_p)$ under the following 2 scenarios:
		\begin{itemize}
			\item[(e)]$\exists\check{\pmb{\theta}}\in B_{\pmb{\xi}}\backslash\pmb{\xi}^*$, s.t. $\ddot{\tilde{\pmb{\xi}}}_p''\xrightarrow{p}\check{\pmb{\xi}}$ and $\exists\epsilon>0$, s.t. $|\ddot{\tilde{\pmb{\alpha}}}_p''-\check{\pmb{\alpha}}|>\epsilon$
			\item[(f)]$\ddot{\tilde{\pmb{\xi}}}_p''\xrightarrow{p}\pmb{\xi}^*$ and $\exists\epsilon>0$, s.t. $|\ddot{\tilde{\pmb{\alpha}}}_p''-\pmb{\alpha}^*|>\epsilon$
		\end{itemize}
		
		For scenario (e), it's natural to arrive that apart from different $\ddot{\tilde{\pmb{\alpha}}}_p''$, there is a $\ddot{\tilde{\pmb{\theta}}}_{pV}'$ from set $V$ in accordance with $\ddot{\tilde{\pmb{\theta}}}_p''$, i.e. $\ddot{\tilde{\pmb{\xi}}}_{pV}'=\ddot{\tilde{\pmb{\xi}}}_p''$, thereby similar arguments as (41) and (42) can be reached, which leads to $F_{n,p}(\ddot{\tilde{\pmb{\theta}}}_p'')\leq F_{n,p}(\hat{\pmb{\theta}}_{L_c})+O_p(1)$.  $G_n(\ddot{\tilde{\pmb{\xi}}}_p'')-G_n(\hat{\pmb{\xi}}_{L_c})=-O_p(m(p))<0$ can also be obtained as discussed in the setting on set $V$, therefore we have $L_{n,p}(\ddot{\tilde{\pmb{\xi}}}_p'')<L_{n,p}(\hat{\pmb{\xi}}_{L_c},\pmb{\alpha}^*)\leq L_{n,p}(\hat{\pmb{\xi}}_p,\hat{\pmb{\alpha}}_p)\ wpt\ 1$. 
		
		For scenario (f), $\exists\epsilon>0$, the location parameters $\mu_k^*$ for $k\in\{1,\cdots,5\}$ are separated with each other with distances larger than $\mathop{min}\limits_{k\in\{1,\cdots,4\}}d_k>2\zeta>\epsilon$, therefore we can reach a result similar as (21):
		\begin{equation}
		F_{n,p}(\ddot{\tilde{\pmb{\theta}}}_p'')=\underset{i=1}{\stackrel{n}{\sum}}\log(\ddot{\tilde{\alpha}}_{pc(i)}'')+\underset{i=1}{\stackrel{n}{\sum}}\underset{t=1}{\stackrel{p}{\sum}}\log[\phi(X_{it}|\ddot{\tilde{\mu}}_{pc(i)}'',(\ddot{\tilde{\sigma}}_{pc(i)}^{''})^2)]+O_p(\frac{n}{e^p})
		\end{equation}
		where $\underset{i=1}{\stackrel{n}{\sum}}\log(\ddot{\tilde{\alpha}}_{pc(i)}'')=\underset{k^*\in K^*}{\stackrel{}{\sum}}n_{k^*} \log(\ddot{\tilde{\alpha}}_{pk^*}'')$, which attains its maximum when $\log(\ddot{\tilde{\alpha}}_{pk^*}'')$ are arranged proportional to $n_k$ for $k^*\in K^*$, $n_k$ is the number of samples belonging to cluster $k$. Obviously under condition (f), $\underset{k^*\in K^*}{\stackrel{}{\sum}}n_{k^*} \log(\ddot{\tilde{\alpha}}_{pk^*}'')<\underset{k^*\in K^*}{\stackrel{}{\sum}}n_{k^*} \log(\alpha_{k^*}^*)$, furthermore we can get $F_{n,p}(\ddot{\tilde{\pmb{\xi}}}_p'',\ddot{\tilde{\pmb{\alpha}}}_p'')<F_{n,p}(\ddot{\tilde{\pmb{\xi}}}_p'',\pmb{\alpha}^*)$, hence  $L_{n,p}(\ddot{\tilde{\pmb{\xi}}}_p'',\ddot{\tilde{\pmb{\alpha}}}_p'')<L_{n,p}(\ddot{\tilde{\pmb{\xi}}}_p'',\pmb{\alpha}^*)\leq L_{n,p}(\hat{\pmb{\theta}}_p)\ wpt\ 1$, which completes the claim under condition (f).\\
		By far we have proved (34), which implies that for $\forall\epsilon>0$, $\forall\pmb{\theta}\in S$, when $p$ is large enough, there is a $\eta>0$, s.t. $L_{n,p}(\pmb{\theta})<L_{n,p}(\hat{\pmb{\theta}}_p)-\eta\ wpt\ 1$, i.e. the event $\{\pmb{\theta}\in S\}$ is contained in the event $\{L_{n,p}(\hat{\pmb{\theta}}_p)<L_{n,p}(\hat{\pmb{\theta}}_p)-\eta\}$ with probability tending to 1, whereas $P(L_{n,p}(\hat{\pmb{\theta}}_p)<L_{n,p}(\hat{\pmb{\theta}}_p)-\eta)=0$, resulting in $P(\pmb{\theta}\in S^c)\xrightarrow{}1$, which coincides with the consistency properties as we claimed in the theorem, so we have completed the full proof.
	\end{proof}
	\subsection{Proof of theorem 2.4}
	\begin{proof}
		We consider the case of the existence of all 5 clusters first and adopt symbol rules defined as in theorem 1, since in this setting bin size $p$ is a fixed constant, for convience we drop symbol $p$ in $F_{n,p}(\mathbf{X}_1,\cdots,\mathbf{X}_n|\pmb{\theta})=F_{n,p}(\pmb{\theta})$, resulting in $F_n(\pmb{\theta})$. For convinence we abbreviate "with probability tending to 1" as "wpt 1".\\
		As mentioned above, we define a set $B$ such that $\forall\pmb{\theta}\in B$, $h(\mathbf{x}|\pmb{\theta})$ is the true underlying distribution. If we define estimators $(\hat{\mu}_{k},\hat{\sigma}_{k}^{2},\hat{\alpha}_{k})$ maximize $\tilde{f}(\mathbf{x_{1}},\cdots,\mathbf{x_{n}}|\pmb{\theta})$ in (2) (i.e. maximize $L_n(\mathbf{x}_1,\cdots,\mathbf{x}_{n}|\pmb{\theta})$) as $\hat{\pmb{\theta}}_n$, we want to first claim that 
		\begin{equation}
		\forall\epsilon>0, \exists N>0, when\ n>N, \mathop{\sup}\limits_{\{\pmb{\theta}:|\pmb{\theta}-\pmb{\theta}^*|>\epsilon\}}L_n(\pmb{\theta})< L_n(\hat{\pmb{\theta}}_n),\ wpt\ 1
		\end{equation}
		To prove (45), we divide the set $\{\pmb{\theta}:|\pmb{\theta}-\pmb{\theta}^*|>\epsilon\}$ into 2 parts:\\
		$(\bigcup\limits_{\check{\pmb{\theta}}\in B{\backslash}\pmb{\theta}^*}\{\pmb{\theta}:|\pmb{\theta}-\check{\pmb{\theta}}|<\epsilon\})\bigcup(\{\pmb{\theta}:|\pmb{\theta}-\pmb{\theta}^*|>\epsilon\}\backslash\bigcup\limits_{\check{\pmb{\theta}}\in B{\backslash}\pmb{\theta}^*}\{\pmb{\theta}:|\pmb{\theta}-\check{\pmb{\theta}}|<\epsilon\})$.\\
		Let $D$ denote the set $\bigcup\limits_{\check{\pmb{\theta}}\in B{\backslash}\pmb{\theta}^*}\{\pmb{\theta}:|\pmb{\theta}-\check{\pmb{\theta}}|<\epsilon\}$, $D^c$ denote its complement in set $\{\pmb{\theta}:|\pmb{\theta}-\pmb{\theta}^*|>\epsilon\}$, then in short we have $\{\pmb{\theta}:|\pmb{\theta}-\pmb{\theta}^*|>\epsilon\}=D\cup D^c$. Next we focus on showing that 
		\begin{equation}
		\mathop{\sup}\limits_{\pmb{\theta}\in D}L_n(\pmb{\theta})>\mathop{\sup}\limits_{\pmb{\theta}\in D^c}L_n(\pmb{\theta}),\ wpt\ 1
		\end{equation}
		Suppose $\tilde{\pmb{\theta}}_n=\mathop{argmax}\limits_{\pmb{\theta}\in D}F_n(\pmb{\theta})$, $\tilde{\tilde{\pmb{\theta}}}_n=\mathop{argmax}\limits_{\pmb{\theta}\in D^c}L_n(\pmb{\theta})$, by the randomicity of $\epsilon$ and continuity of $h(\mathbf{x}|\pmb{\theta})$ in $\pmb{\theta}$, $h(\tilde{\pmb{\theta}}_n)\xrightarrow{p}h(\pmb{\theta})$ for $\pmb{\theta}\in B$; since $\overline{D^c}$ is a closed set, $L_n(\mathbf{x}_1,\cdots,\mathbf{x}_n|\pmb{\theta})$ is continuous in $\pmb{\theta}$, $L_n(\pmb{\theta})$ constrained in $D^c$ must attain its maximum in $\overline{D^c}$ for $\forall\epsilon$ fixed, we must have $h(\mathbf{x}|\tilde{\tilde{\pmb{\theta}}}_n)\nrightarrow h(\mathbf{x}|\pmb{\theta})$ in probability for $\forall\pmb{\theta}\in B$, so by law of large numbers we can obtain
		\begin{equation}
		\begin{split}
		&\frac{1}{n}[F_n(\tilde{\pmb{\theta}}_n)-F_n(\tilde{\tilde{\pmb{\theta}}}_n)]=\int \log\frac{h(\mathbf{x}|\tilde{\pmb{\theta}}_n)}{h(\mathbf{x}|\tilde{\tilde{\pmb{\theta}}}_n)}h(\mathbf{x}|\pmb{\theta}^*)d\mathbf{x}+o_p(1)\\
		&=\int \log\frac{h(\mathbf{x}|\pmb{\theta}^*)}{h(\mathbf{x}|\tilde{\tilde{\pmb{\theta}}}_n)}h(\mathbf{x}|\pmb{\theta}^*)d\mathbf{x}+o_p(1)+o_p(1)=O_p(1)>0
		\end{split}
		\end{equation}
		Combining with the fact that the order of $1/\sigma_{\tau k}^2$ is $O(m(n))$, which is less than $O(n)$, $\forall k=1,\cdots,5$, we can draw the conclusion that
		\begin{equation}
		\begin{split}
		\mathop{\sup}\limits_{\pmb{\theta}\in D}L_n(\pmb{\theta})-\mathop{\sup}\limits_{\pmb{\theta}\in D^c}L_n(\pmb{\theta})&\geq[F_n(\tilde{\pmb{\theta}}_n)-F_n(\tilde{\tilde{\pmb{\theta}}}_n)]+[G_n(\tilde{\pmb{\theta}}_n)-G_n(\tilde{\tilde{\pmb{\theta}}}_n)]\\
		&= O_p(n)-O_p(m(n))>0\ wpt\ 1
		\end{split}
		\end{equation}
		which means that (46) holds, so we can shrink the range of taking supreme of $L_n(\pmb{\theta})$ from $\{\pmb{\theta}:|\pmb{\theta}-\pmb{\theta}^*|>\epsilon\}$ to set $D$.
		
		Next we aim to show that $\mathop{\sup}\limits_{\pmb{\theta}\in D}L_n(\pmb{\theta})\leq L_n(\hat{\pmb{\theta}}_n)$ for $\forall\epsilon>0$ and $n$ sufficiently large. Suppose $\tilde{\pmb{\theta}}_n'=\mathop{argmax}\limits_{\pmb{\theta}\in D}L_n(\pmb{\theta})$, by the fact that $\tilde{\pmb{\theta}}_n'$ lies in $D$ we can conclude the location parameters $\tilde{\mu}_{n1}',\cdots,\tilde{\mu}_{n5}'$ aren't arranged in increasing order, hence only by a location permutation rule $\pi$ such that $\pi(\tilde{\pmb{\theta}}_n')\stackrel{\bigtriangleup}{=}\tilde{\pmb{\theta}}_n^{'\pi}$ can the location parameters  $\{\tilde{\mu}_{nk}^{'\pi}\}_{k=1}^5$ be arranged in increasing order, therefore by lemma 2 
		\begin{equation}
		G_n(\tilde{\pmb{\theta}}_n')-G_n(\tilde{\pmb{\theta}}_n^{'\pi})=-O_p(m(n))
		\end{equation}
		Furthermore, we can see obviously that any location permutation rule of $\tilde{\pmb{\theta}}_n'$ doesn't change the value of $F_n(\tilde{\pmb{\theta}}_n')$, i.e. $F_n(\tilde{\pmb{\theta}}_n')=F_n(\tilde{\pmb{\theta}}_n^{'\pi})$, combining the result we have reached in (49) leads to 
		\begin{equation}
		\begin{split}
		\mathop{\sup}\limits_{\pmb{\theta}\in D}L_n(\pmb{\theta})&=F_n(\tilde{\pmb{\theta}}_n')+G_n(\tilde{\pmb{\theta}}_n')=F_n(\tilde{\pmb{\theta}}_n^{'\pi})+G_n(\tilde{\pmb{\theta}}_n^{'\pi})-O_p(m(n))\\
		&< F_n(\tilde{\pmb{\theta}}_n^{'\pi})+G_n(\tilde{\pmb{\theta}}_n^{'\pi})=L_n(\tilde{\pmb{\theta}}_n^{'\pi})\leq L_n(\hat{\pmb{\theta}}_n)\ wpt\ 1
		\end{split}
		\end{equation}
		adding the fact in (48), we complete the proof of (45).
		
		With (45) holds, there exists for $\forall\epsilon>0$, $\forall\pmb{\theta}$ satisfying $|\pmb{\theta}-\pmb{\theta}^*|>\epsilon$, there is a $\eta>0$ such that $L_n(\pmb{\theta})<L_n(\hat{\pmb{\theta}}_n)-\eta$. Thus, the event $\{|\hat{\pmb{\theta}}_n-\pmb{\theta}^*|>\epsilon\}$ is contained in the event $\{L_n(\hat{\pmb{\theta}}_n)<L_n(\hat{\pmb{\theta}}_n)-\eta\}$ wpt 1, i.e.\\
		\begin{equation}
		P(|\hat{\pmb{\theta}}_n-\pmb{\theta}^*|>\epsilon)\leq P(L_n(\hat{\pmb{\theta}}_n)<L_n(\hat{\pmb{\theta}}_n)-\eta)=0
		\end{equation}
		which implies that $\hat{\pmb{\theta}}_n$ converges in probability to the true parameter $\pmb{\theta}^*$ when 5 clusters all exist.
		
		In the following, we concentrate on the proof under the circumstances of nonexistence of some cluster. For notation simplification, we follow some symbols described in theorem 1 and denote parameters only corresponding to set $K\subseteq\{1,\cdots,5\}$ in $\pmb{\theta}$ as $\pmb{\theta}_K$. $B$ is also defined as a set such that $\forall\pmb{\theta}\in B$, $h(\mathbf{x}|\pmb{\theta})$ is the true underlying distribution. Suppose the true exsiting clusters form a set as $K^*$, if we define a set $J$ as $\{\pmb{\theta}:\exists\epsilon>0,\ s.t.\ |\pmb{\theta}_{K^*}-\pmb{\theta}_{K^*}^*|>\epsilon\ or\ |\alpha_k|>\epsilon\ for\ \exists k\in K^{*c}\ or\ |\mu_k-\tau_k|>\epsilon\ for\ \exists k\in K^{*c}\}$, to accomplish our consistency's proof under this case, we want to first claim that
		\begin{equation}
		\forall\epsilon>0, \exists N>0, when\ n>N, \mathop{\sup}\limits_{\pmb{\theta}\in J}L_n(\pmb{\theta})< L_n(\hat{\pmb{\theta}}_n)\ wpt\ 1
		\end{equation}
		
		As in the case of exsitence of all clusters, we separate the set $J$ into 2 parts: $\{\pmb{\theta}:\{\bigcup\limits_{\check{\pmb{\theta}}\in B}|\pmb{\theta}-\check{\pmb{\theta}}|<\epsilon\}\bigcap J\}\stackrel{\bigtriangleup}{=}M$ and its complement in set $J$ denoted as $J\backslash M$. As long as we keep in mind that any element $\theta$ in set $M$ corresponds to a distribution $h(\mathbf{x}|\pmb{\theta})$ which has arbitrary close distance to the true one, whereas any element in set $J\backslash M$ doesn't. So by similar arguments as in the first circumstance, we can show the conclusion
		\begin{equation}
		\mathop{\sup}\limits_{\pmb{\theta}\in M}L_n(\pmb{\theta})>\mathop{\sup}\limits_{\pmb{\theta}\in J\backslash M}L_n(\pmb{\theta})\ wpt\ 1
		\end{equation}
		In the following we aim to show that $\mathop{\sup}\limits_{\pmb{\theta}\in M}L_n(\pmb{\theta})<L_n(\hat{\pmb{\theta}}_n)$. Suppose $\ddot{\tilde{\pmb{\theta}}}_n=\mathop{argmax}\limits_{\pmb{\theta}\in M}F_n(\pmb{\theta})$, and we define a set  $\Theta_5$ from which $\pmb{\theta}$ can take value as $\pmb{\theta}$ has no constraints apart from $h(\mathbf{x}|\pmb{\theta})$ is a mixture of 5 Gaussian distributions. By lemma 1 we know that the log likelihood under the overfitting setting is equal to the true log likelihood plus a random variable of order $O_p(1)$, so here we can attain the result
		\begin{equation}
		\mathop{\sup}\limits_{\pmb{\theta}\in M}F_n(\pmb{\theta})\leq\mathop{\sup}\limits_{\pmb{\theta}\in\Theta_5}F_n(\pmb{\theta})=F_n(\pmb{\theta}^*)+O_p(1)
		\end{equation}
		Next we aim to show $G_n(\pmb{\theta}^*)-G_n(\ddot{\tilde{\pmb{\theta}}}_n)\gg O_p(1)$, we will discuss it under 2 conditions:
		\begin{itemize}
			\item[(a)]$\exists\epsilon$, $s.t. |\ddot{\tilde{\alpha}}_{nk}|>\epsilon$ for $k\in K_1$, where $|K_1|>|K^*|$, here we use $"|\cdot|"$ to denote cardinality of a set.
			\item[(b)]$\forall\epsilon>0$, $\exists N>0$, when $n>N, |\ddot{\tilde{\alpha}}_{nk}|<\epsilon$ for $k\in K_1^c$, where $|K_1|=|K^*|.$
		\end{itemize}
		
		With regard to condition (a), noting that any $\pmb{\theta}\in M$ refers to a distribution close enough to the true one, therefore $\forall k_1\in K_1\backslash K^*$, $\exists k^*\in K^*$, s.t. $\ddot{\tilde{\mu}}_{nk_1}\xrightarrow{p}\mu_{k^*}^*$, so $\exists\eta$, s.t. $|\ddot{\tilde{\mu}}_{nk_1}-\mu_{k_1}^*|>\eta$, leading to $G_n(\pmb{\theta}^*)-G_n(\ddot{\tilde{\pmb{\theta}}}_n)=O_p(m(n))$.
		
		As for condition (b), by the fact that $\ddot{\tilde{\pmb{\theta}}}_n$ is an element of set $J$, it is not hard to deduce that only by permutation can $\ddot{\tilde{\mu}}_{nk}$ approximate $\mu_k^*$ in arbitrary close distance for $\forall k\in K^*$, this fact results in $G_n(\pmb{\theta}^*)-G_n(\ddot{\tilde{\pmb{\theta}}}_n)=O_p(m(n))$.\\ 
		Combining with the result in (54), we can reach the outcome that $\mathop{\sup}\limits_{\pmb{\theta}\in M}L_n(\pmb{\theta})<L_n(\pmb{\theta}^*)\leq L_n(\hat{\pmb{\theta}}_n)\ wpt\ 1$, adding the fact we have obtained in (53), we complete the proof of (52). With (52) holdes, there exists for $\forall\epsilon>0$, $\forall\pmb{\theta}\in J$, there is a $\eta>0$ such that $L_n(\pmb{\theta})<L_n(\hat{\pmb{\theta}}_n)-\eta\ wpt\ 1$, leading to the fact that the event $\{\pmb{\theta}\in J\}$ is contained in the event $\{L_n(\hat{\pmb{\theta}}_n)<L_n(\hat{\pmb{\theta}}_n)-\eta\}\ wpt\ 1$, i.e.
		\begin{equation}
		P(\pmb{\theta}\in J)\leq P(L_n(\hat{\pmb{\theta}})<L_n(\hat{\pmb{\theta}}_n)-\eta)=0
		\end{equation}
		$P(\pmb{\theta}\in J^c)=1$ represents what we stated in theorem 2 holds, i.e. $(\hat{\mu}_{k},\hat{\sigma}_{k}^{2},\hat{\alpha}_{k})\xrightarrow{p}(\mu_{k}^{*},(\sigma_{k}^{*})^2,\alpha_{k}^{*})$ for $\forall k\in K^*$, $\hat{\alpha}_k\xrightarrow{p}0$, $\hat{\mu}_{k}\xrightarrow{p}\tau_k$ for $\forall k\in K^{*c}$. So by now we have complete the proof of the theorem.
	\end{proof}
	\subsection{Proof of theorem 2.5}
	\begin{proof}
		We adopt the same proof routine as that of theorem 1. It's not hard to deduce that apart from 2 circumstances that we'll discuss in the following, consistency conclusions under other circumstances can be easily derived with the same skills as we described in theorem 1.
		
		Firstly, recall that when the 5 clusters all exist, we separate $H$ into 3 sets $H_1$, $H_2$ and $H_3$ to prove (16). More precisely, we divide $H_1$ into 2 parts $Q$ and $Q^c$, derivation of (18) is the same as before, but derivation of $\mathop{\sup}\limits_{\pmb{\xi}\in Q, \alpha_k>0}L_{n,p}(\pmb{\xi},\pmb{\alpha})<L_{n,p}(\hat{\pmb{\xi}}_p,\hat{\pmb{\alpha}}_p)$ has to be changed.
		
		Following the same symbols as theorem 1, we assume $\tilde{\pmb{\theta}}_p'=(\tilde{\pmb{\xi}}_p',\tilde{\pmb{\alpha}}_p')=\mathop{argmax}\limits_{\pmb{\xi}\in Q,\alpha_k>0}L_{n,p}(\pmb{\xi},\pmb{\alpha})$, by the defination of $Q$, the location parameters $\{\tilde{\mu}_{pk}'\}_{k=1}^5$ of $\tilde{\pmb{\xi}}_p'$ aren't arranged in increasing order, and there exists a location permutation rule $\pi_2$ such that $\pi_2(\tilde{\pmb{\xi}}_p')=\tilde{\pmb{\xi}}_p'^{\pi_2}\xrightarrow{p}\pmb{\xi}^*$, we further denote $\pi_2(\tilde{\pmb{\alpha}}_p')=\tilde{\pmb{\alpha}}_p'^{\pi_2}$, $\pi_2(\tilde{\pmb{\theta}}_p')=\tilde{\pmb{\theta}}_p'^{\pi_2}$. It is obvious that $F_{n,p}(\tilde{\pmb{\theta}}_p')=F_{n,p}(\tilde{\pmb{\theta}}_p'^{\pi_2})$, but because of the location parameters' arrangement in $\tilde{\pmb{\xi}}_p'$, by lemma 2 we have 
		\[
		G_n(\tilde{\pmb{\xi}}_p'^{\pi_2})-G_n(\tilde{\pmb{\xi}}_p')\geq O_p(m(np))>0
		\]
		i.e. $\mathop{\sup}\limits_{\pmb{\xi}\in Q,\alpha_k>0}L_{n,p}(\pmb{\xi},\pmb{\alpha})<L_{n,p}(\tilde{\pmb{\theta}}_p'^{\pi_2})\leq L_{n,p}(\hat{\pmb{\xi}}_p,\hat{\pmb{\alpha}}_p)$ can be reached. Proof of scenarios under $H_2$ and $H_3$ is parallel with that in theorem 1.
		
		Secondly, when there's absence of some cluster, we separate $S$ into 3 sets $V$, $V_1^c$ and $V_2^c$ to prove the correctness of (34). When discussing (34) on set $V$, we consider 2 sub-cases (c) and (d) successively, now in the setting where $n$, $p\rightarrow\infty$, we derive a new proof procedure under case (c).
		
		We still assume $\ddot{\tilde{\pmb{\theta}}}_p'=\mathop{argmax}\limits_{\pmb{\theta}\in V}L_{n,p}(\pmb{\xi},\pmb{\alpha})$, by the defination of $V$ and condition (c), $\forall k^*\in K^*$, there must exist at least one $k\in K_1$ such that $\ddot{\tilde{\pmb{\xi}}}_{pk}'\xrightarrow{p}\pmb{\xi}_{k^*}^*$, and we still use set $D(k^*)$ to denote those $k\in K_1$ with the property $\ddot{\tilde{\pmb{\xi}}}_{pk}'\xrightarrow{p}\pmb{\xi}_{k^*}^*$ holds. Under the setting of this theorem, (39) still can be reached. Now we focus on those $k^*$ with $D(k^*)>1$, by the fact that $|K_1|>|K^*|$, there exists such $k^*$. We pick up a specific term $\underset{c(i)=k^*}{\stackrel{}{\sum}}\log\Big{[}\underset{k\in D(k^*)}{\stackrel{}{\sum}}\alpha_k f(\mathbf{X}_i|\ddot{\tilde{\pmb{\xi}}}_{pk}')+O_p(\frac{1}{e^p})\Big{]}\stackrel{\bigtriangleup}{=}F_{n,p}^{k^*}(\ddot{\tilde{\pmb{\theta}}}_p')$ as described above too. For each sample $i$ belonging to cluster $k^*$, i.e. $c(i)=k^*$, we introduce $k_c^i=\mathop{argmax}\limits_{k\in D(k^*),c(i)=k^*}f(\mathbf{x}_i|\ddot{\tilde{\pmb{\xi}}}_{pk}')$, then we can attain
		\begin{equation}
		\log\Big{[}\underset{k\in D(k^*)}{\stackrel{}{\sum}}\alpha_k f(\mathbf{x}_i|\ddot{\tilde{\pmb{\xi}}}_{pk}')\Big{]}\leq \log\Big{[}\Big{(}\underset{k\in D(k^*)}{\stackrel{}{\sum}}\alpha_k\Big{)} f(\mathbf{x}_i|\ddot{\tilde{\pmb{\xi}}}_{pk_c^i}')\Big{]}
		\end{equation}
		Now we divide the set $\{i|c(i)=k^*\}$ by their label $k_c^i$ and we denote the set $\{i|k_c^i=k_c,c(i)=k^*\}$ as $C_{k_c}^{k^*}$ for $k_c\in D(k^*)$, thus for any realization $\{\mathbf{x}_1,\cdots,\mathbf{x}_n\}$ of $\{\mathbf{X}_1,\cdots,\mathbf{X}_n\}$, these facts can lead to 
		\begin{equation}
		F_{n,p}^{k^*}(\ddot{\tilde{\pmb{\theta}}}_p')\leq\underset{k_c\in D(k^*)}{\stackrel{}{\sum}}\underset{i\in C_{k_c}^{k^*}}{\stackrel{}{\sum}}\log\Big{[}\Big{(}\underset{k\in D(k^*)}{\stackrel{}{\sum}}\alpha_k\Big{)}f(\mathbf{x}_i|\ddot{\tilde{\pmb{\xi}}}_{pk_c}')+O(\frac{1}{e^{p}})\Big{]}
		\end{equation}
		With this in hand, we can also give a new bound of  $F_{n,p}^{k^*}(\ddot{\tilde{\pmb{\theta}}}_p')$ from a hypothesis testing viewpoint. Suppose samples from the same set $C_{k_c}^{k^*}$, $k_c\in D(k^*)$ follow the same normal distribution with parameters indexed by $\ddot{\tilde{\pmb{\xi}}}_{pk_c}'$, consider the following hypothesis testing:
		\begin{equation}
		\begin{split}
		& H_0:\ddot{\tilde{\pmb{\xi}}}_{pk_c}'\  are\ equivalent\  for\ all \ k_c\in D(k^*)\\
		& H_1: \ddot{\tilde{\pmb{\xi}}}_{pk_c}'\ are\ not\ all\ equivalent\ for\ k_c\in D(k^*)
		\end{split}
		\end{equation} 
		The likelihood under $H_1$ is $\underset{k_c\in D(k^*)}{\stackrel{}{\LARGE{\prod}}}\underset{i\in C_{k_c}^{k^*}}{\stackrel{}{\LARGE{\prod}}}f(\mathbf{X}_i|\hat{\pmb{\xi}}_{Lk_c})\stackrel{\bigtriangleup}{=}L_{H_1}$, where $\hat{\pmb{\xi}}_{Lk_c}$ is the likelihood of sample $\{X_{i1},\cdots,X_{ip}\}_{i\in C_{k_c}^{k^*}}$. The likelihood under $H_0$ is $\underset{k_c\in D(k^*)}{\stackrel{}{\LARGE{\prod}}}\underset{i\in C_{k_c}^{k^*}}{\stackrel{}{\LARGE{\prod}}}f(\mathbf{X}_i|\hat{\pmb{\xi}}_{Lc(i)})\stackrel{\bigtriangleup}{=}L_{H_0}$, whereas $\hat{\pmb{\xi}}_{Lc(i)}$ is the likelihood of all sample $\{X_{i1},\cdots,X_{ip}\}$ with $i$ satisfying $c(i)=k^*$. From theory of hypothesis testing we can draw that 
		\begin{equation}
		\log(L_{H_1})-\log(L_{H_0})=O_p(1)
		\end{equation}
		Furthermore, by the assumption that $n/{e^p}\rightarrow 0$, the rightside in (57) can be rewritten as $\underset{c(i)=k^*}{\stackrel{}{\sum}}\log\Big{(}\underset{k\in D(k^*)}{\stackrel{}{\sum}}\alpha_k\Big{)}+\underset{k_c\in D(k^*)}{\stackrel{}{\sum}}\underset{i\in C_{k_c}^{k^*}}{\stackrel{}{\sum}}\log[f(\mathbf{x}_i|\ddot{\tilde{\pmb{\xi}}}_{pk_c}')]+o(1)$, which is obviously less than $\underset{c(i)=k^*}{\stackrel{}{\sum}}\log\Big{(}\underset{k\in D(k^*)}{\stackrel{}{\sum}}\alpha_k\Big{)}+\log(L_{H_1})+o(1)$, here we refer to $L_{H_1}$ as a realization corresponding to those $\{\mathbf{x}_1,\cdots,\mathbf{x}_n\}$ we have mentioned in (57). Combining with (59) we can update a new bound of $F_{n,p}^{k^*}(\ddot{\tilde{\pmb{\theta}}}_p')$:
		\[
		F_{n,p}^{k^*}(\ddot{\tilde{\pmb{\theta}}}_p')\leq\underset{c(i)=k^*}{\stackrel{}{\sum}}\log\Big{(}\underset{k\in D(k^*)}{\stackrel{}{\sum}}\alpha_k\Big{)}+\underset{k_c\in D(k^*)}{\stackrel{}{\sum}}\underset{i\in C_{k_c}^{k^*}}{\stackrel{}{\sum}}\log[f(\mathbf{x}_i|\hat{\pmb{\xi}}_{Lc(i)})]+O(1)
		\]
		If we introduce $\tilde{\pmb{\xi}}_{L}$ with the property that $\forall k^*\in K^*$, $\tilde{\pmb{\xi}}_{Lk^*}=\hat{\pmb{\xi}}_{Lk^*}=\hat{\pmb{\xi}}_{Lc(i)}$ if $c(i)=k^*$, $\forall k\notin K^*$, $\tilde{\pmb{\xi}}_{Lk}=\tau_k$. Then  a new approximation of $F_{n,p}(\ddot{\tilde{\pmb{\theta}}}_p')$ can be obtained
		\begin{equation}
		\begin{split}
		F_{n,p}(\ddot{\tilde{\pmb{\theta}}}_p')&=\underset{k^*\in K^*}{\stackrel{}{\sum}}F_{n,p}^{k^*}(\ddot{\tilde{\pmb{\xi}}}_p')
		\leq\underset{k^*\in K^*}{\stackrel{}{\sum}}\underset{c(i)=k^*}{\stackrel{}{\sum}}\log\Big{(}\underset{k\in D(k^*)}{\stackrel{}{\sum}}\alpha_k\Big{)}+O(1)\\
		&+\underset{k^*\in K^*}{\stackrel{}{\sum}}\underset{k_c\in D(k^*)}{\stackrel{}{\sum}}\underset{i\in C_{k_c}^{k^*}}{\stackrel{}{\sum}}\log[f(\mathbf{x}_i|\hat{\pmb{\xi}}_{Lc(i)})]=F_{n,p}(\tilde{\pmb{\xi}}_L,\ddot{\tilde{\pmb{\alpha}}}_p')+O(1)
		\end{split}
		\end{equation}
		By the property of condition (c), for $\forall k\in K_1\backslash K^*$, $|\ddot{\tilde{\mu}}_{pk}'-\tau_k|=O_p(1)$, adding with lemma 2 can lead to
		\begin{equation}
		G_n(\ddot{\tilde{\pmb{\xi}}}_p')-G_n(\tilde{\pmb{\xi}}_L)=-O_p(m(np))<0
		\end{equation}
		The facts we have reached in (60) and (61) direct to the conclusion we want: $L_{n,p}(\ddot{\tilde{\pmb{\theta}}}_p')<L_{n,p}(\tilde{\pmb{\xi}}_L,\ddot{\tilde{\pmb{\alpha}}}_{n,p}')\leq L_{n,p}(\hat{\pmb{\xi}}_p,\hat{\pmb{\alpha}}_p)\ wpt\ 1$
	\end{proof}
	\subsection{Proof of theorem 2.7}
	\begin{proof}
		The whole process in proving theorem 2 can be completely adpoted in this theorem, so we omit here.
	\end{proof}
	\subsection{Proof of theorem 2.8}
	\begin{proof}
		We adopt similar proof and parameterization skills with that of theorem 2 by Cai et al. (2011). We show only that the type I error tends to 0, proof of the type II error can proceed similarly.
		
		We introduce a substituted testing problem: 
		\begin{equation}
		\begin{split}
		&H_0^{(s)}: \bar{X}_i\sim N(0,\frac{\sigma_1^2}{p}),\quad 1\leq i\leq n\\
		&H_1^{(s)}:\bar{X}_i\sim (1-\epsilon)N(0,\frac{\sigma_1^2}{p})+\epsilon N(A,\frac{\sigma_2^2}{p}),\quad 1\leq i\leq n
		\end{split}
		\end{equation}
		where $\bar{X}_i=\underset{t=1}{\stackrel{p}{\sum}}X_{it}/p$ is the average of the original $p$-dimensional observation $\mathbf{X}_i$. We first conclude that to show the type I error of the likelihood ratio test (LRT) with regard to testing probelm (15) tends to 0, it is sufficient to move on to testing probelm (62). Concretely, it's sufficient to show that under the null hypothesis $H_0^{(s)}$ of (62), the log likelihood ratio $\log(LR_n)=\log(LR_n(\bar{X}_i,\cdots,\bar{X}_n;\epsilon,\sigma_1^2,\sigma_2^2,A,p))\rightarrow-\infty$ in propability as $n$, $p\rightarrow\infty$, $\frac{\log(n)}{p}\rightarrow 0$.
		
		Suppose the log likelihood ratio of testing probelm (15) is $\log(\stackrel{\sim}{LR}_n)=\log(\stackrel{\sim}{LR}_n(\mathbf{X}_i,\cdots,\mathbf{X}_n;\\
		\epsilon,\sigma_1^2,\sigma_2^2,A,p))$, the corresponding rejection region $W_1=\{\log(\stackrel{\sim}{LR}_n(\mathbf{X}_i,\cdots,\mathbf{X}_n;\epsilon,\sigma_1^2,\sigma_2^2,A,p))>0\}$. Analogously, rejection region of the LRT with repect to (62) is $W_2=\{\log(LR_n(\bar{X}_i,\cdots,\bar{X}_n;\\
		\epsilon,\sigma_1^2,\sigma_2^2,A,p))>0\}$. So we can have 2 testing rules related to (15). 
		$$ \varphi_1(\mathbf{X}_1,\cdots,\mathbf{X}_n)=\left\{ \begin{array}{rcl} 1 & & {(\mathbf{X}_1,\cdots,\mathbf{X}_n)\in W_1}\\ 0 & & {(\mathbf{X}_1,\cdots,\mathbf{X}_n)\in W_1^c} \end{array} \right. $$
		$$ \varphi_2(\bar{X}_1,\cdots,\bar{X}_n)=\left\{ \begin{array}{rcl} 1 & & {(\bar{X}_1,\cdots,\bar{X}_n)\in W_2}\\ 0 & & {(\bar{X}_1,\cdots,\bar{X}_n)\in W_2^c} \end{array} \right. $$
		where $W_1^c$ and $W_2^c$ are the complementary set of $W_1$ and $W_2$ separately.
		
		The Neyman-Pearson lemma tells us the optimality of LRT in this test setting, which implies that the type I error of the testing rule $\varphi_1(\mathbf{X}_1,\cdots,\mathbf{X}_n)$ is smaller than that of the testing rule $\varphi_2(\bar{X}_1,\cdots,\bar{X}_n)$. So we have 
		\begin{equation}
		\begin{split}
		P((\mathbf{X}_1,\cdots,\mathbf{X}_n)\in W_1|H_0)&\leq P((\bar{X}_1,\cdots,\bar{X}_n)\in W_2|H_0)\\
		&=P((\bar{X}_1,\cdots,\bar{X}_n)\in W_2|H_0^{(s)})
		\end{split}
		\end{equation}
		The last term in (63) tends to 0 if the claim "$\log(LR_n)\stackrel{p}{\rightarrow}-\infty$ under $H_0^{(s)}$ of (58) as $n$, $p\rightarrow\infty$, $\frac{\log(n)}{p}\rightarrow 0$" holds.
		
		To show $\log(LR_n)\rightarrow-\infty$ in probability, we  conclude that it is sufficient to show that, as $n\rightarrow\infty$, $p\rightarrow\infty$ and $\frac{\log(n)}{p}\rightarrow 0$
		\begin{equation}
		E[\log(LR_n)]\rightarrow -\infty
		\end{equation}
		\begin{equation}
		\frac{var\{\log(LR_n)\}}{(E[\log(LR_n)])^2}\rightarrow 0
		\end{equation}
		Suppose (64) and (65) hold as $n\rightarrow\infty$, $p\rightarrow\infty$ and $\frac{\log(n)}{p}\rightarrow 0$, then $\forall\epsilon>0$, $\forall M>0$, by Chebyshev's inequality we have
		\begin{equation}
		\begin{split}
		P(\log(LR_n)>-M)&=P(\log(LR_n)-E[\log(LR_n)]>-M-E[\log(LR_n)])\\
		&\leq P(|\log(LR_n)-E[\log(LR_n)]|>|-M-E[\log(LR_n)]|)\\
		&\leq\frac{var\{\log(LR_n)\}}{(E[\log(LR_n)]+M)^2}\leq\frac{var\{\log(LR_n)\}}{\frac{1}{2}(E[\log(LR_n)])^2}<\epsilon
		\end{split}
		\end{equation}
		this coincides with $\log(LR_n)\rightarrow-\infty$ in probability.\\
		In the following, we prove (64) and (65) separately. Before that, we introduce some notations first.
		
		Let $\epsilon=\epsilon_n=n^{-\beta}$ for a fixed parameter $0<\beta<1$, $A=A_n(r)=\sqrt{2r\log(n)}$ for $0<r<1$. Denote the density of $N(\mu,\sigma^2)$ by $\phi_{\sigma}(x-\mu)$ and let
		\begin{equation}
		g_{n}(x)=g_n(x;p,r,\sigma_1^2,\sigma_2^2)=\frac{\phi_{\frac{\sigma_2}{\sqrt{p}}}(x-A)}{\phi_{\frac{\sigma_1}{\sqrt{p}}}(x)}
		\end{equation}
		so we have
		\[
		\log(LR_n(\bar{X}_1,\cdots,\bar{X}_n))=\underset{i=1}{\stackrel{n}{\sum}}\log(LR_n(\bar{X}_i))=\underset{i=1}{\stackrel{n}{\sum}}\log(1-\epsilon+\epsilon g_n(\bar{X}_i))
		\]
		Noticing the fact $E(g_n(\bar{X}_i))=1$, we can draw the following approximation:
		\begin{equation}
		\begin{split}
		&E[\log(1-\frac{\epsilon}{1+\epsilon g_n(\bar{X}_i)})+\epsilon g_n(\bar{X_i})]=E[\epsilon g_n(\bar{X}_i)-\frac{\epsilon}{1+\epsilon g_n(\bar{X}_i)}]+O(\epsilon^2)\\
		&=\epsilon E[1-\frac{1}{1+\epsilon g_n(\bar{X}_i)}]+O(\epsilon^2)=\epsilon^2 E[\frac{g_n(\bar{X}_i)}{1+\epsilon g_n(\bar{X}_i)}]+O(\epsilon^2)=O(\epsilon^2)
		\end{split}
		\end{equation}
		If we let $f_n(x)=\log\{1+\epsilon g_n(x)\}-\epsilon g_n(x)$, then $E[\log(LR_n)]$ can be expressed as 
		\begin{equation}
		\begin{split}
		E[\log(LR_n)]&=nE[\log(1-\epsilon+\epsilon g_n(\bar{X}))]\\
		&=nE[\log\{1+\epsilon g_n(\bar{X})\}-\epsilon g_n(\bar{X})]+nO(\epsilon^2)\\
		&=nE[f_n(\bar{X})]+o(1)
		\end{split}
		\end{equation}
		Since there is a constant $c_1\in(0,1)$ and a generic constant $C>0$ s.t. $\log(1+x)\leq c_1x$ for $x>1$ and $\log(1+x)-x\leq-Cx^2$ for $x\leq1$, there exists a constant $C>0$ s.t.
		\begin{equation}
		\begin{split}
		E[f_n(\bar{X})]&\leq-C\{\epsilon E[g_n(\bar{X})\mathbf{1}_{\{\epsilon g_n(\bar{X})>1\}}]+\epsilon^2E[g_n^2(\bar{X})\mathbf{1}_{\{\epsilon g_n(\bar{X})\leq1\}}]\}\\
		&\leq-C\epsilon E[g_n(\bar{X})\mathbf{1}_{\{\epsilon g_n(\bar{X})>1\}}]
		\end{split}
		\end{equation}
		So if we can prove that
		\begin{equation}
		n\epsilon E[g_n(\bar{X})\mathbf{1}_{\{\epsilon g_n(\bar{x})>1\}}]\rightarrow\infty
		\end{equation}
		(64) will hold.\\
		We consider (71) in 2 cases: (a)$\sigma_2^2<\sigma_1^2$ and (b)$\sigma_2^2>\sigma_1^2$. Pay attention to (a) first, by elementary calculation, we can get\\
		\begin{equation}
		g_n(\bar{x})=\frac{\sigma_1}{\sigma_2}exp\{\frac{p(\sigma_2^2-\sigma_1^2)(\bar{x}+\frac{A\sigma_1^2}{\sigma_2^2-\sigma_1^2})^2}{2\sigma_1^2\sigma_2^2}\}exp\{-\frac{A^2p}{2(\sigma_2^2-\sigma_1^2)}\}
		\end{equation}
		Recalling notations $\epsilon=n^{-\beta}$ and $A=\sqrt{2r\log(n)}$, For simplicity, we introduce $D=\frac{2\sigma_1^2\sigma_2^2}{p(\sigma_2^2-\sigma_1^2)}\log(\frac{\sigma_2}{\sigma_1})+\frac{2\sigma_1^2\sigma_2^2}{p(\sigma_2^2-\sigma_1^2)}(\beta+\frac{rp}{\sigma_2^2-\sigma_1^2})\log(n)$, then under case (a), $\bar{x}$ satisfying $\{\epsilon g_n(\bar{x})>1\}$ can be expressed as
		\begin{equation}
		\begin{split} \bar{x}&\in(-\sqrt{D}-\frac{\sigma_1^2\sqrt{2r\log(n)}}{\sigma_2^2-\sigma_1^2},\sqrt{D}-\frac{\sigma_1^2\sqrt{2rlog(n)}}{\sigma_2^2-\sigma_1^2})\stackrel{\bigtriangleup}{=}(A_1,A_2)\\
		&\subseteq(\frac{\sigma_1}{\sigma_1+\sigma_2}\sqrt{2r\log(n)},\frac{\sigma_1}{\sigma_1-\sigma_2}\sqrt{2r\log(n)})\stackrel{\bigtriangleup}{=}(B_1,B_2)
		\end{split}
		\end{equation}
		here $A_1=-\sqrt{D}-\frac{\sigma_1^2\sqrt{2r\log(n)}}{\sigma_2^2-\sigma_1^2}$, $A_2=\sqrt{D}-\frac{\sigma_1^2\sqrt{2r\log(n)}}{\sigma_2^2-\sigma_1^2}$, $B_1=\frac{\sigma_1}{\sigma_1+\sigma_2}\sqrt{2r\log(n)}$, $B_2=\frac{\sigma_1}{\sigma_1-\sigma_2}\sqrt{2r\log(n)}$. The "$\subseteq$" comes from dropping 2 terms $\frac{2\sigma_1^2\sigma_2^2}{p(\sigma_2^2-\sigma_1^2)}\log(\frac{\sigma_2}{\sigma_1})$ and $\frac{2\sigma_1^2\sigma_2^2}{p(\sigma_2^2-\sigma_1^2)}\beta \log(n)$ in $D$, and denote the left term as $D_1=\frac{2r\sigma_1^2\sigma_2^2}{(\sigma_1^2-\sigma_2^2)^2}\log(n)$, we expect when $n$ and $p\rightarrow\infty$, $\frac{\log(n)}{p}\rightarrow 0$, these 2 terms play negligible role compared with the left term $\frac{2r\sigma_1^2\sigma_2^2}{(\sigma_2^2-\sigma_1^2)^2}\log(n)$, i.e.
		\begin{equation}
		\frac{E[g_n(\bar{X})\mathbf{1}_{\{\bar{X}\in(B_1,A_1)\}}]+E[g_n(\bar{X})\mathbf{1}_{\{\bar{X}\in(A_2,B_2)\}}]}{E[g_n(\bar{X})\mathbf{1}_{\{\bar{X}\in(B_1,B_2)\}}]}\rightarrow 0
		\end{equation}
		If (74) holds, then 
		\begin{align*}
		n\epsilon E[g_n(\bar{X})\mathbf{1}_{\{\epsilon g_n(\bar{x})>1\}}]&=n\epsilon E[g_n(\bar{X})\mathbf{1}_{\{\bar{x}\in(A_1,A_2)\}}]\\
		&=n\epsilon E[g_n(\bar{X})\mathbf{1}_{\{\bar{x}\in(B_1,B_2)\}}](1-o(1))
		\end{align*}
		that is if we relax the interval into a more tractable one, the asymptotic outcome doesn't change at all. We prove (74) later and now we move our sight back to (71) on the case $\sigma_2^2<\sigma_1^2$. In this setting, we want to obtain 
		\begin{equation}
		n\epsilon E[g_n(\bar{X})\mathbf{1}_{\{\bar{x}\in(B_1,B_2)\}}]\rightarrow\infty
		\end{equation}
		By Mill's ratio (Wasserman, 2006),
		\begin{align*}
		\bar{\Phi}(\sqrt{2q\log(n)})=PL(n)n^{-q}
		\end{align*}
		here $\bar{\Phi}=1-\Phi$ is the survival function of $N(0,1)$, $PL(n)>0$ is a generic poly-log-term satisfying $\lim \limits_{n\rightarrow\infty}\{PL(n)n^{-\delta}\}=0$ and $\lim \limits_{n\rightarrow\infty}\{PL(n)n^{\delta}\}=0$ for any $\delta>0$. So we can get
		\begin{equation}
		\begin{split}
		n\epsilon E[g_n(\bar{X})\mathbf{1}_{\{\bar{x}\in(B_1,B_2)\}}]&=n^{1-\beta}\int_{B_1}^{B_2}\phi_{\frac{\sigma_2}{\sqrt{p}}}(x-\sqrt{2r\log(n)})dx\\
		&=n^{1-\beta}[1-PL_1(n)n^{-\frac{rp}{(\sigma_1+\sigma_2)^2}}-PL_2(n)n^{-\frac{rp}{(\sigma_1-\sigma_2)^2}}]
		\end{split}
		\end{equation}
		where $PL_1(n)$ and $PL_2(n)$ are 2 generic ploy-log-terms, it is easy to see that when $n, p\rightarrow\infty$, $1-PL_1(n)n^{-\frac{rp}{(\sigma_1+\sigma_2)^2}}-PL_2(n)n^{-\frac{rp}{(\sigma_1-\sigma_2)^2}}\rightarrow 1$, resulting in (75) holds. Furthermore, if (74) holds, we will complete the proof of (71) in the $\sigma_2^2<\sigma_1^2$ case. Notice that 
		\begin{align*}
		E[g_n(\bar{X})\mathbf{1}_{\{\bar{X}\in(A_2,B_2)\}}]=\Phi(\frac{B_2-\sqrt{2r\log(n)}}{\sigma_2/\sqrt{p}})-\Phi(\frac{A_2-\sqrt{2r\log(n)}}{\sigma_2/\sqrt{p}}),\\
		\frac{B_2-A_2}{\sigma_2/\sqrt{p}}=\frac{\sqrt{D_1}-\sqrt{D}}{\sigma_2/\sqrt{p}}=\frac{\sqrt{p}}{\sigma_2}\frac{D_1-D}{\sqrt{D_1}+\sqrt{D}}=\frac{2\sigma_1^2\sigma_2[\log(\frac{\sigma_2}{\sigma_1})+\beta \log(n)]}{\sqrt{p}(\sigma_1^2-\sigma_2^2)(\sqrt{D_1}+\sqrt{D})}.
		\end{align*}
		Since $\sqrt{D}$ and $\sqrt{D_1}$ are all bounded away from $0$,
		\begin{align*} \frac{2\beta\sigma_1^2\sigma_2\log(n)}{\sqrt{p}(\sigma_1^2-\sigma_2^2)(\sqrt{D_1}+\sqrt{D})}\leq \frac{2\beta\sigma_1^2\sigma_2\log(n)}{\sqrt{p}(\sigma_1^2-\sigma_2^2)\sqrt{D_1}}=C_1\sqrt{\frac{\log(n)}{p}}\rightarrow 0
		\end{align*}
		here $C_1$ is a constant having no relationship with $p$ or $n$. By far we can get $\frac{B_2-A_2}{\sigma_2/\sqrt{p}}\rightarrow 0$, i.e. $E[g_n(\bar{X})\mathbf{1}_{\{\bar{X}\in(A_2,B_2)\}}]\rightarrow 0$, proof of the left part $E[g_n(\bar{X})\mathbf{1}_{\{\bar{X}\in(B_1,A_1)\}}]\rightarrow 0$ can be attained in a similar way. Combining the fact we have reached in (76), we have finished the proof of (74).\\
		Next we will focus on (71) under case (b), in this setting, $\bar{x}$ satisfying $\{\epsilon g_n(\bar{x})>1\}$ can be expressed as 
		\begin{equation}
		\begin{split}
		\bar{x}&\in(-\infty,-\sqrt{D}-\frac{\sigma_1^2\sqrt{2r\log(n)}}{\sigma_2^2-\sigma_1^2})\cup(\sqrt{D}-\frac{\sigma_1^2\sqrt{2r\log(n)}}{\sigma_2^2-\sigma_1^2},+\infty)\\
		&\stackrel{\bigtriangleup}{=}(-\infty,A_1)\cup(A_2,+\infty)\\
		&\subseteq(-\infty,\frac{\sigma_1}{\sigma_1-\sigma_2}\sqrt{2r\log(n)})\cup(\frac{\sigma_1}{\sigma_1+\sigma_2}\sqrt{2r\log(n)},+\infty)\\
		&\stackrel{\bigtriangleup}{=}(-\infty,B_2)\cup(B_1,+\infty)
		\end{split}
		\end{equation}
		where the "$\subseteq$" also comes from changing $D$ into $D_1$. With similar proof skills as above, we can also get under case (b), 
		\begin{align*}
		\begin{split}
		n\epsilon E[g_n(\bar{X})\mathbf{1}_{\{\epsilon g_n(\bar{x})>1\}}]&=n\epsilon E[g_n(\bar{X})\mathbf{1}_{\{x\in(-\infty,A_1)\cup(A_2,+\infty)\}}]\\
		&=n\epsilon E[g_n(\bar{X})\mathbf{1}_{\{x\in(-\infty,B_2)\cup(B_1,+\infty)\}}](1-o(1))
		\end{split}
		\end{align*}
		So when $\sigma_2^2>\sigma_1^2$, if we can attain
		\begin{equation}
		n\epsilon E[g_n(\bar{X})\mathbf{1}_{\{x\in(-\infty,B_2)\cup(B_1,+\infty)\}}]\rightarrow\infty
		\end{equation}
		(64) will be fulfilled completely.
		
		By Mill's ratio and some calculations, 
		\begin{equation}
		\begin{split}
		&n\epsilon E[g_n(\bar{X})\mathbf{1}_{\{x\in(-\infty,B_2)\cup(B_1,+\infty)\}}]\\
		&=n^{1-\beta}[\int_{-\infty}^{B_2}\phi_{\frac{\sigma_2}{p}}(x-\sqrt{2r\log(n)})dx+\int_{B_1}^{+\infty}\phi_{\frac{\sigma_2}{p}}(x-\sqrt{2r\log(n)})dx]\\
		&=n^{1-\beta}(1-PL_3(n)n^{-\frac{rp(2\sigma_1-\sigma_2)^2}{\sigma_2^2(\sigma_1-\sigma_2)^2}}-PL_4(n)n^{-\frac{rp}{(\sigma_1+\sigma_2)^2}})
		\end{split}
		\end{equation}
		where $PL_3(n)$ and $PL_4(n)$ are also 2 generic poly-log-terms, when $n,p\rightarrow\infty$, $(1-PL_3(n)n^{-\frac{rp(2\sigma_1-\sigma_2)^2}{\sigma_2^2(\sigma_1-\sigma_2)^2}}-PL_4(n)n^{-\frac{rp}{(\sigma_1+\sigma_2)^2}})\rightarrow1$, resulting in (78) holds. By far we have completed (64), next we show assumption (65).
		
		If we can show that there exists a constant $C>0$ s.t.
		\begin{equation}
		E[\log^2(LR_n(\bar{X}_i))]\leq -CE[\log(LR_n(\bar{X}_i))]
		\end{equation}
		we will have 
		\begin{align*}
		&\frac{var\{\log(LR_n)\}}{(E[\log(LR_n)])^2}=\frac{n(E[\log^2(LR_n(\bar{X}_i))]-(E[\log(LR_n(\bar{X}_i))])^2)}{n^2(E[\log(LR_n(\bar{X}_i))])^2}\\
		&\leq-\frac{1}{n}(1+\frac{C}{E[\log(LR_n(\bar{X}_i))]})=-\frac{1}{n}-\frac{C}{E[\log(LR_n)]}\rightarrow 0
		\end{align*}
		So it is sufficient to prove (80) holds. Since for all $x$, $\log^2(1-\frac{\epsilon}{1+\epsilon g_n(x)})\leq[\frac{\epsilon}{1+\epsilon g_n(x)}]^2\leq \epsilon^2$, we can get
		\begin{align*}
		\log^2\{1-\epsilon+\epsilon g_n(x)\}&=[\log\{1-\frac{\epsilon}{1+\epsilon g_n(x)}\}+\log\{1+\epsilon g_n(x)\}]^2\\
		&\leq2[\log^2\{1-\frac{\epsilon}{1+\epsilon g_n(x)}\}+\log^2\{1+\epsilon g_n(x)\}]\\
		&\leq 2[\epsilon^2+\log^2\{1+\epsilon g_n(x)\}]
		\end{align*}
		resulting in 
		\begin{align*}
		E[\log^2(LR_n(\bar{X}))]\leq E[\log^2\{1+\epsilon g_n(\bar{X})\}]+o(\frac{1}{n})
		\end{align*}
		Furthermore, $\log(1+x)<C\sqrt{x}$ for $x>1$ and $\log(1+x)<x$ for $x>0$,
		\begin{equation}
		E[\log^2\{1+\epsilon g_n(\bar{X})\}]\leq C\{\epsilon E[g_n(\bar{X})\mathbf{1}_{\{\epsilon g_n(\bar{X})>1\}}]+\epsilon^2 E[g_n^2(\bar{X})\mathbf{1}_{\{\epsilon g_n(\bar{X}\leq1)\}}]\}
		\end{equation}
		From (69) and (70) we have 
		\begin{equation}
		\begin{split}
		E[\log(LR_n(\bar{X}))]&=E[f_n(\bar{X})]+o(\frac{1}{n})\\
		&\leq-C\{\epsilon E[g_n(\bar{X})\mathbf{1}_{\{\epsilon g_n(\bar{X})>1\}}]+\epsilon^2E[g_n^2(\bar{X})\mathbf{1}_{\{\epsilon g_n(\bar{X})\leq1\}}]\}
		\end{split}
		\end{equation}
		Combining (81), (82) and comparing with (80), we now reached assumption (65). So the full proof of the theorem is completed.\qedhere
	\end{proof}
	
	\subsection{Proof of theorem 2.9}
	\begin{proof}
		(a) We adpopt the same symbols as described in section 2.6, for simplicity we drop case's symbol "b" and suppose sample size of case is $n$. 
		
		Firstly, we recall some notations. $\hat{\pmb{\theta}}_{b-(b+1)}$, $\hat{\pmb{\theta}}_b$ and $\hat{\pmb{\theta}}_{b+1}$ are estimators such that\\ $\tilde{f}(\mathbf{X}_{1,b-(b+1)},\cdots,\mathbf{X}_{n,b-(b+1)}|\pmb{\theta}_{b-(b+1)})$, $\tilde{f}(\mathbf{X}_{1,b},\cdots,\mathbf{X}_{n,b}|\pmb{\theta}_{b})$ and $\tilde{f}(\mathbf{X}_{1,b+1},\cdots,\mathbf{X}_{n,b+1}|\pmb{\theta}_{b+1})$ attain maximum value, $\hat{\pmb{\theta}}_{b-(b+1)}^{(1)}$ and $\hat{\pmb{\theta}}_{b-(b+1)}^{(2)}$ are separated from $\hat{\pmb{\theta}}_{b-(b+1)}$, corresponding to parameters in the $b$-th bin and ($b+1$)-th bin separately. Concretely, if bin $b$ and $b+1$ are merged, $\mathbf{X}_{i,b-(b+1)}\sim\underset{k=1}{\stackrel{5}{\sum}}\alpha_{b-(b+1),k}N(\pmb{\mu}_{b-(b+1),k},(\pmb{\sigma}_{b-(b+1),k})^2)$, the conditional distribution of $\mathbf{X}_{i,b+1}|\mathbf{X}_{i,b}$ can be expressed as 
		\[
		q(\mathbf{X}_{i,b+1}|\mathbf{X}_{i,b},\pmb{\theta}_{b-(b+1)})=\frac{\underset{k=1}{\stackrel{5}{\sum}}\alpha_{b-(b+1),k}f(\mathbf{X}_{i,b-(b+1)}|\pmb{\mu}_{b-(b+1),k},(\pmb{\sigma}_{b-(b+1),k})^2))}{\underset{k=1}{\stackrel{5}{\sum}}\alpha_{b-(b+1),k}f(\mathbf{X}_{i,b}|\pmb{\mu}_{b-(b+1),k}^{(1)},(\pmb{\sigma}_{b-(b+1),k}^{(1)})^2))}
		\]
		where the denominator is the marginal distribution of $\mathbf{X}_{i,b}$. 
		
		If bin $b$ and $b+1$ can't be merged, the joint distribution of $\mathbf{X}_{i,b}$ and $\mathbf{X}_{i,b+1}$ is \\
		\[
		\begin{split}
        \mathbf{X}_{i,b-(b+1)}&\sim \gamma\Big{[}\underset{k=1}{\stackrel{5}{\sum}}\alpha_{b-(b+1),k}N(\pmb{\mu}_{b-(b+1),k},(\pmb{\sigma}_{b-(b+1),k})^2)\Big{]}\\
        &+(1-\gamma)\Big{[}\underset{\mbox{\tiny$\begin{array}{c}
        		k_2=1\\
        		k_2\neq k_1\\\end{array}$} }{\stackrel{5}{\sum}}\underset{k_1=1}{\stackrel{5}{\sum}}\alpha_{k_1,k_2}N((\pmb{\mu}_{b,k_1},\pmb{\mu}_{b+1,k_2}),((\pmb{\sigma}_{b,k_1})^2,(\pmb{\sigma}_{b+1,k_2})^2))\Big{]}		
        \end{split}	
		\]		
		where $\gamma$ is the proportion of samples which have  the same CN state between bin $b$ and $b+1$, whereas $1-\gamma$ is the proportion of heterogeneous samples between these 2 bins, the 2 bins can be merged if and only if $\lambda=0$. $\alpha_{k_1,k_2}$ is the proportion of heterogeneous samples which have CN state $k_1$ on bin $b$ and have CN state $k_2$ on bin $b+1$. Since we didn't assume batch effect within a CN state, $\pmb{\mu}_{b-(b+1),k}^{(1)}=\pmb{\mu}_{b,k_1}$, $(\pmb{\sigma}_{b-(b+1),k}^{(1)})^2=(\pmb{\sigma}_{b,k_1})^2$ for $k_1=k$; $\pmb{\mu}_{b-(b+1),k}^{(2)}=\pmb{\mu}_{b+1,k_2}$, $(\pmb{\sigma}_{b-(b+1),k}^{(2)})^2=(\pmb{\sigma}_{b+1,k_2})^2$ for $k_2=k$. Under this situation, we denote $\mathbf{X}_{i,b-(b+1)}\sim H(\mathbf{X}_{i,b-(b+1)}|\pmb{\kappa},\lambda)$, $\mathbf{X}_{i,b+1}|\mathbf{X}_{i,b}\sim \tilde{H}(\mathbf{X}_{i,b+1}|\mathbf{X}_{i,b},\pmb{\kappa},\lambda)$ for simplicity, where $\pmb{\kappa}$ denotes all the parameters in the distribution except $\lambda$.
		
		We further assume $\tilde{\pmb{\theta}}_b=\mathop{argmax}\limits_{\pmb{\theta}_b}\underset{i=1}{\stackrel{n}{\sum}}\log[h(\mathbf{X}_{i,b}|\pmb{\theta}_b)]$, where $h(\mathbf{X}_{i,b}|\pmb{\theta}_b)=\underset{i=1}{\stackrel{n}{\sum}}\alpha_{b,k}f(\mathbf{X}_{i,b}|\pmb{\mu}_{b,k},(\pmb{\sigma}_{b,k})^2)$. $(\tilde{\pmb{\kappa}},\tilde{\lambda})=\mathop{argmax}\limits_{\pmb{\kappa},\lambda}\underset{i=1}{\stackrel{n}{\sum}}\log[\tilde{H}(\mathbf{X}_{i,b+1}|\mathbf{X}_{i,b},\pmb{\kappa},\lambda)]$, $(\hat{\pmb{\kappa}},\hat{\lambda})=\mathop{argmax}\limits_{\pmb{\kappa},\lambda}\underset{i=1}{\stackrel{n}{\sum}}\log[H(\mathbf{X}_{i,b-(b+1)}|\pmb{\kappa},\lambda)]$, concretely, we suppose  $\tilde{\pmb{\kappa}}=(\alpha_{b-(b+1),k}^{\tilde{\pmb{\kappa}}},\pmb{\mu}_{b-(b+1),k}^{\tilde{\pmb{\kappa}}},(\pmb{\sigma}_{b-(b+1),k}^{\tilde{\pmb{\kappa}}})^2,\alpha_{k_1,k_2}^{\tilde{\pmb{\kappa}}})$ for $k=1,\cdots,5; k_1=1,\cdots,5,k_2=1,\cdots,5,k_1\neq k_2$.
		
	    If we consider the following test:
		\[
		\begin{split}
		&H_0^{(1)}: \tilde{\alpha}_{b,k}=\alpha_{b-(b+1),k}^{\tilde{\pmb{\kappa}}}+\underset{\mbox{\tiny$\begin{array}{c}
				k_2=1\\
				k_2\neq k\\\end{array}$} }{\stackrel{5}{\sum}}\alpha_{k,k_2}^{\tilde{\pmb{\kappa}}}, \; \tilde{\pmb{\mu}}_{b,k}=\pmb{\mu}_{b-(b+1),k}^{\tilde{\pmb{\kappa}}},\; (\tilde{\pmb{\sigma}}_{b,k})^2=(\pmb{\sigma}_{b-(b+1),k}^{\tilde{\pmb{\kappa}}})^2,  \; 1\leq k\leq 5\\
		&H_1^{(2)}:\tilde{\alpha}_{b,k}\neq\alpha_{b-(b+1),k}^{\tilde{\pmb{\kappa}}}+\underset{\mbox{\tiny$\begin{array}{c}
				k_2=1\\
				k_2\neq k\\\end{array}$} }{\stackrel{5}{\sum}}\alpha_{k,k_2}^{\tilde{\pmb{\kappa}}}\; or \; \tilde{\pmb{\mu}}_{b,k}=\pmb{\mu}_{b-(b+1),k}^{\tilde{\pmb{\kappa}}}\; or \; (\tilde{\pmb{\sigma}}_{b,k})^2=(\pmb{\sigma}_{b-(b+1),k}^{\tilde{\pmb{\kappa}}})^2,  \; \exists k\in\{1,\cdots,5\}
		\end{split}
		\]
		Under the assumption that $\mathbf{X}_{i,b-(b+1)}\sim H(\mathbf{X}_{i,b-(b+1)}|\pmb{\kappa},\lambda)$, which satisfies conditions in $H_0$, the maximum likelihood is $\underset{i=1}{\stackrel{n}{\sum}}\log[H(\mathbf{X}_{i,b-(b+1)}|\hat{\pmb{\kappa}},\hat{\lambda})]$. From properties of likelihood ratio test we can conclude that the difference between $H_0$ and $H_1$ is $O_p(1)$, concretely,
		\begin{equation}
		\underset{i=1}{\stackrel{n}{\sum}}\log[h(\mathbf{X}_{i,b}|\tilde{\pmb{\theta}}_b)]+\underset{i=1}{\stackrel{n}{\sum}}\log[\tilde{H}(\mathbf{X}_{i,b+1}|\mathbf{X}_{i,b},\tilde{\pmb{\kappa}},\tilde{\lambda})]-\underset{i=1}{\stackrel{n}{\sum}}\log[H(\mathbf{X}_{i,b-(b+1)}|\hat{\pmb{\kappa}},\hat{\lambda})]=O_p(1)
		\end{equation}
		
		Under the new hypothesis framework:
		\[H_0^{(2)}:\lambda=0,\;H_1^{(2)}:\lambda\neq 0 \]
		Suppose $\tilde{\tilde{\pmb{\theta}}}_{b-(b+1)}=\underset{i=1}{\stackrel{n}{\sum}}\log[h(\mathbf{X}_{i,b-(b+1)}|\pmb{\theta}_{b-(b+1)})]$, if bin $b$ and $b+1$ can be merged together, which is equivalent to $H_0^{(2)}$ is true, the difference of maximum likelihood under $H_0^{(2)}$ and $H_1^{(2)}$ is $O_p(1)$:
		\begin{equation}
        \underset{i=1}{\stackrel{n}{\sum}}\log[h(\mathbf{X}_{i,b-(b+1)}|\tilde{\tilde{\pmb{\theta}}}_{b-(b+1)})]-\underset{i=1}{\stackrel{n}{\sum}}\log[H(\mathbf{X}_{i,b-(b+1)}|\hat{\pmb{\kappa}},\hat{\lambda})]=O_p(1)		
		\end{equation}
		since $\pmb{\hat{\theta}}_{b-(b+1)}=\mathop{argmax}\limits_{\pmb{\theta}_{b-(b+1)}}\tilde{f}(\mathbf{X}_{1,b-(b+1)},\cdots,\mathbf{X}_{n,b-(b+1)})$, we have
		\begin{equation}
		\underset{i=1}{\stackrel{n}{\sum}}\log[h(\mathbf{X}_{i,b-(b+1)}|\tilde{\tilde{\pmb{\theta}}}_{b-(b+1)})]-\underset{i=1}{\stackrel{n}{\sum}}\log[h(\mathbf{X}_{i,b-(b+1)}|\hat{\pmb{\theta}}_{b-(b+1)})]<O_p(m(np))
		\end{equation}
		here $m(np)$ is any increasing function of $n,p$ with order less than $O(np)$, and $p$ can be length of the merged bin $b-(b+1)$. Similarly,
		\begin{equation}
		0\leq\underset{i=1}{\stackrel{n}{\sum}}\log[h(\mathbf{X}_{i,b}|\tilde{\pmb{\theta}}_{b})]-\underset{i=1}{\stackrel{n}{\sum}}\log[h(\mathbf{X}_{i,b}|\hat{\pmb{\theta}}_{b})]<O_p(m(np))
		\end{equation}
		Combing (83), (84) and (85) we can get
		\begin{equation}
		\underset{i=1}{\stackrel{n}{\sum}}\log[h(\mathbf{X}_{i,b}|\tilde{\pmb{\theta}}_{b})]+\underset{i=1}{\stackrel{n}{\sum}}\log[\tilde{H}(\mathbf{X}_{i,b+1}|\mathbf{X}_{i,b},\tilde{\pmb{\kappa}},\tilde{\lambda})]<\underset{i=1}{\stackrel{n}{\sum}}\log[h(\mathbf{X}_{i,b-(b+1)}|\hat{\pmb{\theta}}_{b-(b+1)})]+O_p(m(np))
		\end{equation}
        by the fact that 
        \[
        \underset{i=1}{\stackrel{n}{\sum}}\log[h(\mathbf{X}_{i,b-(b+1)}|\hat{\pmb{\theta}}_{b-(b+1)})]=\underset{i=1}{\stackrel{n}{\sum}}\log [q(\mathbf{X}_{i,b+1}|\mathbf{X}_{i,b},\hat{\pmb{\theta}}_{b-(b+1)})]+\underset{i=1}{\stackrel{n}{\sum}}\log[h(\mathbf{X}_{i,b}|\hat{\pmb{\theta}}_{b-(b+1)}^{(1)})]
        \]		
        (87) can be reformulated as
        \begin{equation}
        \begin{split}
        &\underset{i=1}{\stackrel{n}{\sum}}\log[h(\mathbf{X}_{i,b}|\tilde{\pmb{\theta}}_{b})]-\underset{i=1}{\stackrel{n}{\sum}}\log[h(\mathbf{X}_{i,b}|\hat{\pmb{\theta}}_{b-(b+1)}^{(1)})]\\
        <&\underset{i=1}{\stackrel{n}{\sum}}\log [q(\mathbf{X}_{i,b+1}|\mathbf{X}_{i,b},\hat{\pmb{\theta}}_{b-(b+1)})]-\underset{i=1}{\stackrel{n}{\sum}}\log[\tilde{H}(\mathbf{X}_{i,b+1}|\mathbf{X}_{i,b},\tilde{\pmb{\kappa}},\tilde{\lambda})]+O_p(m(np))\\
        \leq&O_p(m(np))
        \end{split}
        \end{equation}
        Combining (86) and (88) we can reach the conclusion:
        \[
        \underset{i=1}{\stackrel{n}{\sum}}\log[h(\mathbf{X}_{i,b}|\hat{\pmb{\theta}}_b)]-\underset{i=1}{\stackrel{n}{\sum}}\log[h(\mathbf{X}_{i,b}|\hat{\pmb{\theta}}_{b-(b+1)}^{(1)})]\leq O_p(m(np))
        \]        
		
		By far we have completed circumstance of case sample, for control sample, the proof skill is completely the same, so the order claimed in the theorem can be reached.
		
		(b) The idea is completely the same as proposition 1. We follow symbols in (a), $\hat{\pmb{\theta}}_b$ and $\hat{\pmb{\theta}}_{b+1}$ are estimators of bin $b$ and $b+1$ before merging, $\hat{\pmb{\theta}}_{b-(b+1)}$ are estimators after merging. So value of $M_{b,b+1}$ is denoted as:
		\[
		\begin{split}
		M_{b,b+1}=&\underset{i=1}{\stackrel{n}{\sum}}\log\big{[}h(\mathbf{X}_{i,b}|\hat{\pmb{\theta}}_b)\big{]}-\underset{i=1}{\stackrel{n}{\sum}}\log\big{[}h(\mathbf{X}_{i,b}|\hat{\pmb{\theta}}_{b-(b+1)}^{(1)})\big{]}\\
		&+\underset{i=1}{\stackrel{n}{\sum}}\log\big{[}h(\mathbf{X}_{i,b+1}|\hat{\pmb{\theta}}_{b+1})\big{]}-\underset{i=1}{\stackrel{n}{\sum}}\log\big{[}h(\mathbf{X}_{i,b+1}|\hat{\pmb{\theta}}_{b-(b+1)}^{(2)})\big{]}
		\end{split}
		\]
		By law of large numbers and consistency of $\hat{\pmb{\theta}}_b$, $\hat{\pmb{\theta}}_{b+1}$, i.e. $\hat{\pmb{\theta}}_{b}\xrightarrow{p}\pmb{\theta}_{b}$, $\hat{\pmb{\theta}}_{b+1}\xrightarrow{p}\pmb{\theta}_{b+1}$, we have
		\[
		\frac{1}{n}\underset{i=1}{\stackrel{n}{\sum}}\Big{\{}\log\big{[}h(\mathbf{X}_{i,b}|\hat{\pmb{\theta}}_b)\big{]}-\log\big{[}h(\mathbf{X}_{i,b}|\hat{\pmb{\theta}}_{b-(b+1)}^{(1)})\big{]}\Big{\}}=\int \log \frac{h(\mathbf{X}_b|\pmb{\theta}_b)}{h(\mathbf{X}_b|\hat{\pmb{\theta}}_{b-(b+1)}^{(1)})}h(\mathbf{X}_b|\pmb{\theta}_{b})d\mathbf{X}_b+o_p(1)
		\]
		\[
		\begin{split}
		&\frac{1}{n}\underset{i=1}{\stackrel{n}{\sum}}\Big{\{}\log\big{[}h(\mathbf{X}_{i,b+1}|\hat{\pmb{\theta}}_{b+1})\big{]}-\log\big{[}h(\mathbf{X}_{i,b+1}|\hat{\pmb{\theta}}_{b-(b+1)}^{(2)})\big{]}\Big{\}}\\
		=&\int \log \frac{h(\mathbf{X}_{b+1}|\pmb{\theta}_{b+1})}{h(\mathbf{X}_{b+1}|\hat{\pmb{\theta}}_{b-(b+1)}^{(2)})}h(\mathbf{X}_{b+1}|\pmb{\theta}_{b+1})d\mathbf{X_{b+1}}+o_p(1)
		\end{split}
		\]
		Since $\exists c_b>0$,$\exists k\in\{1,\cdots,5\}$, s.t. $|\alpha_{bk}-\alpha_{(b+1),k}|>c_b$, there must be $|\alpha_{b,k}-\alpha_{b-(b+1),k}|>\frac{c_b}{2}$ or $|\alpha_{b+1,k}-\alpha_{b-(b+1),k}|>\frac{c_b}{2}$, resulting in one of the 2 K-L divergence above attain order $O_p(1)$, i.e. the order of $M_{b,b+1}$ is $O_p(n)$
	\end{proof}
	
	\subsection{Proof of lemma 2.2}
	\begin{proof}
		For convinence we introduce the symbol $\pi(\{1,\cdots,5\})=\{\pi(1),\cdots,\pi(5)\}$, $\mu_k^*-\tau_k=\zeta_k$ with $|\zeta_k|<\zeta$ for $k=1,\cdots,5$, and assume $e_k=\zeta_{k+1}-\zeta_k$ for $k\in\{1,\cdots,4\}$, then by the setting in lemma 2, $|e_k|<\mathop{min}\limits_{k\in\{1,\cdots,4\}}d_k$ for $k=1,\cdots,4$. Since the prior variance $\sigma_{\tau k}^2$ for $k\in\{1,\cdots,5\}$ are the same, (7) is equivalent to 
		\begin{align*}
		\underset{k=1}{\stackrel{5}{\sum}}(\mu_k^*-\tau_k)^2<\underset{k=1}{\stackrel{5}{\sum}}(\mu_{\pi(k)}^*-\tau_k)^2
		\end{align*}
		which is also the same as 
		\begin{equation}
		\begin{split}
		\underset{k=1}{\stackrel{5}{\sum}}(\tau_k+\zeta_k)\tau_k>\underset{k=1}{\stackrel{5}{\sum}}(\tau_{\pi(k)}+\zeta_{\pi(k)})\tau_k\\
		i.e.\ \underset{k=1}{\stackrel{5}{\sum}}(\tau_k-\tau_{\pi(k)})\tau_k>\underset{k=1}{\stackrel{5}{\sum}}(\zeta_{\pi(k)}-\zeta_k)\tau_k
		\end{split}
		\end{equation}
		if we introduce $d_k$ and $e_k$ to substitute $\tau_k-\tau_{\pi(k)}$ and $\zeta_{\pi(k)}-\zeta_k$, we will have
		\begin{equation}
		\underset{k=1}{\stackrel{5}{\sum}}[\underset{j=k}{\stackrel{\pi(k)-1}{\sum}}d_j(-1)^{\mathbf{1}_{\{\pi(k)>k\}}}]\tau_k>\underset{k=1}{\stackrel{5}{\sum}}[\underset{j=k}{\stackrel{\pi(k)-1}{\sum}}-e_j(-1)^{\mathbf{1}_{\{\pi(k)>k\}}}]\tau_k
		\end{equation}
		noticing that $\tau_k=\tau_1+\underset{l=1}{\stackrel{k-1}{\sum}}d_l$ and $\underset{k=1}{\stackrel{5}{\sum}}\tau_k^2>\underset{k=1}{\stackrel{5}{\sum}}\tau_k\tau_{\pi(k)}$ for any nonidentical location permutation rule $\pi$, which is because the arrangement of $\{\tau_k\}_{k=1}^5$ is in increasing order, then $\underset{k=1}{\stackrel{5}{\sum}}(\tau_k-\tau_{\pi(k)})\tau_k$ is the summation of some $d_{k_1}d_{k_2}$ for $k_1,k_2\in\{1,\cdots,5\}$, from this, there must exists 5 functions of $\tau_1,\cdots,\tau_5$ which can be expressed as $v_k(\tau_1,\cdots,\tau_5)>0$ for $k=1,\cdots,5$ such that 
		\begin{equation}
		\underset{k=1}{\stackrel{5}{\sum}}[\underset{j=k}{\stackrel{\pi(k)-1}{\sum}}d_j(-1)^{\mathbf{1}_{\{\pi(k)>k\}}}]\tau_k=\underset{k=1}{\stackrel{5}{\sum}}d_k v_k(\tau_1,\cdots,\tau_5)
		\end{equation}
		Since the structures of the two sides of inequality in (90) are the same, we can also get a similar result as (91)
		\begin{equation}
		\underset{k=1}{\stackrel{5}{\sum}}[\underset{j=k}{\stackrel{\pi(k)-1}{\sum}}-e_j(-1)^{\mathbf{1}_{\{\pi(k)>k\}}}]\tau_k=\underset{k=1}{\stackrel{5}{\sum}}-e_k v_k(\tau_1,\cdots,\tau_5)
		\end{equation}
		Combining the fact that $|e_j|<\mathop{min}\limits_{k\in\{1,\cdots,4\}}d_k$ and $v_k(\tau_1,\cdots,\tau_5)>0$ for $k=1,\cdots,5$,\\
		$\underset{k=1}{\stackrel{5}{\sum}}-e_k v_k(\tau_1,\cdots,\tau_5)<\underset{k=1}{\stackrel{5}{\sum}}d_k v_k(\tau_1,\cdots,\tau_5)$, which implies (89) and (90) hold, that is what we aim to prove.
	\end{proof}
	\subsection{Proof of proposition 2.6}
	\begin{proof}
		In our proof we drop the bin's index $b$ for simplicity. Suppose case samples $\mathbf{X}_i\sim\underset{k=1}{\stackrel{5}{\sum}}\alpha_{k}^d f(\mathbf{X}|\pmb{\mu}_k^{d},(\pmb{\sigma}_k^d)^2)$, control samples $\mathbf{Y}_j\sim\underset{k=1}{\stackrel{5}{\sum}}\alpha_k^c f(\mathbf{Y}|\pmb{\mu}_k^c,(\pmb{\sigma}_k^c)^2)$, where $f(\mathbf{x}|\pmb{\mu}_k^{d},(\pmb{\sigma}_k^d)^2)$ is Gaussian distribution with mean $\pmb{\mu}_k^{d}$ and covariance matrix $Diag((\pmb{\sigma}_k^d)^2)$ valued on $\mathbf{x}$, interpretation of $f(\mathbf{y}|\pmb{\mu}_k^c,(\pmb{\sigma}_k^c)^2)$  is similar. We further suppose estimators of case and control used to calculate likelihood under $H_1$ are $(\hat{\alpha}_k^d,\hat{\pmb{\mu}}_k^d,(\hat{\pmb{\sigma}}_k^d)^2)_{k=1}^5\stackrel{\bigtriangleup}{=}\hat{\pmb{\theta}}_{H_1}^d$ and $(\hat{\alpha}_k^c,\hat{\pmb{\mu}}_k^c,(\hat{\pmb{\sigma}}_k^c)^2)_{k=1}^5\stackrel{\bigtriangleup}{=}\hat{\pmb{\theta}}_{H_1}^c$, which can be obtained by maximizing (9) and control's analogous form, these parameters have consistency by theorem 4, i.e. $\hat{\pmb{\theta}}_{H_1}^d\xrightarrow{p}\pmb{\theta}_{H_1}^d$, $\hat{\pmb{\theta}}_{H_1}^c\xrightarrow{p}\pmb{\theta}_{H_1}^c$. If estimators of proportion parameters under $H_0$ are denoted as $\hat{\alpha}_{kH_0}$, $k=1,\cdots,5$, and the overall parameters of case and control under $H_0$ are $(\hat{\alpha}_{kH_0},\hat{\pmb{\mu}}_k^d,(\hat{\pmb{\sigma}}_k^d)^2)_{k=1}^5\stackrel{\bigtriangleup}{=}\hat{\pmb{\theta}}_{H_0}^d$ and $(\hat{\alpha}_{kH_0},\hat{\pmb{\mu}}_k^c,(\hat{\pmb{\sigma}}_k^c)^2)_{k=1}^5\stackrel{\bigtriangleup}{=}\hat{\pmb{\theta}}_{H_0}^c$, we can write the likelihood ratio value as:
		\[
		\Lambda=\underset{i=1}{\stackrel{N_1}{\sum}}\log\big{[}h(\mathbf{X}_i|\hat{\pmb{\theta}}_{H_1}^d)\big{]}-\underset{i=1}{\stackrel{N_1}{\sum}}\log\big{[}h(\mathbf{X}_i|\hat{\pmb{\theta}}_{H_0}^d)\big{]}+\underset{j=1}{\stackrel{N_2}{\sum}}\log\big{[}h(\mathbf{Y}_j|\hat{\pmb{\theta}}_{H_1}^c)\big{]}-\underset{j=1}{\stackrel{N_2}{\sum}}\log\big{[}h(\mathbf{Y}_j|\hat{\pmb{\theta}}_{H_0}^c)\big{]}
		\]
		where $h(\mathbf{X}|\hat{\pmb{\theta}}_{H_1}^d)$ is the mixture Gaussian density function $\underset{k=1}{\stackrel{5}{\sum}}\hat{\alpha}_k^d f(\mathbf{X}|\hat{\pmb{\mu}}_k^d,(\hat{\pmb{\sigma}}_k^d)^2)$, other 3 functions $h$ have similar interpretation.
		
		By law of large numbers and consistency of $\hat{\pmb{\theta}}_{H_1}^d$, $\hat{\pmb{\theta}}_{H_1}^c$, we have
		\begin{equation}
		\begin{split}
		\frac{1}{N_1}\underset{i=1}{\stackrel{N_1}{\sum}}\Big{\{}\log\big{[}h(\mathbf{X}_i|\hat{\pmb{\theta}}_{H_1}^d)\big{]}-\log\big{[}h(\mathbf{X}_i|\hat{\pmb{\theta}}_{H_0}^d)\big{]}\Big{\}}=\int \log \frac{h(\mathbf{X}|\pmb{\theta}_{H_1}^d)}{h(\mathbf{X}|\hat{\pmb{\theta}}_{H_0}^d)}h(\mathbf{X}|\pmb{\theta}_{H_1}^d)d\mathbf{X}+o_p(1)\\
		\frac{1}{N_2}\underset{j=1}{\stackrel{N_2}{\sum}}\Big{\{}\log\big{[}h(\mathbf{Y}_j|\hat{\pmb{\theta}}_{H_1}^c)\big{]}-\log\big{[}h(\mathbf{Y}_j|\hat{\pmb{\theta}}_{H_0}^c)\big{]}\Big{\}}=\int \log \frac{h(\mathbf{Y}|\pmb{\theta}_{H_1}^c)}{h(\mathbf{Y}|\hat{\pmb{\theta}}_{H_0}^c)}h(\mathbf{Y}|\pmb{\theta}_{H_1}^c)d\mathbf{Y}+o_p(1)
		\end{split}
		\end{equation}
		
		By the assumption that $\exists c_b>0$, $\exists k\in\{1,\cdots,5\}$, s.t. $|\alpha_k^d-\alpha_k^c|>c_b$, we can conclude $|\alpha_k^d-\hat{\alpha}_{kH_0}|>\frac{c_b}{2}$ or $|\alpha_k^c-\hat{\alpha}_{kH_0}|>\frac{c_b}{2}$, leading to one of the 2 K-L divergences in (87) attain order $O_p(1)$, so the likelihood ratio value $\Lambda=O_p(N_1)$ or $\Lambda=O_p(N_2)$, which means that the type II error of our test tends to 0.
	\end{proof}

\end{document}